\documentclass[aps,superscriptaddress,nofootinbib,eqsecnum,tightenlines,floatfix]{revtex4}
\usepackage{epsfig,rotating}
\usepackage{amsmath,amssymb}
\usepackage{dsfont}
\numberwithin{equation}{section}
\usepackage{graphics}
\usepackage{rotate}
\usepackage{psfrag}
\begin{document}
\newcommand{\D}{\displaystyle} 
\newcommand{\T}{\textstyle} 
\newcommand{\SC}{\scriptstyle} 
\newcommand{\SSC}{\scriptscriptstyle} 
\newcommand{\be}{\begin{equation}}
\newcommand{\ee}{\end{equation}}
\newcommand{\avg}[1]{\langle #1 \rangle}
\newcommand{\vx}{{\boldsymbol{x}}}
\newcommand{\rv}{{\boldsymbol{r}}}
\newcommand{\vq}{\ensuremath{\vec{q}}}
\newcommand{\pv}{\ensuremath{\vec{p}}}
\newcommand{\bp}{\ensuremath{\mathbf{p}}}
\newcommand{\bq}{\ensuremath{\mathbf{q}}}
\newcommand{\br}{\ensuremath{\mathbf{r}}}
\def\AJ{{\it Astron. J.} }
\def\ARAA{{\it Annual Rev. of Astron. \& Astrophys.} }
\def\ApJ{{\it Astrophys. J.} }
\def\ApJL{{\it Astrophys. J. Letters} }
\def\ApJS{{\it Astrophys. J. Suppl.} }
\def\ApP{{\it Astropart. Phys.} }
\def\AA{{\it Astron. \& Astroph.} }
\def\AAR{{\it Astron. \& Astroph. Rev.} }
\def\AAL{{\it Astron. \& Astroph. Letters} }
\def\AASu{{\it Astron. \& Astroph. Suppl.} }
\def\AN{{\it Astron. Nachr.} }
\def\IJMP{{\it Int. J. of Mod. Phys.} }
\def\JGR{{\it Journ. of Geophys. Res.}}
\def\JHEP{{\it Journ. of High En. Phys.} }
\def\JPhG{{\it Journ. of Physics} {\bf G} }
\def\MNRAS{{\it Month. Not. Roy. Astr. Soc.} }
\def\Nature{{\it Nature} }
\def\NewAR{{\it New Astron. Rev.} }
\def\NJPh{{\it New Journ. of Phys.} }
\def\PASP{{\it Publ. Astron. Soc. Pac.} }
\def\PhFl{{\it Phys. of Fluids} }
\def\PLB{{\it Phys. Lett.}{\bf B} }
\def\PhysRep{{\it Phys. Rep.} }
\def\PR{{\it Phys. Rev.} }
\def\PRD{{\it Phys. Rev.} {\bf D} }
\def\PRL{{\it Phys. Rev. Letters} }
\def\RMP{{\it Rev. Mod. Phys.} }
\def\Science{{\it Science} }
\def\ZfA{{\it Zeitschr. f{\"u}r Astrophys.} }
\def\ZfN{{\it Zeitschr. f{\"u}r Naturforsch.} }
\def\etal{{\it et al.}}
\hyphenation{mono-chro-matic sour-ces Wein-berg
chang-es Strah-lung dis-tri-bu-tion com-po-si-tion elec-tro-mag-ne-tic
ex-tra-galactic ap-prox-i-ma-tion nu-cle-o-syn-the-sis re-spec-tive-ly
su-per-nova su-per-novae su-per-nova-shocks con-vec-tive down-wards
es-ti-ma-ted frag-ments grav-i-ta-tion-al-ly el-e-ments me-di-um
ob-ser-va-tions tur-bul-ence sec-ond-ary in-ter-action
in-ter-stellar spall-ation ar-gu-ment de-pen-dence sig-nif-i-cant-ly
in-flu-enc-ed par-ti-cle sim-plic-i-ty nu-cle-ar smash-es iso-topes
in-ject-ed in-di-vid-u-al nor-mal-iza-tion lon-ger con-stant
sta-tion-ary sta-tion-ar-i-ty spec-trum pro-por-tion-al cos-mic
re-turn ob-ser-va-tion-al es-ti-mate switch-over grav-i-ta-tion-al
super-galactic com-po-nent com-po-nents prob-a-bly cos-mo-log-ical-ly
Kron-berg Berk-huij-sen}

\def\refalt@jnl#1{{\rm #1}}

\def\aj{\refalt@jnl{AJ}}                   
\def\araa{\refalt@jnl{ARA\&A}}             
\def\apj{\refalt@jnl{ApJ}}                 
\def\apjl{\refalt@jnl{ApJ}}                
\def\apjs{\refalt@jnl{ApJS}}               
\def\ao{\refalt@jnl{Appl.~Opt.}}           
\def\apss{\refalt@jnl{Ap\&SS}}             
\def\aap{\refalt@jnl{A\&A}}                
\def\aapr{\refalt@jnl{A\&A~Rev.}}          
\def\aaps{\refalt@jnl{A\&AS}}              
\def\azh{\refalt@jnl{AZh}}                 
\def\baas{\refalt@jnl{BAAS}}               
\def\jrasc{\refalt@jnl{JRASC}}             
\def\memras{\refalt@jnl{MmRAS}}            
\def\mnras{\refalt@jnl{MNRAS}}             
\def\pra{\refalt@jnl{Phys.~Rev.~A}}        
\def\prb{\refalt@jnl{Phys.~Rev.~B}}        
\def\prc{\refalt@jnl{Phys.~Rev.~C}}        
\def\prd{\refalt@jnl{Phys.~Rev.~D}}        
\def\pre{\refalt@jnl{Phys.~Rev.~E}}        
\def\prl{\refalt@jnl{Phys.~Rev.~Lett.}}    
\def\pasp{\refalt@jnl{PASP}}               
\def\pasj{\refalt@jnl{PASJ}}               
\def\qjras{\refalt@jnl{QJRAS}}             
\def\skytel{\refalt@jnl{S\&T}}             
\def\solphys{\refalt@jnl{Sol.~Phys.}}      
\def\sovast{\refalt@jnl{Soviet~Ast.}}      
\def\ssr{\refalt@jnl{Space~Sci.~Rev.}}     
\def\zap{\refalt@jnl{ZAp}}                 
\def\nat{\refalt@jnl{Nature}}              
\def\iaucirc{\refalt@jnl{IAU~Circ.}}       
\def\aplett{\refalt@jnl{Astrophys.~Lett.}} 
\def\apspr{\refalt@jnl{Astrophys.~Space~Phys.~Res.}}
\def\bain{\refalt@jnl{Bull.~Astron.~Inst.~Netherlands}}
\def\fcp{\refalt@jnl{Fund.~Cosmic~Phys.}}  
\def\gca{\refalt@jnl{Geochim.~Cosmochim.~Acta}}   
\def\grl{\refalt@jnl{Geophys.~Res.~Lett.}} 
\def\jcp{\refalt@jnl{J.~Chem.~Phys.}}      
\def\jgr{\refalt@jnl{J.~Geophys.~Res.}}    
\def\jqsrt{\refalt@jnl{J.~Quant.~Spec.~Radiat.~Transf.}}
\def\memsai{\refalt@jnl{Mem.~Soc.~Astron.~Italiana}}
\def\nphysa{\refalt@jnl{Nucl.~Phys.~A}}   
\def\physrep{\refalt@jnl{Phys.~Rep.}}   
\def\physscr{\refalt@jnl{Phys.~Scr}}   
\def\planss{\refalt@jnl{Planet.~Space~Sci.}}   
\def\procspie{\refalt@jnl{Proc.~SPIE}}   

\let\astap=\aap
\let\apjlett=\apjl
\let\apjsupp=\apjs
\let\applopt=\ao

\title{\Large Towards the Chalonge 17th Paris Cosmology Colloquium 2013:

\medskip

HIGHLIGHTS and CONCLUSIONS

of the Chalonge 16th Paris Cosmology Colloquium 2012 

\medskip

\Large THE NEW STANDARD MODEL OF THE UNIVERSE: LAMBDA WARM DARK MATTER $\Lambda$WDM.
THEORY AND OBSERVATIONS

\medskip
  
Ecole Internationale d'Astrophysique Daniel Chalonge

Observatoire de Paris

in the historic Perrault building, 25-27 July 2012.}

\author{\Large \bf   H. J. de Vega$^{(a,b)}$,  M.C. Falvella$^{(c)}$, N. G. Sanchez$^{(b)}$}

\date{\today}

\affiliation{$^{(a)}$ LPTHE, Universit\'e
Pierre et Marie Curie (Paris VI) et Denis Diderot (Paris VII),
Laboratoire Associ\'e au CNRS UMR 7589, Tour 24, 5\`eme. \'etage, 
Boite 126, 4, Place Jussieu, 75252 Paris, Cedex 05, France. \\
$^{(b)}$ Observatoire de Paris,
LERMA. Laboratoire Associ\'e au CNRS UMR 8112.
 \\61, Avenue de l'Observatoire, 75014 Paris, France.\\
$^{(c)}$ Italian Space Agency and MIUR, Viale Liegi n.26,
00198 Rome, Italy.}

\begin{abstract}

The Chalonge 16th Paris Cosmology Colloquium (25-27 July 2012) in the historic 
Paris Observatory's Perrault building, combined in the
Chalonge School spirit real cosmological/astrophysical data and hard theory
predictive approach connected to them in the framework of the Warm Dark Matter Standard Model of the Universe:Theory and Observations : 
News and reviews from Atacama Cosmology Telescope (ACT) and WMAP, 
South Pole Telescope (SPT), {\it Herschel}, QUIET, {\it Planck}, JWST, UFFO, KATRIN and MARE experiments; 
astrophysics, particle physics and nuclear physics warm dark matter (WDM) searches and galactic observations, related theory and simulations,
with the aim of synthesis, progress and clarification.  Peter Biermann, Carlo Burigana, 
Christopher Conselice, Asantha Cooray, Hector J. de Vega,  Carlo Giunti \& Marco Laveder, 
John Kormendi \&  K. C. Freeman,  Ernest Ma, John C. Mather, Lyman Page, George F. Smoot,  Norma G. Sanchez, 
present here their highlights of the Colloquium.  $\Lambda$WDM (Warm Dark Matter) is progressing 
impressively 
over $\Lambda$CDM whose galactic scale crisis and scientific decline are staggering. 
  Summary and conclusions by H. J. de Vega, 
M. C. Falvella and N. G. Sanchez stress among other points:  

\medskip
 
{\bf (I)} Data confirm primordial CMB gaussianity. Inflation effective theory predicts
negligible primordial non-gaussianity, negligible scalar index running  
and a tensor to scalar ratio $ r \sim 0.05-0.04 $ at reach/border line of next CMB observations;
the present data with this theory clearly prefer new inflation $(V''< 0)$ and the `cosmic banana' region in the $r$ vs $n_s$
diagram; early fast-roll inflation is generic and provides the lowest multipoles depression and 
TE oscillations.
Impressive CMB results were reported i.e from ACT and WMAP, SPT, QUIET, with 9 clear peaks in the CMB power spectrum,  cosmic parameter degeneracies resolved, ACT primordial 
helium observation and ACT CMB lensing detection. Analysis of neutrino species and masses with CMB data is highlighted as a hot topic besides the B-mode results. PIXIE (Primordial Inflation Explorer) plans B-mode detection to limit $ r < 10^{-3} $ at more than 5 sigma and test inflation. PIXIE also allows to test keV warm dark-matter and reionization. Gamma ray bursts probes of the early universe -re-ionization era- have been highlighted. \\

{\bf (II)} Significant progresses are made in dSph galaxy observations: Very low star formation rates 
and very low baryonic feedback are found. Baryon feedback does not affect DM structure, CDM does not 
produce realistic galaxies with realistic feedback and star formation recipes. Cusped profiles are 
strongly disfavoured. Cored (non cusped) DM halos and warm (keV scale mass) DM are strongly favored from 
theory and astrophysical observations, WDM naturally produces the observed small scale structures; 
keV sterile neutrinos of mass  between $2$ and $4$ keV are the most serious WDM candidates. Progresses in $\Lambda$WDM simulations were reported. Wimps (heavier than 1 GeV) are strongly disfavoured by galaxy observations. In addition,  
the observed cosmic ray positron and electron excesses are explained by natural astrophysical 
processes, while annihilating/decaying cold dark matter models, besides being highly tailored, are strongly disfavored.\\
  
Inside galaxy cores, below $ \sim 100$ pc, quantum WDM effects are important.
Quantum WDM calculations (Thomas-Fermi approach) provide galaxy cores and their sizes, 
galaxy masses, velocity dispersions and density profiles in remarkable agreement with the observations.
All evidences point to a dark matter particle mass around 2 keV.
Baryons, which represent 16\% of DM, are expected to give a correction to pure WDM results.\\

{\bf (III)} Sterile neutrinos with keV scale mass (the main WDM candidate) could be detected in 
beta decay for Tritium and Renium and in the electron capture in Holmiun.
Decay into X rays could be detected observing DM
dominated galaxies, and the distortion of the black-body CMB spectrum.
The effective number of neutrinos N$_{\rm eff}$ measured by WMAP9 and Planck satellites
is compatible with one or two Majorana sterile neutrinos in the eV mass scale.
The WDM contribution to  N$_{\rm eff}$ is of the order $ \sim 0.01 $ and therefore
too small to be measurable with the CMB (anisotropy) data.

\medskip

Putting all together, evidence that $\Lambda$CDM and its proposed baryon cures do not work at small scales is staggering. Increasing and impressive evidence favour a fermionic DM particle mass   
of about 2 keV which naturally produce galaxy observations, cored density profiles 
and their sizes. Quantum WDM effects are important, particularly for wdarf galaxies. Overall,  $\Lambda$WDM  and keV scale DM particles deserve dedicated
astronomical and laboratory experimental searches, theoretical work and numerical simulations. KATRIN experiment in the future could adapt its set-up to look to keV scale sterile neutrinos. It will be a a fantastic discovery to detect dark matter in a beta decay.  Photos of the Colloquium are included.

\end{abstract}

\maketitle

\newpage

\tableofcontents

\newpage

\section{Purpose of the Colloquium, Context and Introduction}

The main aim of the series `Paris Cosmology Colloquia', in the framework of the International 
School of Astrophysics  {\bf `Daniel Chalonge'}, is to put together real data : cosmological, 
astrophysical, particle physics, nuclear physics data, and hard theory predictive approach 
connected to them in the framework of the   Standard Model of the Universe. 

\medskip

The Chalonge Paris Cosmology Colloquia 
bring together physicists, astrophysicists and astronomers from the world over. Each year these 
Colloquia are more attended and appreciated both by PhD students, post-docs, senior participants 
and lecturers. 
The format of the Colloquia is intended to act as a true working meeting and laboratory of ideas, 
and allow easy and fruitful mutual contacts and communication.

\medskip

The discussion sessions are as important as the lectures themselves.

\bigskip

The {\bf 16th Paris Cosmology Colloquium 2012}  was devoted to `THE NEW STANDARD MODEL OF THE UNIVERSE : LAMBDA WARM DARK MATTER $\Lambda$WDM: THEORY AND OBSERVATIONS'. 

\bigskip

The  Colloquium took  place during three full days  (Wednesday July 25, Thursday 26 and 
Friday July 27) at the parisian campus of  Paris Observatory (HQ), in the historic 
Perrault building.

\bigskip

Never as in this period, the Golden 
Age of Cosmology, the major subjects of the Daniel Chalonge School were so timely and in 
full development: Recently, Warm (keV scale) Dark Matter (WDM) emerged impressively over CDM (Cold Dark Matter) as the leading Dark Matter candidate. Astronomical evidence that Cold Dark Matter ($\Lambda$CDM) and its proposed tailored cures , namely CDM + baryons, do not work at small scales is staggering. $\Lambda$WDM solves naturally the problems of $\Lambda$CDM and agrees remarkably well with the observations at small as well as large and cosmological scales. $\Lambda$ WDM numerical simulations naturally  {\it agree} with observations at all scales, in contrast to $\Lambda$ CDM simulations which only agree at large scales.

\bigskip

In the context of the new Dark Matter situation represented by Warm (keV scale) Dark Matter which implies novelties in the astrophysical, cosmological and keV particle and nuclear physics context, this 16th Paris Colloquium 2012 was devoted to the $\Lambda$WDM Standard Model of the Universe.

\bigskip

An {\bf Open Session} took place before the end of the Colloquium with three Lectures :  The James Webb Space Telescope (JWST) by John Mather, Science PI of the JWST,  the Ultra-Fast Flash Observatory (UFFO) Pathfinder by George Smoot, and the discovery of the present acceleration in the expansion of the Universe by Brian Schmidt, the three lecturers being laureats of the Nobel prize of physics  and of the Daniel Chalonge Medal.

\medskip

The {\bf Daniel Chalonge Medal 2012} was awarded to Brian Schmidt during the Open Session of the Colloquium for his brilliant accomplishment discoverying the present accelerated expansion of the Universe, compatible with a cosmological constant, and for his contributions as lecturer in the Chalonge School and formation of young researchers and students in modern cosmology and the construction of the Standard Model of the Universe, formidable task at the center of the Chalonge School activity. See section V devoted to the Daniel Chalonge Medal 2012 in these Highlights.

\bigskip

\begin{center}

{\bf The Main topics of this Colloquium included} :  

\end{center}

\medskip

Observational and theoretical progress on the nature of dark matter: keV scale Warm Dark Matter.
The new quantum WDM effects and dwarfs galaxies\\ 
\medskip
Dark energy: cosmological constant: the quantum energy of the cosmological vacuum. \\
\medskip
Large and small scale structure formation in agreement with observations at large scales and small (galactic) scales.\\
\medskip
The ever increasing problems and confusing tailored cures of $\Lambda$CDM and $\Lambda$CDM + baryons.\\
\medskip
The emergence of Warm (keV scale) Dark Matter from theory and observations.\\ 
\medskip
Neutrinos in astrophysics and cosmology.\\
\medskip
The new serious dark matter candidate: Sterile neutrinos at the keV scale. \\ 
\medskip
Neutrino mass bounds from cosmological data and from high precision beta decay experiments.\\
\medskip
The active neutrino mass absolute scale with KATRIN. The search for sterile neutrinos.\\
\medskip
The analysis of the CMB+LSS+SN data with the effective (Ginsburg-Landau) effective theory of inflation: New Inflation (double well inflaton potential) strongly favored by the CMB + LSS + SN data.\\ 
\medskip
The presence of the lower bound for the primordial gravitons (non vanishing tensor to scalar ratio r) with the present CMB + LSS + SN data. CMB polarization and forecasts for Planck. \\
\medskip
CMB measurements. WMAP. The Atacama Cosmology Telescope. QUIET.{\it Planck}\\ 
\medskip
Recent Extragalactic Survey results from {\it Herschel}.\\
\medskip
The James Webb Space Telescope: mission and science\\
\medskip
The Ultra-Fast Flash Observatory (UFFO)\\

\bigskip

\begin{center}

{\bf CONTEXT, CDM CRISIS and CDM DECLINE:} 

\end{center}

\bigskip

On large cosmological scales, CDM agrees in general with observations but CDM does not agree with observations on galaxy scales and small scales. Over most of twenty years, increasing number of cyclic arguments and ad-hoc mechanisms or recipes were-and continue to be- introduced in the CDM galaxy scale simulations, in trying to deal with the CDM small scale crisis: Cusped profiles and overabundance of substructures are predicted by CDM. Too many satellites are predicted by CDM simulations while cored profiles and no such overabundant substructures are seen by astronomical observations. Galaxy formation within CDM is increasingly confusing and in despite of the proposed cures, does not agree with galaxy observations.

\medskip

A host of ad-hoc mechanisms are proposed and advocated to cure the CDM problems. `Baryon and supernovae feedbacks',
non circular motions, triaxiality, mergers, `cusps hidden in cores', 
`strippings' are some of such mechanisms tailored or exagerated for the purpose of obtaining the desired result without having a well established physical basis. For example, the strong "baryon and 
supernovae feedback" introduced to transform the CDM cusps into cores in baryon+CDM simulations
corresponds to a large star formation rate contradicting the observations.

\medskip

On the CDM particle physics side, 
the problems are no less critical: So far, all the {\it dedicated} experimental searches after most of 
twenty five years to find the theoretically proposed CDM particle candidate (WIMP) have {\bf failed}. The
CDM indirect searches (invoking CDM annihilation) to explain cosmic ray positron excesses, are in crisis as 
well, as wimp annihilation models are plagued with growing tailoring or fine tuning, and in any case, 
such cosmic rays excesses are well explained and reproduced by natural astrophysical process and sources, 
as non-linear acceleration, shocks and magnetic winds around massive explosion stars, quasars, or the 
interaction of the primary cosmic-ray protons with the interstellar medium producing positrons. The so-called 
and repeatedealy invoked `wimp miracle' is nothing but been able to solve one equation with three unknowns  
(mass, decoupling temperature, and annhiliation cross section) within WIMP models
theoretically motivated by SUSY model building twenty years ago
(SUSY was very fashionable at that time and believed a popular motivation for many proposals).

\medskip

After more than twenty five years -and as often in big-sized science-, CDM research has by now its own internal 
inertia: growing simulations involve large super-computers and large number of people working with, CDM particle 
wimp search involve large and longtime planned experiments, huge number of people, (and huge budgets); one should not be surprised in 
principle, if a fast strategic change would not yet operate in the CDM and wimp research, although
they would progressively decline. 

\medskip

Similar situation happens in the CDM+baryon (super)computer simulations, 
in which tailored models fail to reproduce observations and ever increasing `baryon complexity' 
is advocated as the reason of such a failure, becoming each time``more and more complex'' and so complicated with no hope of reaching conclusion at any time,  `scape to the future'. Namely, in CDM dominated galaxies,  baryons with complexes environments and feedbacks need to make the main work against the undiserable features produced by CDM. Basically, baryon effects need to be produced and overwelheimed to `destroy' and elliminate the wrong galactic CDM features. Moreover, the contradiction appears: 
if the cure to the main CDM effects in galaxies needs to destroy the main CDM effects: Why then to use CDM ? 

\bigskip

In contrast to the CDM situation, the WDM research situation is progressing fast, the subject is new and WDM essentially {\it works}, naturally reproducing the observations over all the scales, small as well as large and cosmological scales ($\Lambda$WDM). $N$-body CDM simulations {\bf fail} to produce the observed structures for {\bf small} scales less than some kpc.
Both $N$-body WDM and CDM simulations yield {\bf identical and correct} structures 
for scales larger than some kpc.
At intermediate scales WDM give the {\bf correct abundance} of substructures.  {\bf Quantum} fermionic WDM  reproduces:  the observed cores and their sizes, the observed galaxy masses and their sizes, the phase space density and velocity dispersions from the dwarfs to the larger galaxies (elliptical, spirals). Dwarf galaxies are quantum macroscopic WDM objects. WDM became a hot topic and the subject of many doctoral and post-doctoral researches.

\bigskip

\begin{center}

{\bf SUMMARY and CONCLUSIONS:}

\end{center}

\bigskip

This 16th Paris Colloquium 2012 addressed the last progresses made in the NEW STANDARD MODEL OF THE 
UNIVERSE-WARM DARK MATTER with both theory and observations. In the tradition of the Chalonge School, 
an effort of clarification and synthesis is  made by combining in a conceptual framework, theory, 
analytical, observational and numerical simulation results which reproduce observations, both in 
astrophysics and particle and nuclear physics, keV sterile neutrinos being
today the more serious candidates for Dark Matter.

\bigskip

\begin{center} 

{\bf The subject have been approached in a fourfold coherent way:}

\end{center}

\bigskip

(I) Conceptual context, the standard cosmological model includes inflation

\bigskip

(II) CMB observations and astronomical observations,  cosmological, large, intermediate and  small galactic scales, linked to structure formation at the different scales.

\bigskip

(III) WDM theory, models and WDM numerical simulations which reproduce observations at large and small (galactic) scales. The new quantum WDM effects and dwarf galaxies as bounded quantum macroscopic objects
(degenerate WDM fermions and gravity).

\bigskip

(IV) WDM particle candidates, keV sterile neutrinos: particle models and astrophysical constraints on them.
Experimental WDM keV sterile neutrino searches.

\bigskip

Peter Biermann, Carlo Burigana, Christopher Conselice, Asantha Cooray, Hector J. de Vega,  Carlo Giunti \& Marco Laveder, John Kormendi \&  K. C. Freeman,  Ernest Ma, John C. Mather, Lyman Page, George F. Smoot,  Norma G. Sanchez, present here their highlights of the Colloquium. 

\medskip

Summary and conclusions are presented at the end by H. J. de Vega, M. C. Falvella and N. G. Sanchez in 
three subsections: 

\medskip

\begin{center}

A. General view and clarifying remarks. 

\medskip

B. Conclusions. 

\medskip

C. The present context and future in the Dark Matter research and Galaxy formation

\bigskip

The Summary of the Conclussions stress among other points :

\end{center}

\bigskip

{\bf (I)} Data confirm primordial CMB gaussianity. Inflation effective theory predicts
negligible primordial non-gaussianity, negligible scalar index running  
and a tensor to scalar ratio $ r \sim 0.05-0.04 $ at reach/border line of next CMB observations;
the present data with this theory clearly prefer new inflation and the `cosmic banana' region in the $r$ vs $n_s$
diagram; early fast-roll inflation is generic and provides the lowest multipoles depression and TE oscillations.
Impressive CMB data were reported, i.e  from ACT and WMAP, SPT, QUIET, the CMB power spectrum displays  $9$ clear peaks, various cosmic parameter degeneracies are resolved, ACT observed primordial helium and detected the CMB lensing, CMB alone provides evidence for dark energy. Analysis of neutrino species and masses with CMB data is highlighted besides the B-mode results. The Primordial Inflation Explorer (PIXIE) is planned to test inflation and to robustly detect CMB tensor (B-mode) polarisation (primordial gravitational waves) to limit $r < 10^{-3}$ at more than 5 sigma. PIXIE also allows to test keV warm dark-matter and reionization. Gamma ray bursts appear as a rich and interesting avenue -both theoretical and experimental- to study the early universe in the re-ionization era.
 
\bigskip
 
{\bf (II)} Significant progresses are made in dSph galaxy observations: Very low star formation rates and very low baryonic feedback are found. Baryon feedback does not affect DM structure, CDM does not produce realistic galaxies with realistic feedback and star formation recipes. Very precise kinematics of stars in dSph is probing directly DM density profiles: cusped profiles are excluded at the $95 \%  CL$. Cored (non cusped) DM halos and warm (keV scale mass) DM are strongly favored from theory and astrophysical observations, they naturally produce the observed small scale structures; keV sterile neutrinos of mass  between $1$ and $4$ keV say are the most serious candidates. Progresses in computing the distribution function for WDM sterile neutrinos as well as $\Lambda$WDM simulations are going on. Wimps (heavier than 1 GeV) are strongly disfavoured combining theory with galaxy observations. In addition,  
the observed cosmic ray positron and electron excesses are explained by natural astrophysical processes, while annihilating/decaying cold dark matter models are highly tailored for the purpose and strongly  disfavored.

\bigskip

{\bf (III)} The main galaxy magnitudes: halo radius, galaxy masses and velocity dispersions
computed for quantum keV WDM fermions are
fully consistent with all the observations for all types of galaxies. 
Namely, fermionic WDM treated quantum mechanically, as it must be, is able to reproduce
the observed galaxy cores and their sizes, for the broad variety of galaxies, which is highly remarkably.
Therefore, the effect of including 
baryons is expected to be a correction to the pure WDM results, consistent with the fact that dark matter 
is in average six times more abundant than baryons. WDM quantum effects play a fundamental role in the inner galaxy regions. The quantum effects however are negligeable for CDM: the heavy (GeV) wimps behave classically. The quantum pressure of the WDM fermions solves the core size problem and provides the correct observed galaxy masses and sizes covering from the compact dwarfs to the larger and dilute galaxies, spirals, ellipticals. Dwarf galaxies are natural macroscopic quantum objects supported against gravity by the WDM quantum pressure. 

\bigskip

{\bf (IV)} All evidences point to a dark matter particle mass around 2 keV.
Sterile neutrinos with keV scale mass (the main WDM candidate) could be detected in 
beta decay for Tritium and Renium and in the electron capture in Holmiun.
Decay into X rays could be detected observing DM
dominated galaxies, and the distortion of the black-body CMB spectrum.
The effective number of neutrinos, N$_{\rm eff}$ measured by WMAP9 and Planck satellites
is compatible with one or two Majorana sterile neutrinos in the eV mass scale.
The WDM contribution to  N$_{\rm eff}$ is of the order $ \sim 0.01 $ and therefore
too small to be measurable by CMB observations.

\medskip

So far, {\bf not a single valid} objection arose against WDM.

\bigskip

Putting all together, evidence that $\Lambda$CDM does not work at small scales is staggering. Impressive evidence points that DM particles have a mass in the keV scale and that those keV scale particles naturally produce the small scale structures observed in galaxies. Wimps (DM particles heavier than 1 GeV) are strongly disfavoured combining theory with galaxy astronomical observations. keV scale sterile neutrinos are the most serious DM candidates and deserve dedicated  experimental searchs and simulations. Astrophysical constraints  put the sterile neutrinos mass in the range $ 1 < m < 4$ keV favoring a mass about $2$ keV.   $\Lambda$WDM simulations with 1 keV particles reproduce remarkable well the observations, the distribution of Milky Way satellites in the range above $\sim 40$ kpc, sizes of local minivoids and velocity functions. 

\bigskip

Overall, $\Lambda$WDM  and WDM sterile neutrinos of few keV mass deserve dedicated
astronomical and laboratory experimental searches, theoretical work and simulations. MARE -and hopefully an adapted KATRIN- experiment could provide a sterile neutrino signal. 

\medskip

The experimental search for serious WDM particle candidates (sterile neutrinos) 
appears urgent and important: it will be a fantastic discovery to detect dark matter in a beta decay. 
There is a formidable WDM work to perform ahead of us, these highlights point some of the directions where 
it is worthwhile to put the effort.

\bigskip

\begin{center}

{\bf FORMAT OF THE COLLOQUIUM AND EXHIBITION}

\end{center}

\bigskip

All Lectures are  plenary and followed by a discussion. 
Enough time is provided to the discussions.  

\begin{center}

Informations of the Colloquium are available on

\medskip
 
 {\bf http://www.chalonge.obspm.fr/colloque2012.html}

\end {center}

\bigskip

Informations on the previous Paris Cosmology Colloquia and  
on the Chalonge school events are available at  

\begin{center}

 {\bf http://chalonge.obspm.fr}

(lecturers, lists of participants, lecture files and photos during the Colloquia).

\end {center}

\bigskip

This Paris Chalonge Colloquia series started in 1994 at the Observatoire de Paris. 
The series cover selected topics of high current interest in the interplay between 
cosmology, astrophysics and fundamental physics. 
The purpose of this series is an updated understanding, from a fundamental, conceptual and unifying view, of the progress 
and current problems in the early universe, cosmic microwave background radiation, large and small scale 
structure and neutrinos in astrophysics and the interplay between them. Emphasis is given to the 
mutual impact of fundamental physics, astrophysics and cosmology, both at theoretical and experimental 
-or observational- levels. 

\bigskip

Deep understanding, clarification, synthesis, a careful interdisciplinarity within a 
fundamental physics and unifying approach, are goals of this series of Colloquia.

\bigskip

Sessions leave enough time for private discussions and to enjoy 
the beautiful parisian campus of Observatoire de Paris (built on orders from Colbert and to 
plans by Claude Perrault from 1667 to 1672).

\bigskip

Sessions take  place in the Cassini Hall, on the meridean of Paris, in 'Salle du Conseil'
(Council Room) under the portraits of Laplace, Le Verrier, Lalande, Arago, Delambre and Louis XIV, and in the 
'Grande Galerie' (the Great Gallery), in the historic Perrault building ('B\^atiment Perrault') 
of Observatoire de Paris HQ.

\bigskip

An {\bf Exhibition} at the `Grand Galerie' (Great Gallery) and "Salle Cassini" (Cassini Hall) 
retraced the 20 years of activity of the Chalonge School and {\bf "The Golden Days" in 
`Astrofundamental Physics': The Construction of the Standard Model Of the Universe}, 
as well as {\bf `The World High Altitude Observatories Network: World Cultural and Scientific Heritage'}.

\medskip
 
The books and proceedings of the School since its creation, as well as historic 
Daniel Chalonge material, Chalonge instruments and original plasters of the Daniel Chalonge Medal were on exhibition at the Grande Galerie.

\bigskip

The exhibitions were prepared by Maria Cristina Falvella, Alba Zanini, Fran\c{c}ois Sevre, 
Nicole Letourneur and Norma G. Sanchez.

\medskip

Photos and Images by : Kathleen Blumenfeld, Letterio Pomara, Jean Mouette,
Fran\c{c}ois Sevre, Fran\c{c}ois Colas, Nadia Blumenfeld, Gerard Servajean,
Sylvain Cnudde, graphic design by Emmanuel Vergnaud. 

\bigskip

More information on the Colloquia of this series can be found in the Proceedings
(H.J. de Vega and N. Sanchez, Editors) published by World Scientific Co. since 1994 and by 
Observatoire de Paris, and the Chalonge School Courses published by World Scientific Co 
and by Kluwer Publ Co. since 1991.

\bigskip

We address here the recent turning point in the research of Dark Matter represented by 
Warm Dark Matter (WDM) putting together astrophysical, cosmological, particle and nuclear physics WDM,
astronomical observations, theory and WDM numerical simulations which naturally reproduce the observations, 
as well as the experimental search for the WDM particle candidates (sterile neutrinos).

\bigskip

This 16th Paris Chalonge Colloquium 2012 enlarges, strentghs and unifies with new topics, lecturers and deep discussions the issues discussed 
and pre-viewed in the Warm Dark Matter Chalonge Meudon Workshop 2012 in June 2012.

\bigskip

The Highlights and Conclusions of the previous recent Colloquia, in particular the 15th Paris Chalonge Colloquium 2011 and of the Chalonge Meudon Workshops 2011 and 2012 are a useful and complementary introduction to these Highlights and can be read with benefit. They are available at:

\begin{center}

http://arxiv.org/abs/1203.3562\\
\medskip
http://arxiv.org/abs/1305.7452\\
\medskip
http://arxiv.org/abs/1109.3187\\
\medskip
http://arxiv.org/abs/1009.3494\\

\medskip

\end{center}

\bigskip

We want to express our grateful thanks to all the sponsors of the Colloquium,  
to all the lecturers for their excellent and polished presentations, to all the 
lecturers and participants for their active participation and their contribution to 
the outstanding discussions and lively atmosphere, to the assistants, secretaries and 
all collaborators of the Chalonge School, who made this event so harmonious, wonderful and successful . 

\bigskip

We are pleased to give you appointment at the next Colloquium of this series:

\begin{center}

\bf{The 17th Paris Cosmology Colloquium 2013 devoted to 

\bigskip

THE NEW STANDARD MODEL OF THE UNIVERSE: LAMBDA WARM DARK MATTER ($\Lambda$WDM) THEORY versus OBSERVATIONS

\bigskip

Observatoire de Paris, historic Perrault building,  24, 25, 26 JULY 2013.

\bigskip

http://www.chalonge.obspm.fr/colloque2013.html}

\end{center}

\bigskip

\begin{center}
                                   
With Compliments and kind regards,

\bigskip  
                                            
{\bf Hector J de Vega, $ \; $ Maria Cristina Falvella,  $ \; $  Norma G Sanchez}

\end{center}

\begin{figure}[ht]
\includegraphics[height=14cm,width=18cm]{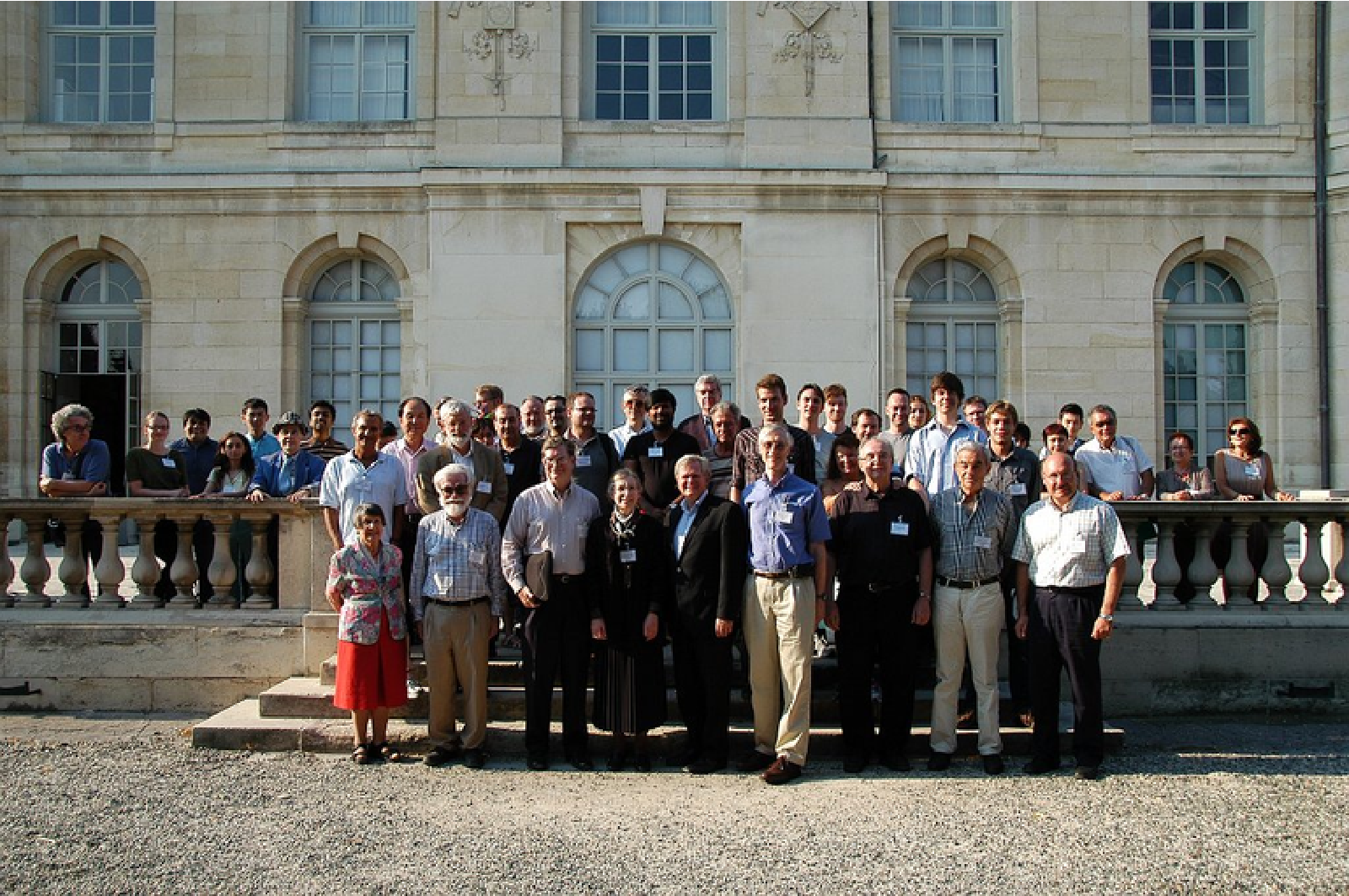}
\caption{Group Photo}
\end{figure}

\begin{figure}[ht]
\includegraphics[height=24cm,width=18cm]{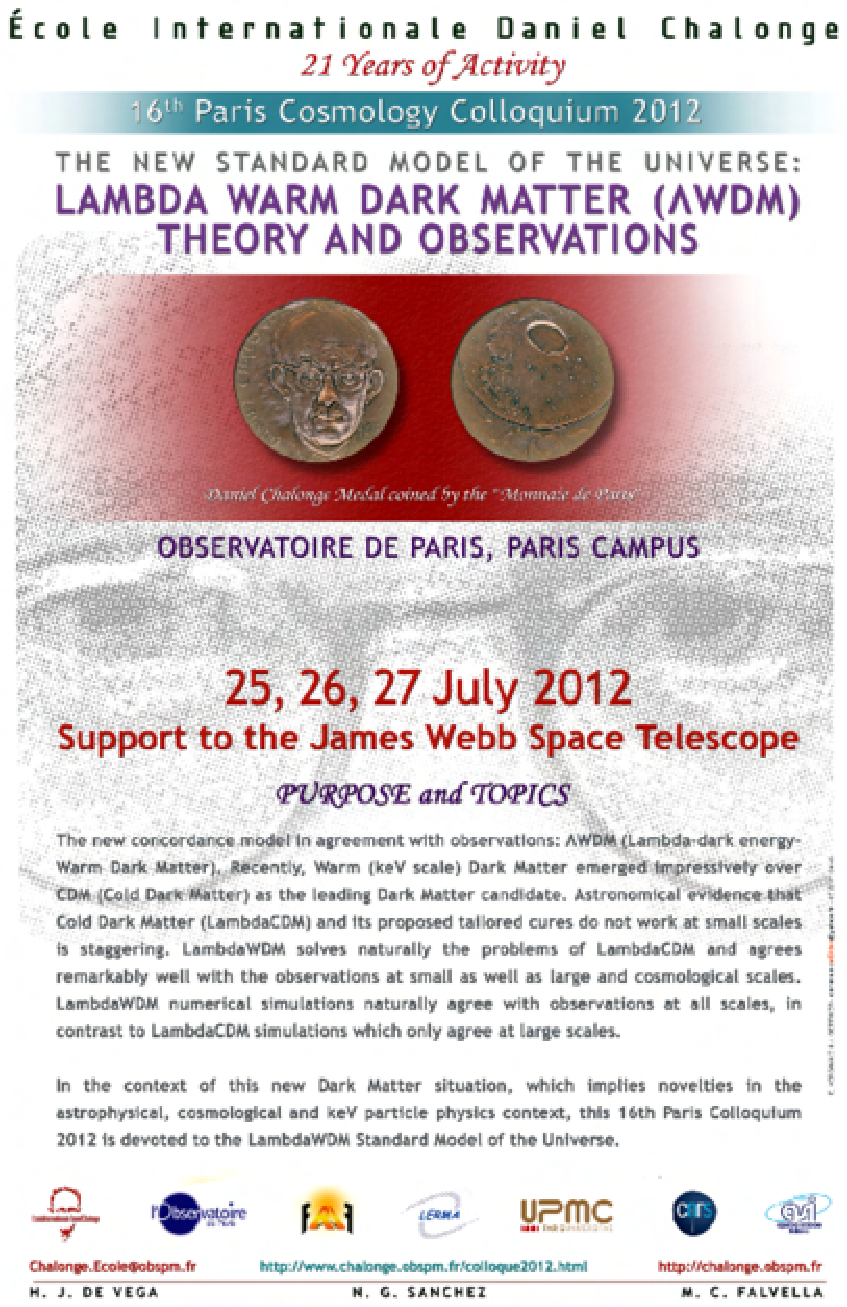}
\caption{Poster of the Conference}
\end{figure}

\newpage

\section{Programme and Lecturers}

\begin{itemize}
\item{{\bf Peter BIERMANN} (Chalonge Medal, MPI-Bonn, Germany \& Univ of Alabama, Tuscaloosa, USA)\\
Warm dark matter: cosmological and astrophysical signatures.}
\item{{\bf Carlo BURIGANA \& Reno MANDOLESI} (INAF-IASF, Bologna, Italy)\\
Results from the PLANCK satellite and their science implications}
\item{{\bf Christopher J. CONSELICE} (University of Nottingham, School of Physics and Astronomy, UK)\\
Observational overview on galaxy formation and evolution}
\item{{\bf Asantha COORAY} (University of California,Irvine, USA)\\
Recent cosmological results with Herschel Space Observatory.}
\item{{\bf Sudeep DAS} (Berkeley Center Cosmological Physics BCCP-LNBL, Berkeley, USA)\\
CMB lensing research and results: a status report}
\item{{\bf Hector J. DE VEGA} (CNRS LPTHE Univ de Paris VI UPMC, France)\\
The Standard Model of the Universe: Warm Dark Matter and galaxy structure.}
\item{{\bf Carlo GIUNTI} (INFN di Torino \& Universita, Turin, Italy) \& {\bf Marco LAVEDER} 
(Padova University, Italy) Sterile Neutrinos: Phenomenology and Fits}
\item{{\bf John KORMENDY} (Texas Univ., Dept Astron., Austin TX, USA)\\
Scaling Laws for Dark Matter Halos in Late-Type and Dwarf Spheroidal Galaxies}
\item{{\bf Anthony N. LASENBY} (Chalonge Medal, Cavendish Laboratory, Cambridge, UK)\\
The CMB in the Standard Model of the Universe: A Status Report}
\item{{\bf Ernest MA} (Univ California Riverside, IAS Hong Kong and IAS Singapore)\\
The KeV WDM particle in a left-right extension of the Standard Model of particle physics}
\item{{\bf John C.MATHER} (Chalonge Medal, Nobel prize of Physics, 
NASA Goddard Space Flight Center, Greenbelt, MD, USA) Open Lecture: The James Webb Space Telescope}
\item{{\bf F\'elix MIRABEL} (CEA-Saclay, France \& IAFE-Buenos Aires, Argentina)\\
Re-ionization of the Universe}
\item{{\bf QUIET Collaboration} (Hogan Nguyen spokeperson, Fermilab, IL, USA)\\
Last results on the B-mode CMB polarization from QUIET.}
\item{{\bf Lyman PAGE} (Princeton Univ. Dept. of Physics, NJ, USA)\\
ACT results and WMAP 9-year results and their cosmological implications}
\item{{\bf Rafael REBOLO} (Inst. Astrofisico Canarias, Tenerife, Spain)\\
CMB polarization and Science with the QUIJOTE Experiment}
\item{{\bf Norma G. SANCHEZ} (CNRS LERMA Observatoire de Paris, France)\\
The Standard Model of the Universe:Warm Dark Matter in agreement with observations}
\item{{\bf Brian P. SCHMIDT} Nobel prize of Physics (Research School Astron. \& Astroph. 
Australian Nat Univ, Weston Creek, Australia)\\
Open Lecture: Discovery of the Accelerated Expansion of the Universe}
\item{{\bf Aldo SERENELLI} (Inst. de Ciencies de l'Espai ICE-CSIC, Barcelone Bellaterra, Spain)\\
The solar abundance problem and solar models.}
\item{{\bf George F. SMOOT III}  Chalonge Medal, Nobel prize of Physics (BCCP LBL Berkeley,IEU Seoul, 
Univ Paris Diderot USA)\\
Open Lecture: News from UFFO (Ultra Fast Flash Observatory) Pathfinder: Science and Launch}
\item{{\bf Michel Ange TOGNINI} (Brigadier General French Air Force, Astronaut, France)\\
Open Lecture : Profession Astronaut}

\item{{\bf Matthew WALKER} (Harvard-Smithsonian Center for Astrophysics, Cambridge, MA, USA)\\
Observations of cored density profiles in galaxies and related results}

\item{{\bf Christian WEINHEIMER} (Institut f\"ur Kernphysik Universit\"at M\"unster, M\"unster, Germany)\\
Absolute scale of the active neutrino mass and the search of sterile neutrinos}
\end{itemize}

\section{Award of the Daniel Chalonge Medal 2012}

\bigskip

The Daniel Chalonge Medal 2012 has been awarded to Professor {\bf Dr Brian P. SCHMIDT}.   

\bigskip 

The International School of Astrophysics "Daniel Chalonge" has awarded the Daniel Chalonge Medal 2012 to Professor Brian P. Schmidt, Nobel Laureate of physics 2011 for the outstanding results of the discovery of the  present accelerated expansion of the Universe.

\bigskip

The medal was presented to Brian Schmidt on the 27th July  2012 during the last and open session of the 16th Paris Cosmology Colloquium at the Paris Observatory (Observatoire de Paris), attended by about hundred participants from over the world, among them cosmologists John Mather and George Smoot,  both laureates with the Nobel prize in Physics 2006 and with the Daniel Chalonge Medal, Chalonge Medal laureates Anthony Lasenby and Peter Biermann, and French test pilot, Brigadier General in the French Air Force, and former CNES and ESA Head astronaut Michel Tognini.  Brian Schmidt brilliantly exposed the discovery of the accelerated expansion of the Universe. He showed the successive steps and efforts in measuring and interpreting the unexpected results found by him and his team in the project of observing the most distant supernovae.

\bigskip

The medal was awarded to Brian Schmidt in recognition to his huge contribution to the modern cosmology and in particular for his outstanding effort in promoting and leading key missions for the study of the Universe, deeply discussed in the frame of the International Astrophysics School D. Chalonge.  Brian Schmidt was lecturer in several courses of the School, and from the very beginning in the discovery of the acceleration of the Universe. The scientific activity of Brian Schmidt, recognized with the Nobel Prize in 2011, has motivated the realization of very important projects in cosmology and the training and formation of young physicists and astrophysicists over the world.

\bigskip

With the Chalonge Medal, the International School of Astrophysics Daniel Chalonge confirms its great appreciation to Dr. Brian Schmidt and his contribution to the construction of the Standard Model of the Universe, at the center of the Chalonge School programmes.  This recognizes the outstanding scientific value and impact of the discovery of the dark energy component, compatible with the cosmological constant, which constitutes about three quarters of the content of the Universe.

\bigskip

The Chalonge Medal, coined exclusively for the Chalonge School by the prestigious Hotel de la Monnaie de Paris (the French Mint), is a surprise award and only nine Chalonge medals have been awarded in the 21 year school history.

The list of the awarded Chalonge Medals is the following:

\bigskip

\begin {itemize}

\item {1991: Subramanyan Chandrasekhar, Nobel prize of physics.}
\item {1992: Bruno Pontecorvo.}
\item {2006: George Smoot, Nobel prize of physics.}
\item {2007: Carlos Frenk.}
\item {2008: Anthony Lasenby.}
\item {2008: Bernard Sadoulet.}
\item {2009: Peter Biermann.}
\item {2011: John Mather, Nobel prize of physics.}
\item {2012: Brian Schmidt, Nobel prize of physics.}

\end{itemize}

\bigskip

The Medal aknowledges science with great intellectual endeavour and a human face. True and healthy science. Outstanding gentleperson scientists. Scientists recipients of the Daniel Chalonge Medal are Ambassadors of the School.

\bigskip

{\bf See the announcement, full history, photo gallery and links at:}\\

\bigskip

http://chalonge.obspm.fr

\bigskip

The Daniel Chalonge Medal 2012:

\bigskip

http://chalonge.obspm.fr/Medal$_{}$Chalonge2012.pdf

\bigskip
Brian Schmidt's Open Lecture 2012:

\bigskip

http://www.chalonge.obspm.fr/Paris12$_{}$Schmidt.pdf

\bigskip

Photo Gallery:

\bigskip

http://chalonge.obspm.fr/albumopensession2012/index.html

\bigskip

Open Session 2012 :

\bigskip

http://www.planetastronomy.com/special/2012-special/27jul/chalonge-cosmo.htm

\bigskip

Archives Daniel Daniel Chalonge:  

\bigskip

http://chalonge.obspm.fr/Archives$_{}$Daniel$_{}$Chalonge.html

\bigskip

\section{Highlights by the Lecturers}

\begin{center}
  
More informations on the Colloquium Lectures are at: 

\bigskip

{\bf http://www.chalonge.obspm.fr/colloque2012.html}

\end{center}

\subsection{Peter L. BIERMANN}

\vskip -0.3cm

\begin{center}

MPI for Radioastronomy, Bonn, Germany;  Dept. of Phys. \& Astr., Univ. of Alabama, Tuscaloosa, AL, USA

\bigskip

{\bf ASTROPHYSICS OF WARM DARK MATTER and ITS SIGNATURES} 

\end{center}

\bigskip

Dark matter has been first detected 1933 (Zwicky) and basically behaves
like a non-EM-interacting gravitational gas of particles.  Particle
physics suggests with an elegant argument that there could be a lightest
supersymmetric particle, which is a dark matter candidate, possibly
visible via decay in odd properties of energetic particles and photons:
Such discoveries were made: (i) an upturn in the CR-positron fraction,
(ii) an upturn in the CR-electron spectrum, (iii) a cosmic ray anisotropy
in data from Tibet, SuperK, Milagro, and now with IceCube, (iv) a flat
radio emission component near the Galactic Center (WMAP haze), (v) a
corresponding IC component in gamma rays (Fermi haze and Fermi bubble),
and a flat $\gamma$-spectrum at the Galactic Center (Fermi), (vi) the 511
keV annihilation line also near the Galactic Center, (vii) an upturn in
the CR-spectra of all elements from Helium, with a hint of an upturn for
Hydrogen, (viii) a derivation of the interaction grammages separately for
CR-protons and CR-heavy nuclei, and ix) the complete cosmic ray spectrum
with a steep powerlaw leading to a dip near $3 \, 10^{18}$ eV in terms
of $E^{3} \, (d N/d E)$, then a broad bump near $5 \, 10^{19}$ eV,
turning down towards $10^{21}$ eV (KASCADE, IceTop, KASCADE-Grande,
Auger).  

\bigskip

All these features can be quantitatively explained, eliminating
any argument from supersymmetry.  These explanations build on the action
of cosmic rays accelerated in the magnetic winds of very massive stars,
when they explode (Biermann et al. 2009 - 2012): this work is based on
predictions from 1993 (Biermann 1993, Biermann \& Cassinelli 1993,
Biermann \& Strom 1993, Stanev et al 1993; review at ICRC Calgary 1993);
this approach is older and simpler than adding WR-star supernova
CR-contributions with pulsar wind nebula CR-contributions, is also
simpler than using magnetic field enhancement in ISM-shocks, and also
simpler than using the decay of a postulated particle; this approach
also gave quantitative predictions from 1993 which can now be tested.

\bigskip

This concept gives an explanation for the cosmic ray spectrum as
Galactic plus one extragalactic source, Cen A (Gopal-Krishna et al.
2010, Biermann \& de Souza 2012).  The data do not require any extra
source population below the MWBG induced turnoff - commonly referred to
as the GZK-limit: Greisen (1966), Zatsepin \& Kuzmin (1966); in fact the
cut-off observed in the spectrum may well derive from an energy limit in
the source due to spatial limits.  All this is possible, since the
magnetic horizon appears to be quite small (consistent with the
cosmological MHD simulations of Ryu et al. 2008).  It also entails that
Cen A is our highest energy physics laboratory accessible to direct
observations of charged particles.

\bigskip

All this allows to go back to galaxy
data to derive the key properties of the dark matter particle: Work by
Tremaine \& Gunn (1979), Hogan \& Dalcanton (2000), Gilmore et al. (from
2006, 2009), Strigari et al. (2008), Boyanovsky et al. (2008), Gentile
et al. (2009) and de Vega \& Sanchez (2010 - 2012), de Vega et al.
(2012), Destri et al. (2012) clearly points to a keV particle, with high
probability constrained to within 2 - 4 keV Fermion, without using any
specific model.  

\medskip

A right-handed neutrino is a Fermion candidate to be
this particle (e.g. Kusenko \& Segre 1997; Fuller et al. 2003; Kusenko
2004; also see Kusenko et al. 2010, and Lindner et al. 2010; for a
review see Kusenko 2009; Biermann \& Kusenko 2006; Stasielak et al.
2007; Loewenstein et al. 2009). This particle has the advantage to allow
star formation very early, near redshift 80, and so also allows the
formation of supermassive black holes: they possibly formed out of
agglomerating massive stars, in the gravitational potential well of the
first DM clumps, whose mass in turn is determined by the properties of
the DM particle (weakly degenerate Fermion galaxy cores).  

\medskip

The supermassive star gives rise to a large HII region, possibly dominating
the Thomson depth observed by WMAP.  This black hole formation can be
thought of as leading to a highly energetic supernova remnant, a Hyper
Nova Remnant (HNR).  Black holes in turn also merge, but in this manner
start their mergers at masses of a few million solar masses, about ten
percent of the baryonic mass inside the initial dark matter clumps, and
at a fraction of $10^{-4.5}$ of the baryonic mass in their sphere of
influence to the next such black hole.  This readily explains the
supermassive black hole mass function (Caramete \& Biermann 2010).  

\bigskip

The action of the formation of the first super-massive black holes allows a
possible path to determine the dark matter particle mass, under the
proviso that it is a right-handed neutrino, as advocated by some (e.g.,
Kusenko 2009):  
\medskip
(a)  Determine the Galactic radio background spectrum, and check for residual all-sky emission\\ 
\medskip
(b)  Determine the extragalactic radio background spectrum, if possible (Kogut et al. 2011)\\
\medskip
(c)  Match it with various models, such a the Hyper Nova Remnants radio emission\\
\medskip
(d)Such a match implies angular fine structure of this emission on the sky,
which may be detectable\\ 
\medskip
(e)  Determine the Thomson depth through to
recombination, match it with the HII regions, HNRs, the action by X-rays
from the early stellar black holes, or any other model, and determine,
if possible its angular structure on the sky - each have their specific
signature in size and Thomson depth\\
\medskip
(f)  If there is a residual Thomson
depth which is not structured, then all the normal mechanisms fail due
to their spatial distribution, including the HII regions and HNRs, and
only a very distributed source of ionization could explain it\\
\medskip
(g) The strength of the residual Thomson depth directly scales with the action
of the decay of a dark matter particle such as a right-handed neutrino:
This gives the mass of the particle, given sufficient accuracy.\\  

\bigskip

Our conclusion is that a right-handed neutrino of a mass of a few keV is the
most interesting candidate to constitute dark matter.  Its mass
determination seems feasible.

\bigskip

{\bf References}

\begin{description}

\item[1]  Zwicky, F., {\it Helvetica Physica Acta} {\bf 6}, 110 (1933)

\medskip

\item[2]  Biermann, P. L., Becker, J. K., Meli, A., Rhode, W., 		
	Seo, E.-	S., Stanev, T., \PRL {\bf 103}, 061101 (2009); arXiv:0903.4048

\medskip

\item[3]  Biermann, P.L., Becker, J.K., Caceres, G., Meli, A., Seo, E.-S.,
\& Stanev, T., \ApJL {\bf 710}, L53 - L57 (2010); arXiv:0910.1197

\medskip

\item[4]  Gopal-Krishna, Biermann, P.L., de Souza, V., Wiita, P.J., 
ApJL {\bf 720}, L155 - L158 (2010); arXiv:1006.5022
 
\medskip

\item[5]  Biermann, P.L., Becker, J.K., Dreyer, J., Meli, A., Seo, E.-S., \& Stanev, T., \ApJ  {\bf 725}  184 - 187 (2010); arXiv: 1009.5592

\medskip

\item[6]  Biermann, P.L., \& de Souza, V., eprint arXiv: 1106.0625 (2011)

\medskip

\item[7] Biermann, P.L., \AA {\bf 271}, 649 (1993); astro-ph/9301008

\medskip

\item[8] Biermann, P.L., \& Cassinelli, J.P.,  \AA {\bf 277},
  691 (1993); astro-ph/9305003

\medskip

\item[9] Biermann, P.L., \& Strom, R.G., \AA {\bf 275}, 659 (1993); astro-ph/9303013

\medskip  

\item[10]  Stanev, T., Biermann, P.L., \& Gaisser, T.K., 
  \AA {\bf 274}, 902 (1993); astro-ph/9303006

\medskip

\item[11]  Biermann, P.L., invited plenary lecture at 23rd Internat. Conf. 
on Cosmic Rays, in Proc. ``Invited, Rapporteur and Highlight papers''; Eds. 
D. A. Leahy et al., World Scientific, Singapore, p. 45 (1995)
 
 \medskip

\item[12]  Greisen, K., \PRL  {\bf 16}, 748 (1966)  

\medskip

\item[13]  Zatsepin, G. T., Kuz'min, V. A., {\it Zh. E.T.F. Pis'ma Redaktsiiu} 
{\bf 4}, p.114 (1966); transl. {\it J. of Exp. and Theor. Physics Lett.} 
{\bf 4}, 78 (1966)

\medskip
\
item[14]  Ryu, D., Kang, H., Cho, J., Das, S.,\Science {\bf 320}, 909 (2008)

\medskip

\item[15]  Das, S., Kang, H., Ryu, D., Cho, J., \ApJ  {\bf 682}, 29 (2008)
ltra-High-Energy Protons through the 

\medskip

\item[16]  Dalcanton, J. J., Hogan, C. J., \ApJ {\bf 561}, 35 - 45 (2001); 
arXiv:astro-ph/0004381

\medskip

\item[17]  Gilmore, G.,  et al., eprint  astro-ph/0703308 (2007)

\medskip

\item[18]  Wyse, R., \& Gilmore, G.,  eprint arXiv/0708.1492 (2007)

\medskip

\item[19]  Strigari, L. E. \etal., eprint astro-ph/0603775 (2006)

\medskip

\item[20] Boyanovsky, D., de Vega, H. J., Sanchez, N. G., \PRD  {\bf 77}, id. 043518 (2008)

\medskip

\item[21]  Gentile, G., Famaey, B., Zhao, H., Salucci, P., \Nature {\bf 461},  627  (2009)

\medskip

\item[22]  de Vega, H. J., \& Sanchez, N. G., \MNRAS {\bf 404}, 885 (2010); arXiv:0901.0922 (2009)
	
\medskip

\item[23]  de Vega, H. J., \& Sanchez, N. G., eprint arXiv:0907.0006 (2009)
 
\medskip

\item[24] Kusenko, A., Segre, G., \PLB  {\bf 396}, 197 (1997)

\medskip

\item[25] Fuller, G. M., Kusenko, A., Mocioiu, I., Pascoli, S., \PRD {\bf 68}, id. 103002 (2003)

\medskip

\item[26] Kusenko, A., \IJMP {\bf D 13}, 2065 (2004)

\medskip

\item[27]  Kusenko, A., {\it Phys. Rep.} {\bf 481}, 1 (2009)

\medskip

\item[28] Biermann, P. L., \& Kusenko, A., \PRL {\bf 96}, 091301 (2006); astro-ph/0601004

\medskip

\item[29]  Stasielak, J., Biermann, P.L:, \& Kusenko, A., \ApJ {\bf 654}, 290-303 (2007); 
   astro-ph/0606435

\medskip

\item[30]  Loewenstein, M., Kusenko, A., Biermann, P.L., \ApJ {\bf 700}, 426 - 435 (2009); arXiv:0812.2710
\medskip

\item[31]  Kusenko, A., Int.\ J.\ Mod.\ Phys.\ D {\bf 13}, 2065 (2004); astro-ph/0409521

\medskip

\item[32]  Kusenko, A., Takahashi, F., Yanagida, T. T., \PLB {\bf 693}, 144 (2010)

\medskip

\item[33] Adulpravitchai, A., Gu, P.-H., Lindner, M., \PRD  {\bf 82}, id. 073013 (2010)

\medskip

\item[34]  Caramete, L.I., Biermann, P.L., \AA {\bf 521}, id.A55 (2010); arXiv:0908.2764

\medskip

\item[35]  Biermann, P. L., Becker, J. K., Caramete, A. Curutiu, L., Engel, R., Falcke, H., 
Gergely, L. A., Isar, P. G., Maris, I. C., Meli, A., Kampert, K. -H., Stanev, T., 
Tascau, O., Zier, C., invited review for the conference CRIS2008, Malfa, 
Salina Island, Italy, Ed. A. Insolia,{\it Nucl. Phys. B, Proc. Suppl.} 
{\bf 190}, 61 - 78 (2009); arXiv: 0811.1848v3
\item[36] Caramete, L.I., Tascau, O., Biermann. P.L., \& Stanev, T., submitted \AA (2022); arXiv:1106.5109

\medskip

\item[37]  Caramete, L.I., Biermann, P.L., submitted (2011); arXiv:1107.2244

\end {description}

\bigskip

\subsection{ {\it Planck} Collaboration, presented by Carlo BURIGANA}

\vskip -0.3cm

\begin{center}

INAF-IASF Bologna, Via Piero Gobetti 101, I-40129, Bologna, Italy\\
Dipartimento di Fisica, Universit\`a degli Studi di Ferrara, Via Giuseppe Saragat 1, I-44100 Ferrara, Italy

\bigskip

{\bf Results from the  {\it Planck} satellite and their scientific implications} 

\end{center}

\medskip

At the date of mid July 2012, the {\it Planck} (http://www.esa.int/Planck) cosmic microwave background (CMB) anisotropy probe, launched into space on 14 May 2009 at 13:12:02 UTC, by an Ariane 5
ECA launcher, from the Guiana Space Centre, Kourou, French Guiana, is still successfully operating. The spacecraft accumulated data with its two instruments, the High Frequency Instrument (HFI) {\bf [1]}, based on bolometers working between 100 and 857 GHz, and the Low Frequency Instrument (LFI) {\bf [2]}, based on radiometers working between 30 and 70 GHz, up to the consumption of the cryogenic liquids on January 2012, achieving $\simeq 29.5$ months of integration, corresponding to about five complete sky surveys. A further 12 months extension is on-going for observations with LFI only, cooled down with the cryogenic system provided by HFI. Moreover, {\it Planck\/} is sensitive to linear polarization up to $353\,$GHz.
A summary of the {\it Planck} performance is provided in the table. See {\bf [3]} for descriptions of the {\it Planck\/} scientific programme.

\bigskip

CMB maps are contaminated by a significant level of foreground emission of both Galactic and extragalactic origin. 
As for polarization, the most critical Galactic diffuse foregrounds are certainly synchrotron and thermal dust emission,
whereas free-free emission gives a negligible contribution. Other components, like spinning dust and haze, are still poorly known in polarization, while {\it Planck\/} temperature data recently allowed us to improve their comprehension. 
Synchrotron emission is the dominant Galactic foreground signal
at low frequencies, up to $\sim 60$ GHz, where dust emission starts to dominate. At the  {\it Planck\/} frequency channel of 70 GHz Galactic foregrounds are at their minimum level, at
least at angular scales above $\sim $ one degree.
External galaxies are critical only at high $\ell$, and extragalactic radio sources are likely the most crucial in polarization up to frequencies $\sim 200$ GHz, the most suitable for CMB anisotropy experiments. 

Undetected extragalactic point sources contaminate the CMB APS depending on their detection threshold typically around $\simeq 0.1$ Jy. 
Except at very high multipoles, their residual is likely significantly below that coming from
Galactic foregrounds.

\bigskip

Thanks to its great sensitivity and resolution on the whole sky and to its wide frequency coverage that allows a substantial progress in foreground modeling and removal, {\it Planck\/}  will open a new era in our understanding of the Universe and of its astrophysical structures. {\it Planck\/} will improve the accuracy of current measures of a wide set of cosmological parameters by a factor from $\sim 3$ to $\sim 10$ and will characterize the geometry of the Universe with unprecedented accuracy. {\it Planck\/} will put light on many of the open issues in the connection between the early stages of the Universe and the evolution of the cosmic structures, from the characterization of primordial conditions  and perturbations, to the late phases of cosmological reionization.

\bigskip

The {\it Planck\/} perspectives for some crucial selected topics linking cosmology to fundamental physics (the neutrino masses and effective species number, the primordial helium abundance, various physics fundamental constants, the parity property of CMB maps and its connection with CPT symmetry with emphasis to the Cosmic Birefringence, the detection of the stochastic field of gravitational waves) will also show how {\it Planck\/}represents an extremely powerful {\it fundamental and particle physics laboratory}. Some of the above analyses will be carried out mainly through a precise measure of CMB anisotropy angular power spectrum (APS) in temperat ure, polarization and in their 
correlations, whereas others, in particular those related to the geometry of the Universe and to the research of non-Gaussianity signatures, are based on the exploitation of the anisotropy pattern. The most ambitious goal is the possible detection of the so-called B-mode APS.

\bigskip

The first publications of the main cosmological (i.e. properly based on {\it Planck} CMB maps) implications are expected in early 2013, together with the delivery of a first set of {\it Planck\/} maps and cosmological products based on the first 15 months of data. They will be mainly based on temperature data. 

\bigskip

Waiting for these products, a
first multifrequency view of the {\it Planck} astrophysical sky has been presented on
January 2011.
The first scientific results, the so-called {\it Planck}
Early Papers (PEPs) (http://www.sciops.esa.int/index.php?project=PLANCK\&page=Planck\_Published\_Papers), 
have been published by Astronomy and Astrophysics (EDP sciences) in the dedicated Volume 536 (December 2011). They describe the instrument performance in flight including
thermal behaviour (PEPs I--IV), the LFI and HFI data analysis pipelines (PEPs V--VI), 
the main astrophysical results about Galactic science (PEPs XIX--XXV),
extragalactic sources and far-IR background (PEPs XIII--XVIII),
and Sunyaev-Zel'dovich effects and cluster properties  (PEPs VIII--XII and XXVI).
These papers have complemented by a subsequent work on blazars
based on a combination of high energy and {\it Planck} observations (see {\bf [4]}).
The first {\it Planck} sky maps are at the basis of the construction of the {\it Planck\/} Early Release Compact Source Catalog (ERCSC) (see PEP VII
and {\it The Explanatory Supplement to the Planck Early Release Compact Source Catalogue}), the first {\it Planck} product delivered to the scientific community.

\bigskip

Of course, accumulating sky surveys and refining data analysis, the {\it Planck\/}
sensitivity to point sources will significantly improve in next years. The forthcoming {\it Planck\/} Legacy Catalog of Compact Sources (PCCS), to be released in early 2013 and to be updated in the subsequent years, will represent one of the major {\it Planck} products relevant for multi-frequency studies of compact or point--like sources.

\bigskip

A further set of astrophysical results has been presented in occasion of the Conference {\it Astrophysics
from radio to sub-millimeter wavelengths: the {\it Planck\/} view and other
experiments} held in Bologna on 13-17 February 2012 (http://www.iasfbo.inaf.it/events/planck-2012/).
Several articles, constituting the set of so-called {\it Planck} Intermediate Papers (PIPs), have been already submitted in 2012 (PIPs I-X, at the date of October 2012, see again
http://www.sciops.esa.int/index.php?project=PLANCK\&page=Planck\_Published\_Papers) and others are in preparation. 
{\it Planck} papers will be roughly divided into two wide categories: those mainly based only on total intensity data and those requiring well established (and, typically, with a longer time schedule) polarization data. 
Among the various topics addressed by PIPs, the currently available papers mainly focus on cluster physics and their cosmological relevance, derived from {\it Planck} data through the analysis of the Sunyaev-Zel'dovich effect and in combination with X-ray data from XMM-Newton and the microwave ground-based interferometer AMI, and on the improvement achieved on number counts and spectral indices of extragalactic sources, infrared and radio sources (see {\bf [5]} for a review on these topics). 
Regarding Galactic science, using precise full-sky observations from {\it Planck}, and applying several methods of component separation, the emission from the Galactic haze at microwave wavelengths has been 
identified and characterized. Finally, we remember the presentation of the map of CO Galactic emission that can be used to search for faint CO emission associated with the dark-gas and to guide ground-based follow-up observations.

\bigskip

\begin{table}[!ht]
  \caption{{\it Planck\/} performances. The average sensitivity, $\delta$T/T,
per  FWHM$^2$ resolution element (FWHM is reported in arcmin) is
given in CMB temperature units (i.e. equivalent thermodynamic
temperature) for for 29.5 (plus 12 for LFI) months of integration.. The white noise (per
frequency channel for LFI and per detector for HFI) in 1~sec of
integration (NET,  in  $\mu$K $\cdot \sqrt{{\rm s}}$) is also given
in CMB temperature units. The other used acronyms are: DT = detector
technology, N of R (or B) = number of radiometers (or bolometers),
EB = effective bandwidth (in GHz). Adapted from {\bf [1,2]} and consistent with PEPs III and IV. 
Note that at 100 GHz all bolometers are polarized, thus the temperature measure is
derived combining data from polarized bolometers.
Of course, the real sensitivity of the whole mission will have to
include also the potential residuals of systematic effects. 
The {\it Planck\/} mission has been designed to suppress potential systematic effects down to $\sim \mu K$ level or below.}

\bigskip

    \begin{tabular}{l c c c}
\hline
LFI & & &\\
\hline
    Frequency (GHz) &   $30\,$  & $44\,$    & $70\,$ \\
\hline
    InP DT  & MIC   & MIC   & MMIC \\

    FWHM    & 33.34 &   26.81 & 13.03 \\
    N of R (or feeds)   & 4 (2) & 6 (3) & 12 (6) \\
    EB  & 6 & 8.8   & 14 \\
    NET & 159 & 197 & 158 \\
    $\delta$T/T [$\mu$K/K] (in $T$) & 2.04    & 3.14 & 5.17 \\
    $\delta$T/T [$\mu$K/K] (in $P$) & 2.88    & 4.44 & 7.32 \\

\end{tabular}
    \begin{tabular}{l c c}
\hline
HFI & & \\
\hline
    Frequency (GHz)     & $100\,$   & $143\,$    \\
\hline

         FWHM  in $T$ ($P$)     & 9.6 (9.6) & 7.1 (6.9)  \\
            N of B in $T$ ($P$) & (8) & 4 (8)  \\
    EB  in $T$ ($P$) & 33 (33) & 43 (46)   \\
    NET in $T$ ($P$) & 100  (100) & 62 (82)  \\

    $\delta$T/T [$\mu$K/K] in $T$ ($P$) &  2.04 (3.31)  & 1.56 (2.83) \\
\hline
\end{tabular}
    \begin{tabular}{l c c c c}
\hline
HFI & & \\
\hline
    Frequency (GHz) &    $217\,$  & $ 353 \,$ & & \\
\hline

    FWHM     in $T$ ($P$) & 4.6 (4.6) &  4.7 (4.6) & & \\
    N of B in $T$ ($P$)   & 4 (8) & 4 (8) & & \\
    EB  in $T$ ($P$) &  72 (63)     & 99 (102) & & \\
    NET  in $T$ ($P$) & 91 (132) & 277 (404) & & \\

    $\delta$T/T [$\mu$K/K] in $T$ ($P$) & 3.31 (6.24)  & 13.7 (26.2) & & \\
\hline
\end{tabular}
    \begin{tabular}{l c c}
\hline
HFI & & \\
\hline 
    Frequency (GHz) &    $545\,$  & $ 857 \,$ \\
\hline

    FWHM     in $T$ & 4.7 &  4.3 \\
    N of B in $T$   & 4 & 4  \\
    EB  in $T$ & 169    & 257 \\
    NET in $T$ & 2000 & 91000 \\

    $\delta$T/T [$\mu$K/K] in $T$   & 103    & 4134 \\
\hline
\end{tabular}
\label{table:sens}
\end{table}

\bigskip

{{\small{\noindent
{\bf Acknowledgements --} {CB acknowledges the support by the ASI/INAF Agreement I/072/09/0 for the {\it Planck\/}  LFI Activity of Phase E2 and by MIUR through PRIN 2009 grant n. 2009XZ54H2.
{\it Planck} is a project of the European Space Agency - ESA - with instruments provided by
two scientific Consortia funded by ESA member states (in particular the lead countries: France and Italy) with
contributions from NASA (USA), and telescope reflectors provided in a collaboration between ESA and a scientific
Consortium led and funded by Denmark.}}}}

\bigskip

{\bf References}

\begin{description}

\item[1] N. Mandolesi, et al., 2010, A\&A 520, A3

\medskip

\item[2] J.-M. Lamarre, et al., 2010, A\&A 520, A9

\medskip

\item[3] {\it Planck\/} Collaboration, 2005, ESA publ. ESA-SCI(2005)/01, arXiv:0604069; J. Tauber, et al., 2010, A\&A 520, A1

\medskip

\item[4] P. Giommi, et al., 2011, A\&A 541, A160

\medskip

\item[5] Toffolatti L., Burigana C., Arg\"ueso F., Diego J.M., Extragalactic Compact Sources in the  {\it Planck\/} Sky and their Cosmological Implications, (2012), in Open Questions in Cosmology, InTechOpen (www.intechopen.com), Scient. Ed. Gonzalo Olmo (Valencia, Spain), in press.

\end{description}

\newpage

\subsection{Christopher J. CONSELICE}

\vskip -0.3cm

\begin{center}

Centre for Particle Theory and Astronomy, University of Nottingham, NG7 2RD\\

\bigskip

{\bf Galaxy Formation in an Observational and Cosmological Context} 

\end{center}

\medskip

The formation of galaxies is a difficult problem, and we are still debating  
the exact history and the physics responsible for establishing the modern  
galaxy population.  One of the issues concerning galaxy formation that is  
now becoming clear is that this formation and evolution happens in a
cosmological context, which ultimately implies that galaxy assembly 
involves the structure of the universe, as well as the properties of the 
dark matter.

\bigskip

One the major current ideas regarding this is while the exact nature of 
dark matter is not known, the major dominant paradigm is that the dark 
matter is cold, an idea which is successful at reproducing the large scale 
features of the universe.  As a result theory can now predict how 
galaxies form within this context. In fact, there are now many simulations
 and models for how galaxies form within the Cold Dark matter paradigm, 
however most of these simulations to date have not been able to reproduce 
the abundances or the formation history of massive galaxies 
(e.g., Conselice et al. 2007; Bertone \& Conselice 2009; Marchesini et al. 
2010; Guo et al. 2011).

\medskip

The summary of this is as follows: when examining the abundances of massive 
galaxies, typically those with M$_{*} > 10^{11}  \, M_{\odot} $, the number densities 
up to $z \sim 1-2$ are similar to what is found in the local universe 
(e.g., Conselice et al. 2011; Mortlock et al. 2011).  While in CDM models 
galaxy formation happens hierarchically in the sense that galaxies form 
largely through mergers (both major and minor), the exact role of this 
process is not yet clear (e.g., Conselice et al. 2003). Therefore to 
study the property of the the most massive galaxies and their formation 
we must go to higher redshifts, namely at $z > 2$.  

\bigskip

Recent results have accomplish this by examining the most massive systems 
at $1.5 < z < 3$ with a Hubble Space Telescope survey of 82 massive galaxies 
called the GOODS NICMOS Survey (GNS).  Recent results from 
this survey have shown that the formation modes for the most massive 
galaxies are dominated by mergers (Bluck et al. 2009; 2011) and the 
accretion of gas from the intergalactic medium (Conselice et al. 2012) with
the bulk of this formation occurring at $z > 1$ (Mortlock et al. 2011). 

\bigskip

The star formation rate for massive galaxies with M$_{*} > 10^{11} \, M_{\odot} $ 
is also relatively constant over the epoch $1 < z < 3$ (Bauer et al. 2011), 
increasing the stellar mass within massive galaxies by  
112$\pm$42\%, thereby approximately doubling it.
When examining the likely gas masses within these systems at $z \sim 3$ 
using the inverse Schmidt-Kennicutt relation, it can easily be shown that 
not enough gas mass is present to sustain this star formation, by a factor 
of roughly five.   

\bigskip

As we have made the first measurement of the 
minor merger history for this galaxies within GNS (Bluck et al. 2012), 
we are able to make a measured calculation of how much gas is likely 
brought in with these.  The result is an imbalance between the amount 
of gas needed to sustain the star formation we observe and the amount
present in-situ within the galaxy plus that brought in from mergers, suggesting
that gas accretion from the intergalactic medium is an important process
for driving baryons into galaxies.  Overall we calculate that the
gas accretion rate is 78$\pm 40 \, M_{\odot} $ per year over the epoch
$1 < z < 3$ within these massive systems (Conselice et al. 2012).

\bigskip

These results show that major and minor mergers are roughly equally 
important in the galaxy formation process for systems with 
M$_{*} > 10^{11} \, M_{\odot}$ (Bluck et al. 2012).   Gas accretion, as measured 
using the method above, is even more important than mergers 
(e.g., Conselice et al. 2012).  The ratio between the amount of stellar mass 
created from gas brought in from cold gas accretion and mergers 
(both major+minor) is 60 to 40\%, with gas accretion the slightly dominant 
method.  Theoretical results predicting this ratio are inconsistent -- with 
either mergers or accretion the dominant method, with significant variation 
(e.g., Conselice et al. 2012).

\bigskip

As mentioned above, the number densities of these massive galaxies are 
underpredicted in galaxy formation models at $z > 2$ 
(e.g., Guo et al. 2010).    To further investigate this problem requires 
that we investigate how the formation process of 
galaxies occurs, and whether models can reproduce this.  One of major 
methods for doing this is to investigate how well CDM models can reproduce 
the formation history of galaxies as seen through processes such as merging.   

\bigskip

For example, Bertone \& Conselice (2009) compare the merger history of 
galaxies to the predictions from the Millennium simulation.  This 
comparison shows that the Millennium simulation underpredicts the number of 
major mergers by a similar order of magnitude (factor of 10) that it 
underpredicts the abundances of galaxies.  The reasons for this are 
unclear, but may relate to either the underlying cosmological 
assumptions, or the way in which baryons are implement in these simulations.  
Future work, including investigating the underlying role of dark matter, and 
other cosmological assumptions, will have to be included in future research 
on how massive galaxies form.

\bigskip

{\bf References}

\begin{description}

\item[1] Bertone, S., Conselice, C.J. 2009, MNRAS, 396, 2345

\medskip

\item[2] Bauer, A.E. et al. 2011, MNRAS, 417, 289

\medskip

\item[3] Bluck, A., et al. 2009, MNRAS, 394, 51L

\medskip

\item[4] Bluck, A., et al. 2012, MNRAS, 747, 34

\medskip

\item[5] Conselice, C.J., Bershady, M.A., Dickinson, M., Papovich, C. 2003, AJ, 126, 1183

\medskip

\item[6] Conselice, C.J., et al. 2007, MNRAS, 381, 962

\medskip

\item[7] Conselice, C.J., et al. 2011, MNRAS, 413, 80

\medskip

\item[8] Conselice, C.J., et al. 2012, arXiv:1206.6995

\medskip

\item[9] Guo, Q., et al. 2011, MNRAS, 413, 101

\medskip

\item[10] Marchesini, D., et al. 2010, ApJ, 725, 1277

\medskip

\item[11] Mortlock, A. et al. 2011, MNRAS, 413, 2845

\end{description}

\bigskip

\subsection{Asantha COORAY}

\vskip -0.3cm

\begin{center}

Department of Physics and Astronomy, University of California, Irvine, CA 92697 USA.

\bigskip

{\bf Summary of Extragalactic Survey Results from the Herschel Space Observatory}

\end{center}

\medskip

In my talk I summarized some of the recent results published by extragalactic survey consortia, mainly Herschel HerMES and ATLAS collaborations,
related to properties of the distant dusty starburst galaxies, gravitational lensing at sub-mm wavelengths, clustering, and cosmic infrared background anisotropies,
among others.

In particulat, I summarized the results of Cooray et al. (2010) [1] related to Herschel source clustering, which finds that the halo mass
scale is few times $10^{12}$ solar masses for the bright sub-mm galaxies with flux densities above 30 mJy at 250 microns.

\bigskip

Related to gravitational lensing I showed results from Fu et al. (2012) [2], a Planck-detected lensed SMG in Herschel-ATLAS data and
Wardlow et al. (2012) [3], lensing rate statistics from HerMES. The latter paper presents a complete sample of lensed SMGs down to flux
densities of 80 mJy at 500 microns and finds the surface density of the lensed SMGs to be about 0.15 deg$^{-2}$ when the flux density is
about 100 mJy at 500 microns. All of those sources, apart from those identified with nearby galaxies or radio-loud blazars, are gravitationally
lensed and the efficiency of their identification is close to 100\%.  Other results related to Herschel lensed galaxies are presented in Refs. [4,5].

\medskip

With Herschel surveys covering more than 1000 deg.$^2$, the general expection
is that Herschel surveys will lead to a clear sample of about 250 lensed galaxies, with lensed galaxies mainly at $z \sim 1$ to 4. A few are expected
to be at $z > 6$, consistent with existing results. In terms of the foreground lenses, we expect few to be at $z \sim 1$ to 2, allowing us to
extend the existing mass and density profile mesurements of galaxies to high redshifts than possible with optically-selected lensed galaxies, such as
in SDSS.

\bigskip

Related to CIB anisotropies and the angular power spectrum, I discussed results from Amblard et al. [6] that measured the CIB power spectrum
in detail, with a clear detection of the 1-halo term, with HerMES data in the Lockman-SWIRE field. Those measurements have been interpreted
in terms of the halo model for the faint source distribution with an efficient halo mass scale for sub-mm galaxies of $3 \times 10^{11}$ solar masses.
I presented results from a new halo model for CIB anisitropy power spectrum in De Bernardis \& Cooray [7] that fixes some of the issues in the earlier
simple models. This includes a faint-end slope for the satellites that is steeper than 1, inconsistent with the satellite halo occupation distribution 
in the optical wavelengths.

\bigskip

It is generally challenging to cross-identify sub-mm galaxies at other wavelengths due to poor spatial resolution, and large confusion noise floor,
of Herschel SPIRE imaging data. Some early results from my research group, discussed at the meeting, are presented in Kim et al. [8].

\bigskip

{\bf References}

\begin{description}

\item[1] Cooray, A. et al. 2010, A\&A Herschel Special Issue, 518, L18

\medskip

\item[2] Fu, H. et al. 2012, ApJ, 753, 134

\medskip

\item[3] Wardlow, J. et al. 2012, arXiv.org:1205.3778

\medskip

\item[4] Bussmann, S. et al. 2012, ApJ, 756, 134

\medskip

\item[5] Gonzalez-Nuevo, J, 2012, ApJ, 749, 65

\medskip

\item[6] Amblard, A. et al. 2011, Nature, 470, 510

\medskip

\item[7] De Bernardis, F. \& Cooray, A. 2012, arXiv.org:1206.1324

\medskip

\item[8] Kim, S. et al. 2012, ApJ, 756, 28

\end{description}

\bigskip

\subsection{Hector J. DE VEGA and Norma G. SANCHEZ}

\vskip -0.3cm

\begin{center}

H.J.dV: LPTHE, CNRS/Universit\'e Paris VI-P. \& M. Curie \& Observatoire de Paris, Paris, France\\
\medskip
N.G.S: LERMA, CNRS/Observatoire de Paris, Paris, France

\medskip

{\bf Predictions of the Effective Theory of Inflation in the Standard Model of the 
Universe and the CMB+LSS data analysis}

\end {center}

Inflation is today a part of the Standard Model of the Universe supported by
the cosmic microwave background (CMB) and large scale structure (LSS)
datasets. Inflation solves the horizon and flatness problems and
naturally generates  density fluctuations that seed LSS and CMB anisotropies,
and tensor perturbations (primordial gravitational waves). 
Inflation theory is
based on a scalar field  $ \varphi $ (the inflaton) whose potential
is fairly flat leading to a slow-roll evolution. 

\vskip 0.1 cm

We focus here on the following new aspects of inflation. \\
\medskip
We present the
effective theory of inflation \`a la {\bf Ginsburg-Landau} in which
the inflaton potential is a polynomial in the field $ \varphi $ and has
the universal form $ V(\varphi) = N \; M^4 \;
w(\varphi/[\sqrt{N}\; M_{Pl}]) $, where $ w = {\cal O}(1) , \;
M \ll M_{Pl} $ is the scale of inflation and  $ N \sim 60 $ is the number
of efolds since the cosmologically relevant modes exit the horizon till
inflation ends. 

\medskip

The slow-roll expansion becomes a systematic $ 1/N $ expansion and
the inflaton couplings become {\bf naturally small} as powers of the ratio
$ (M / M_{Pl})^2 $. 

\medskip

The spectral index and the ratio of tensor/scalar
fluctuations are $ n_s - 1 = {\cal O}(1/N), \; r = {\cal O}(1/N) $ while
the running index turns to be $ d n_s/d \ln k =  {\cal O}(1/N^2) $
and therefore can be neglected. 

\medskip

The {\bf energy scale of inflation }$ M \sim 0.7
\times 10^{16}$ GeV turns to be completely determined by the amplitude of the scalar 
adiabatic fluctuations [1-2].

\begin{figure}[ht]
\includegraphics[height=6cm,width=10cm]{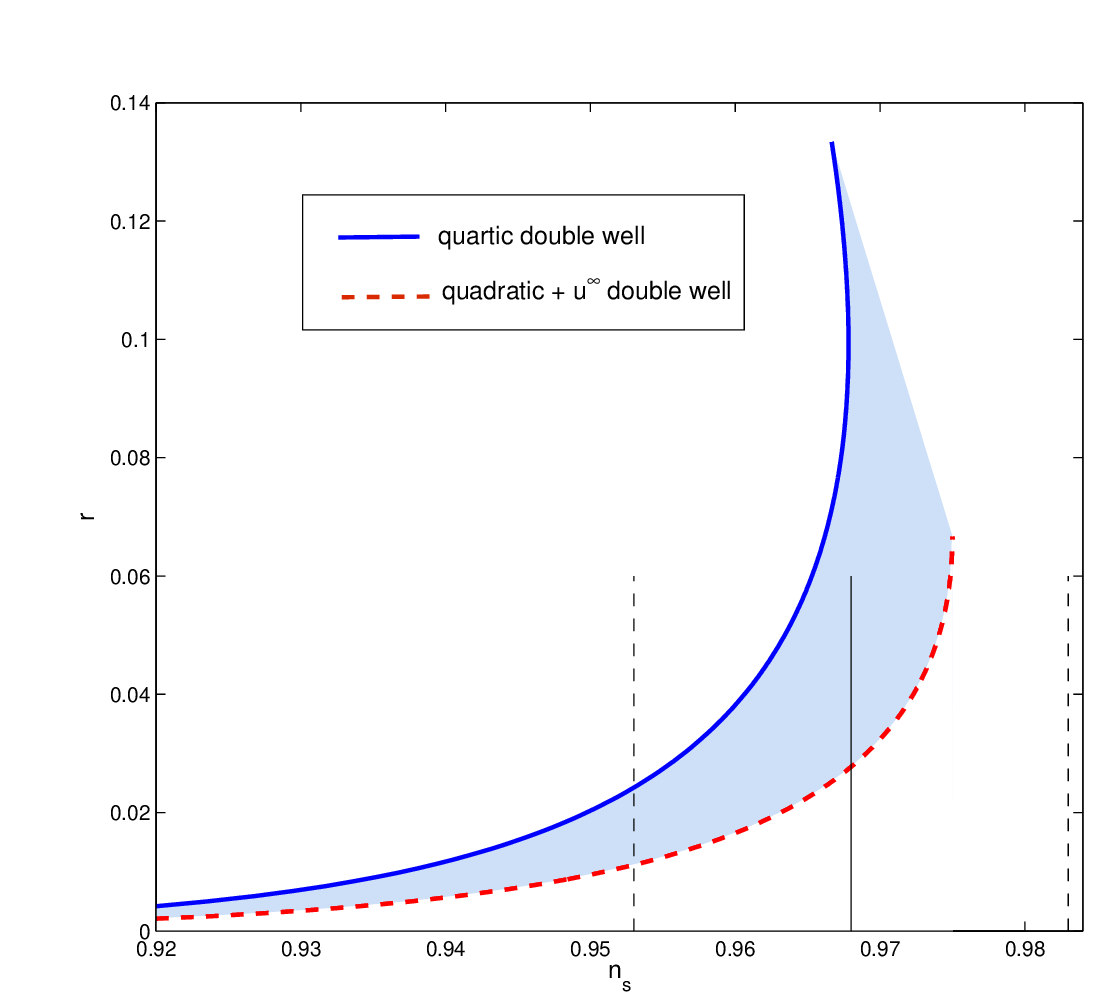}
\caption{The universal banana region $ \cal B $ in the $ (n_s, r) $-plane
  setting $ N = 60 $. The upper border of the region $ \cal B $ corresponds to
  the fourth order double--well potential (new inflation). The lower border is
  described by the potential $ V(\varphi) = \frac12{ m^2} \,
  \left(\frac{m^2}{\lambda} - \varphi^2\right) $ for $ \varphi^2 < m^2/\lambda $
  and $ V(\varphi) = \infty $ for $ \varphi^2 > m^2/\lambda $ [4].  We
  display in the vertical full line the observed value $ n_s = 0.968 \pm
  0.015 $ using the WMAP+BAO+SN data set.  The broken vertical lines delimit
  the $ \pm 1 \, \sigma$ region.}
\label{banana}
\end{figure}

\bigskip

A complete analytic study plus the
Monte Carlo Markov Chains (MCMC) analysis of the available
CMB+LSS data (including WMAP5) with fourth degree trinomial potentials
showed [1-3]: 

\begin {itemize}

\item {{\bf(a)} the {\bf spontaneous breaking} of the
$ \varphi \to - \varphi $ symmetry of the inflaton potential.} 

\vskip 0.1 cm

\item {{\bf(b)} a {\bf lower bound} for $ r $ in new inflation:
$ r > 0.023 \; (95\% \; {\rm CL}) $ and $ r > 0.046 \;  (68\% \;
{\rm CL}) $. }

\vskip 0.1 cm

\item {{\bf(c)} The preferred inflation potential is a {\bf double
well}, even function of the field with a moderate quartic coupling
yielding as most probable values: $ n_s \simeq 0.964 ,\; r\simeq
0.051 $. This value for $ r $ is within reach of forthcoming CMB
observations. }

\vskip 0.1 cm

\item {{\bf(d)} The present data in the effective theory of
inflation clearly {\bf prefer new inflation}. }

\vskip 0.1 cm

\item { {\bf(e)} Study of higher degree
inflaton potentials show that terms of degree higher than four do not
affect the fit in a significant way. In addition, horizon exit happens for
$ \varphi/[\sqrt{N} \; M_{Pl}] \sim 0.9 $ making higher order terms
in the potential $ w $ negligible [4]. }

\item { {\bf(f)} Within the Ginsburg-Landau potentials in new inflation,
$ n_s$ and  $r$ in the $(n_s, r)$ plane are within the universal banana region 
fig. \ref{banana} and $ r $ is in the range $ 0.021 < r < 0.053 $ [4].}

\end {itemize}

We summarize the physical effects of
{\bf generic} initial conditions (different from Bunch-Davies) on the
scalar and tensor perturbations during slow-roll and
introduce the transfer function $ D(k) $ which encodes the observable
initial conditions effects on the power spectra.
These effects are more prominent in the \emph{low}
CMB multipoles: a change in the initial conditions during slow roll can
account for the observed CMB {\bf quadrupole suppression} [1].

\vskip 0.2 cm

Slow-roll inflation is generically preceded by a
short {\bf fast-roll} stage. Bunch-Davies initial conditions are the
natural initial conditions for the fast-roll perturbations. 

\vskip 0.2 cm

The characteristic time scale of the fast-roll era turns to be 
$ t_1= (1/m) \; \sqrt{V(0)/[3 \; M^4] } \sim 10^4 \;  t_{Planck} $.
The {\bf whole} evolution of the fluctuations along the 
decelerated and inflationary 
fast-roll and slow-roll eras is computed in ref. [5]. 

\vskip 0.2 cm

The Bunch-Davies initial conditions 
(BDic) are generalized for the fast-roll case in which the potential felt by the 
fluctuations can never be neglected. The fluctuations feel 
a {\bf singular attractive} potential near the $ t = t_* $ singularity 
(as in the case of a particle in a central singular potential) with {\bf exactly} the 
{\bf critical} strength ($ -1/4 $) allowing the fall to the centre.

\vskip 0.2 cm

The power spectrum gets {\bf dynamically modified} by the effect of the 
fast-roll eras and the choice of BDic at a finite time, 
through the transfer function $ D(k) $. The power spectrum
vanishes at $ k = 0 . \; D(k) $ presents a first peak for $ k \sim 2/\eta_0 $
($ \eta_0 $ being the conformal initial time), 
then oscillates with decreasing amplitude and vanishes asymptotically for $ k \to \infty $.
The transfer function $ D(k) $ affects the {\bf low} CMB multipoles $ C_{\ell} $:
the change $ \Delta C_{\ell}/ C_{\ell} $ for $ 1 \leq \ell \leq 5 $ is computed in [5] as a
function of the starting instant of the fluctuations $ t_0 $.
CMB quadrupole observations indicate large {\bf suppressions} which
are well reproduced for the range  $ t_0 - t_\ast \gtrsim 0.05/m \simeq 10100 \; t_{Planck} $.

\vskip 0.2 cm

A MCMC analysis of the WMAP+SDSS data {\bf including fast-roll} shows that the quadrupole
mode exits the horizon about 0.2 efold before fast-roll ends and its
amplitude gets suppressed. In addition, fast-roll fixes the {\bf initial
inflation redshift} to be $ z_{init} = 0.9 \times 10^{56} $ and
the {\bf total number} of efolds of inflation to be $ N_{tot} \simeq 64 $ [1,3].
Fast-roll fits the TT, the TE and the EE
modes well reproducing the quadrupole supression. 

\vskip 0.2 cm

A thorough study of the
{\bf quantum loop corrections} reveals that they are very small and controlled by
powers of $(H /M_{Pl})^2 \sim {10}^{-9} $, {\bf a conclusion that validates the
reliability of the effective theory of inflation [1].} 

\vskip 0.2 cm

This work [1-4] shows how powerful is
the Ginsburg-Landau effective theory of inflation in predicting
observables that are being or will soon be contrasted to observations.

\vskip 0.2 cm

The Planck satellite provides with unprecedented accuracy 
the primary CMB anisotropies.
The Standard Model of the Universe (including inflation) provides the context to analyze the CMB 
and other data. The Planck performance for $ r $ related to the primordial $ B $ 
mode polarization, will depend on the quality of the data analysis.

\vskip 0.2 cm

The Ginsburg Landau approach to inflation allows to take high benefit of the CMB data.\_

We evaluate the Planck precision to the recovery of cosmological
parameters within a reasonable toy model for residuals of systematic
effects of instrumental and astrophysical origin based on publicly
available information.
We use and test two relevant models: the $\Lambda$CDM$r$ model,
i. e. the standard $\Lambda$CDM model augmented by  $ r $, and the
$\Lambda$CDM$r$T model, where the scalar spectral index, $ n_s $, and $ r $
are related through the theoretical `banana-shaped' curve $ r = r(n_s) $
coming from the double-well inflaton potential (upper boundary of the banana region
fig. \ref{banana}. In the latter case,
$ r = r(n_s) $ is imposed as a hard constraint in the MCMC data
analysis. We take into account the white noise sensitivity of Planck in the 70,
100 and 143 GHz channels as well as the residuals from systematics
errors and foregrounds. Foreground residuals turn to affect
only the cosmological parameters sensitive to the B modes [6].

\vskip 0.2 cm

In the Ginsburg-Landau inflation approach, better measurements on $ n_s $, 
as well as on TE and EE modes will improve 
the prediction on $ r $ even if a detection of B modes is still
lacking [6].

\vskip 0.2 cm

Forecasted B mode detection probability by
the most sensitive HFI-143 channel:
At the level of foreground residual equal to
30\% of our toy model, only a 68\% CL detectiof $ r $ is very likely.
For a 95\% CL detection the level of
foreground residual should be reduced to 10\%
or lower of the adopted toy model. The possibility to detect
$ r $ is borderline [6].

\vskip 0.2 cm

In Table \ref{etivspl} we compare the predictions of the effective theory of inflation (ETI)
[1-4] with the first Planck data release in 2013. The Planck satellite confirms so far
all our predictions.

\begin{table}
\begin{tabular}{|c|c|c|} \hline  
 Quantity  &  ETI Prediction & Planck 2013 \\ \hline 
Spectral index $ 1 - n_s $ & order $ 1/N = 0.02 $ & 0.04 \\ \hline  
Running $ d n_s/d ln k $ &  order $ 1/N^2 = 0.0004 $ & $ < 0.01 $ \\ \hline  
Non-Gaussianity $ f_{NL} $ & order $ 1/N = 0.02 $ & $ < 6 $  \\ \hline  
 & ETI + WMAP+LSS &  \\ \hline 
tensor/scalar ratio $r$ &  r = 0.04-0.05 & $ < 0.11 $  \\ \hline 
inflaton potential &  &  \\ 
curvature $ V''(0) $   & $ V''(0) < 0 $ & $ V''(0) < 0 $ \\ \hline 
\end{tabular}
\caption{ETI: effective theory of inflation [1-4].
ETI + WMAP+LSS means the MCMC analysis combining the ETI
with WMAP and LSS data. 
Such analysis called for an inflaton potential with negative curvature
at horizon exit. The double well potential is favoured
(new inflation). The Planck satellite confirms the ETI predictions.}
\label{etivspl}
\end{table}

\begin{figure}[ht]
  \includegraphics[height=5cm]{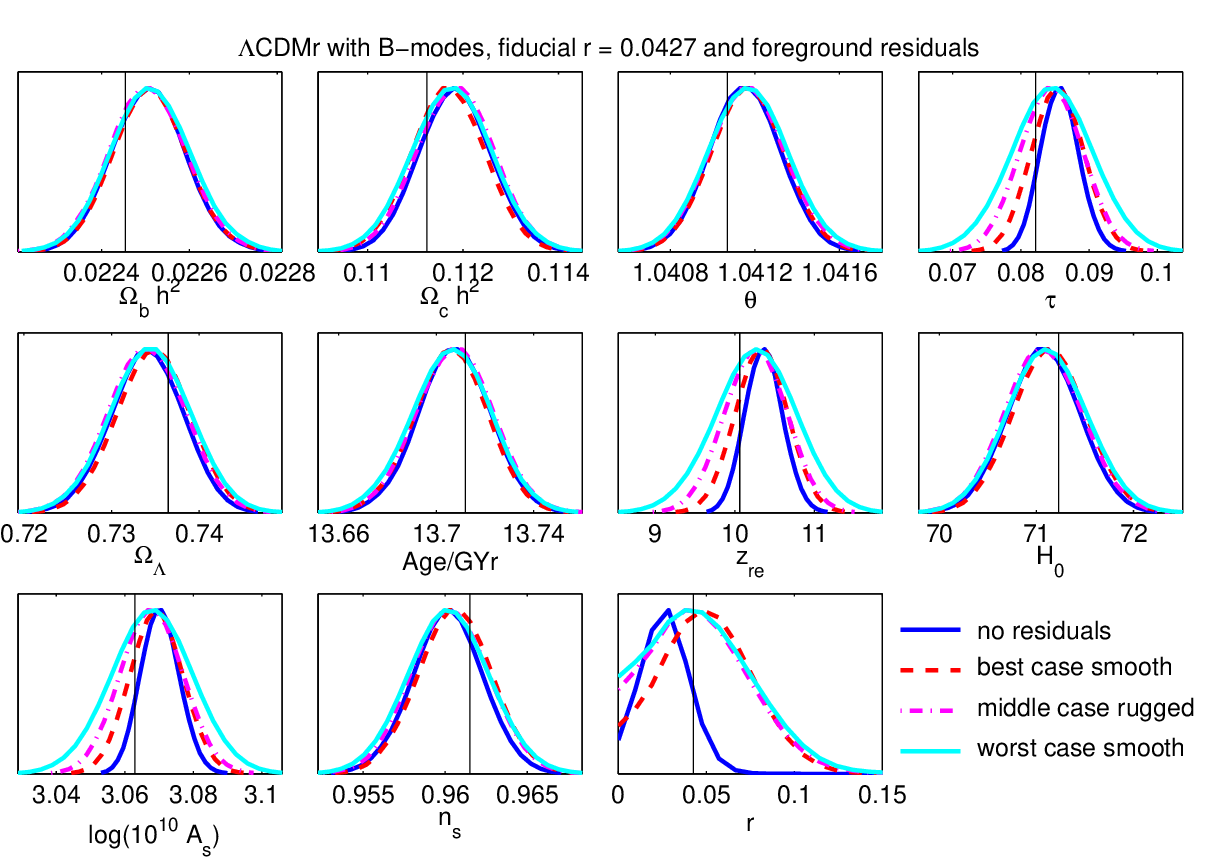}\\
  \includegraphics[height=5cm]{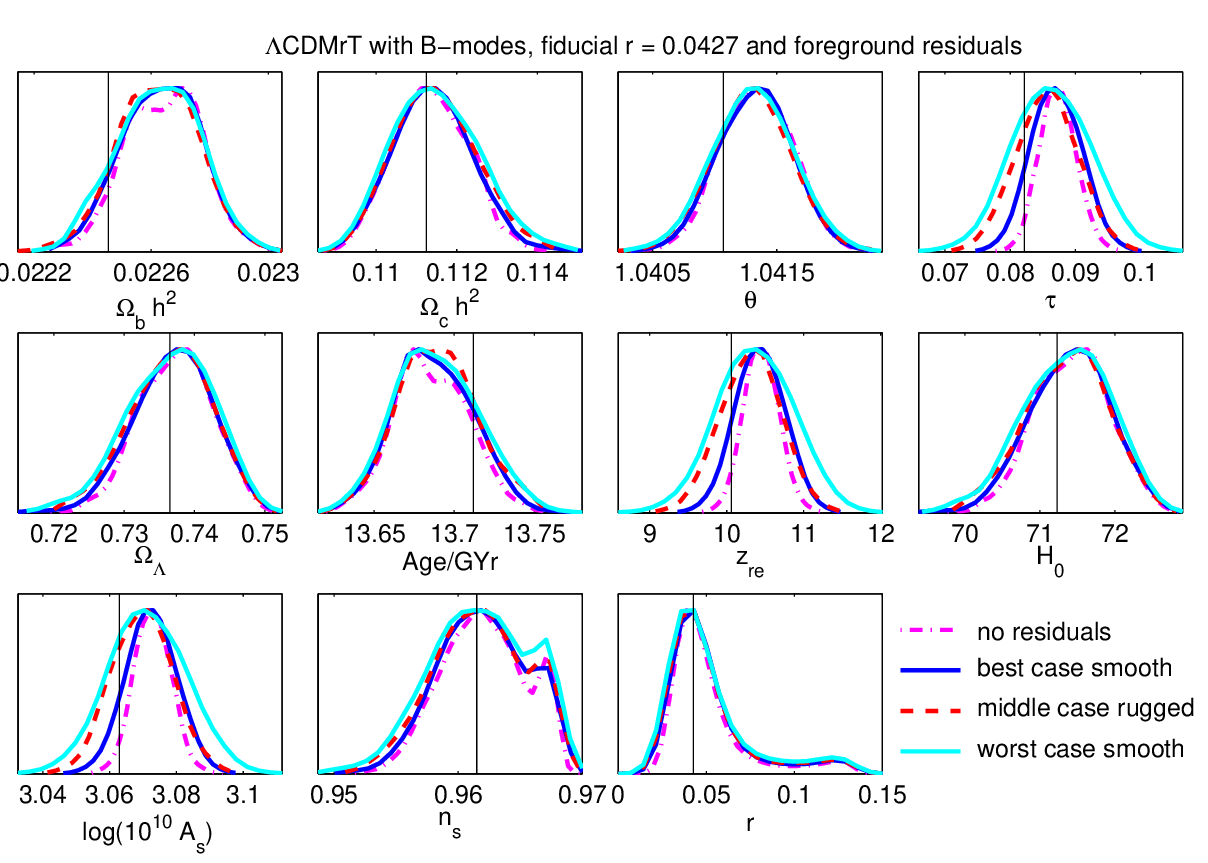}
  \caption{Forecasts for Planck [6]. 
Upper panel: Cumulative $3-$channel marginalized likelihood distributions,
    including $ B $ modes and foreground residuals, of the cosmological
parameters for the $\Lambda$CDM$r$ model.  The fiducial ratio is 
$ r = 0.0427 $. Lower panel:
Cumulative marginalized likelihoods from the three channels for the
    cosmological parameters for the $\Lambda$CDM$r$T model including $ B $ modes
    and fiducial ratio $ r = 0.0427 $ and the foreground residuals. 
We plot the
distributions  in four cases: (a) without residuals, (b) best case smooth:
with 30\% of the toy model residuals in the $TE$ and $E$ modes.
(c) worst case smooth: with the toy model residuals in the $TE$ and $E$ modes.
(d) with 65\% of the toy model 
residuals in the $TE$ and $E$ modes 
and $ 88 \mu K^2 $ in the $T$ modes rugged by 
Gaussian fluctuations of $ 30 \% $ relative strength. }
  \label{ufa7}
\end{figure}

\bigskip

{\bf References}

\bigskip
 
[1]  Review article: D. Boyanovsky, C. Destri, H. J. de Vega, N. G. Sanchez

Int. J. Mod. Phys. A24, 3669-3864 (2009) and  author's references therein.

\medskip

[2] C. Destri, H. J. de Vega, N. Sanchez,
Phys. Rev. D77, 043509 (2008), astro-ph/0703417.

\medskip

[3] C. Destri, H. J. de Vega, N. G. Sanchez, arXiv:0804.2387.
Phys. Rev. D 78, 023013 (2008).

\medskip

[4] C. Destri, H. J. de Vega, N. G. Sanchez, arXiv:0906.4102, 
Annals of Physics, {\bf 326}, 578 (2011).\\
\noindent D. Boyanovsky, H. J. de Vega, C. M. Ho et N. G. Sanchez, 
Phys. Rev.  D75, 123504 (2007).

\medskip

[5] C. Destri, H. J. de Vega, N. G. Sanchez, 
Phys. Rev. D81, 063520 (2010).

\medskip

[6] C. Burigana, C. Destri, H. J. de Vega, A. Gruppuso, N. Mandolesi, P. Natoli, N. G. Sanchez,
arXiv:1003.6108, Astrophysical Journal, 724, 588 (2010).

\bigskip

\subsection{Hector J. DE VEGA and Norma G. SANCHEZ}

\vskip -0.3cm

\begin{center}

HJdV: LPTHE, CNRS/Universit\'e Paris VI-P. \& M. Curie \& Observatoire de Paris.\\
\medskip
NGS: Observatoire de Paris, LERMA \& CNRS

\bigskip

{\bf Quantum WDM fermions and gravitation determine the observed galaxy structures.} 

\end{center}

Warm dark matter (WDM) means DM particles with mass $ m $ in the keV scale.
For large scales, for structures beyond $ \sim 100$ kpc, WDM and CDM yield identical results 
which agree with observations. For intermediate scales, WDM gives the correct abundance of substructures.
Inside galaxy cores, below $ \sim 100$ pc, $N$-body classical physics simulations 
are incorrect for WDM because at such scales quantum WDM effects are important.
Quantum calculations (Thomas-Fermi approach) for the WDM fermions provide galaxy cores, 
galaxy masses, velocity dispersions and density profiles in agreement with the observations.
All evidences point to a dark matter particle mass around 2 keV.
Baryons, which represent 16\% of DM, are expected to give a correction to pure WDM results.

\bigskip

The detection of the DM particle depends upon the particle physics model.
Sterile neutrinos with keV scale mass (the main WDM candidate) can be detected in 
beta decay for Tritium and Renium and in the electron capture in Holmiun.
The sterile neutrino decay into X rays can be detected observing DM
dominated galaxies and through the distortion of the black-body CMB spectrum.
The effective number of neutrinos, N$_{\rm eff}$ measured by WMAP9 and Planck satellites
is compatible with one or two Majorana sterile neutrinos in the eV mass scale.
The WDM contribution to  N$_{\rm eff}$ is of the order $ \sim 0.01 $ and therefore
too small to be measurable by CMB observations.

So far, {\bf not a single valid} objection arose against WDM.

\bigskip

The characteristic length scale after DM decoupling
is the {\bf free streaming scale (or Jeans' scale)}. 
Solving the linear Boltzmann-Vlasov equations for ultrarelativistic decoupling yields (see [1]),
\be\label{fs}
r_{Jeans} = 57.2 \, {\rm kpc}
\; \frac{\rm keV}{m} \; \left(\frac{100}{g_d}\right)^{\! \frac13} \; , 
\ee
where $ g_d $ equals the number of UR degrees of freedom at decoupling. 
DM particles can {\bf freely} propagate over distances of the order of the free streaming scale.
Therefore, structures at scales smaller or of the order of $ r_{Jeans} $ are {\bf erased}
for a given value of $ m $.

\bigskip

The observed size of the DM galaxy substructures is in the 
$ \sim 1 - 100 $ kpc scale. Therefore, eq.(\ref{fs}) indicates that $ m $ 
should be in the keV scale. That is, Warm Dark Matter particles.
This indication is confirmed by the phase-space density observations in galaxies
[2] and further relevant evidence from galaxy observations.

\bigskip

CDM particles with $ m \sim 100$ GeV have $ r_{Jeans} \sim 0.1 $ pc.
Hence CDM structures keep forming till scales as small as the solar system.
This result from the linear regime is confirmed  as a {\bf robust result} by
$N$-body CDM simulations. However, it has {\bf never been observed} in the sky. 
Adding baryons to CDM does not cure this serious problem. 
There is {\bf over abundance} of small structures in CDM and in CDM+baryons
(also called the satellite problem). CDM has {\bf many serious} conflicts with observations as:
\begin{itemize}
\item{Galaxies grow through merging in CDM.
Observations show that galaxy mergers are {\bf rare} ($ <10 \% $).}
\item{Pure-disk galaxies (bulgeless) are observed whose formation 
through CDM is unexplained}.
\item{CDM predicts {\bf cusped} density profiles: $ \rho(r) \sim 1/r $ for small $ r $.
Observations show {\bf cored } profiles. 
Adding baryons to CDM models does not eliminate cusps [3].}
\end{itemize}
Both $N$-body WDM and CDM simulations yield {\bf identical and correct} structures 
for scales larger than some kpc.
At intermediate scales WDM give the {\bf correct abundance} of substructures.

{\vskip 0.2cm} 

Inside galaxy cores, below  $ \sim 100$ pc, $N$-body classical physics simulations 
are incorrect for WDM because quantum effects are important in WDM at these scales.
WDM predicts correct structures for small scales (below kpc) when its {\bf quantum} nature is
taken into account [4].

{\vskip 0.2 cm} 

All searches of CDM particles (wimps) look for $ m \gtrsim 1 $ GeV.
The fact that the DM mass is in the keV scale explains why no detection 
has been reached.
Moreover, past, present and future reports of signals of such CDM experiments
{\bf cannot be due to DM detection} because the DM particle mass is in the keV scale.
The inconclusive signals in such experiments should be originated by phenomena
of other kinds. Notice that the supposed wimp detection signals
contradict each other supporting the notion that these  signals are {\bf unrelated to any DM
detection}.

{\vskip 0.2cm}

Positron excess in cosmic rays are unrelated to DM physics but to astrophysical
sources and astrophysical mechanisms and can be explained by them [6].

\bigskip

In order to determine whether a physical system has a classical or quantum nature
one has to compare the average distance between particles with their
de Broglie wavelength. The de Broglie wavelength of DM particles in galaxies is
\be\label{LdB}
\lambda_{dB}  = \frac{\hbar}{m \; v} \quad , \quad {\rm and ~ the ~ average ~ interparticle ~ distance,}
\quad d = \left( \frac{m}{\rho_h} \right)^{\! \! \frac13} \; ,
\ee
where $ v $ is the velocity dispersion and $ \rho_h $ is the average density in the  galaxy core.
The value of the ratio $ {\cal R} $ indicates whether the system is classical or quantum
$ {\cal R} \equiv \lambda_{dB}/d $. From eqs.(\ref{LdB}), $ \cal R $ can be expressed as [4]
$ {\cal R} = \hbar \; \left(Q_h/m^4\right)^{\! \! \frac13} \; . $
where $  Q_h \equiv \rho_h/\sigma^3 $ is  the phase-space density.
Using now the observed  values of $ Q_h $ from Table \ref{pgal} yields $ \cal R $ in the range
\be\label{quant}
 2 \times 10^{-3}  \; \left( \displaystyle \frac{\rm keV}{m}\right)^{\! \frac43}
< {\cal R} < 1.4 \; \left( \displaystyle \frac{\rm keV}{m}\right)^{\! \frac43}
\ee
The larger value of $ \cal R $ is for ultracompact dwarfs while the smaller value of $ \cal R $ 
is for big spirals.

\medskip 

The ratio $ \cal R $ around unity clearly implies a macroscopic quantum object.
Eq.(\ref{quant}) clearly shows {\bf solely from observations} 
that compact dwarf galaxies are natural macroscopic quantum objects for WDM [4].

\medskip

We see from eq.(\ref{quant}) that for CDM, that is for $ m\gtrsim $ GeV,
$ {\cal R}_{CDM} \lesssim 10^{-8} $, and therefore quantum effects are negligible in CDM.

\vskip 0.1 cm 

We estimate the quantum pressure in galaxies using the Pauli principle together
with the Heisenberg relations and show that dwarf galaxies are supported by the
fermionic {\it WDM quantum pressure} [4].

\bigskip

\begin{table}
\begin{tabular}{|c|c|c|c|c|c|} \hline  
 Galaxy  & $ \displaystyle \frac{r_h}{\rm pc} $ & $  \displaystyle \frac{v}{\frac{\rm km}{\rm s}} $
& $ \displaystyle  \frac{\hbar^{\frac32} \;\sqrt{Q_h}}{({\rm keV})^2} $ & 
$ \rho(0)/\displaystyle \frac{M_\odot}{({\rm pc})^3} $ & $ \displaystyle \frac{M_h}{10^6 \; M_\odot} $
\\ \hline 
Willman 1 & 19 & $ 4 $ & $ 0.85 $ & $ 6.3 $ & $ 0.029 $
\\ \hline  
 Segue 1 & 48 & $ 4 $ & $ 1.3 $ & $ 2.5 $ & $ 1.93 $ \\ \hline  
  Leo IV & 400 & $ 3.3 $ & $ 0.2 $ & $ .19 $ & $ 200 $ \\ \hline  
Canis Venatici II & 245 & $ 4.6 $ & $ 0.2 $   & $ 0.49 $ & $ 4.8 $
\\ \hline  
Coma-Berenices & 123 & $ 4.6 $  & $ 0.42 $   & $ 2.09 $  & $ 0.14 $
\\ \hline  
 Leo II & 320 & $ 6.6 $ & $ 0.093 $  & $ 0.34 $ & $ 36.6 $
\\  \hline  
 Leo T & 170 & $ 7.8 $ &  $ 0.12 $  & $ 0.79 $ & $ 12.9 $
\\ \hline  
 Hercules & 387 & $ 5.1 $ &  $ 0.078 $  & $ 0.1 $ & $ 25.1 $
\\ \hline  
 Carina & 424 & $ 6.4 $ & $ 0.075 $  & $ 0.15 $ & $ 32.2 $
\\ \hline 
 Ursa Major I & 504 & 7.6  &  $ 0.066 $  & $ 0.25 $ & $ 33.2 $
\\ \hline  
 Draco & 305 & $ 10.1 $ &  $ 0.06 $  & $ 0.5 $ & $ 26.5 $
\\ \hline  
 Leo I & 518  & $ 9 $ &  $ 0.048 $  & $ 0.22 $ & $ 96 $
\\ \hline  
 Sculptor & 480  & $ 9 $ & $ 0.05 $  & $ 0.25  $ & $ 78.8 $
\\ \hline 
 Bo\"otes I & 362 & $ 9 $ & $ 0.058 $  & $ 0.38 $ & $ 43.2 $
\\ \hline  
 Canis Venatici I & 1220  & $ 7.6 $ & $ 0.037 $ & $ 0.08 $ & $ 344 $
\\ \hline  
Sextans & 1290 & $ 7.1 $ & $ 0.021 $ & $ 0.02 $ & $ 116 $
\\ \hline 
 Ursa Minor & 750 & $ 11.5 $ & $ 0.028 $  & $ 0.16 $ & $ 193 $
\\ \hline  
 Fornax  & 1730 & $ 10.7 $ & $ 0.016 $  & $ 0.053  $ & $ 1750 $
\\  \hline  
 NGC 185  & 450 & $ 31 $ & $ 0.033 $ & $ 4.09 $ & $ 975 $
\\ \hline  
 NGC 855  & 1063 & $ 58 $ & $ 0.01 $ & $ 2.64 $ & $ 8340 $
\\ \hline  
  Small Spiral  & 5100  & $ 40.7 $ & $ 0.0018 $ & $ 0.029 $ & $ 6900 $
\\ \hline  
NGC 4478 & 1890 & $ 147 $ & $ 0.003 $ & $ 3.7 $ & $ 6.55 \times 10^4 $
\\ \hline  
 Medium Spiral & $ 1.9 \times 10^4 $ & $ 76.2 $ & $ 3.7 \times 10^{-4} $ & $ 0.0076 $ & $ 1.01 \times 10^5 $
\\ \hline  
 NGC 731 & 6160 & $ 163 $ & $ 9.27 \times 10^{-4} $ & $ 0.47 $ & $ 2.87 \times 10^5 $
\\ \hline 
 NGC 3853   & 5220 & $ 198 $ & $ 8.8 \times 10^{-4} $ & $ 0.77 $  
& $ 2.87 \times 10^5 $ \\ \hline 
NGC 499  & 7700 &  $ 274 $ & $ 5.9 \times 10^{-4} $ & 
$ 0.91 $ & $ 1.09 \times 10^6 $ \\   \hline 
Large Spiral & $ 5.9 \times 10^4 $ & $ 125 $ & $ 0.96 \times 10^{-4} $ & $ 2.3 \times 10^{-3} $ & 
$ 1. \times 10^6 $ \\ \hline  
\end{tabular}
\caption{Observed values $ r_h $, velocity dispersion $ v, \;  \sqrt{Q_h}, \; \rho(0)$ 
and $ M_h $ covering from ultracompact galaxies to large spiral galaxies. 
The phase space density is larger for smaller galaxies, both in mass and size.
Notice that the phase space density is obtained
from the stars velocity dispersion which is expected to be smaller than the DM  velocity dispersion.
Therefore, the reported $ Q_h $ are in fact upper bounds to the true values.}
\label{pgal}
\end{table}

\bigskip

We treat in ref. [4] the self-gravitating fermionic DM in the Thomas-Fermi approximation, in which
the central quantity to derive is the DM chemical potential $ \mu(r) $,
namely the free energy per particle, a measure of the gravitation potential through an additive constant, ie $ \mu(0) $.  We consider the spherical symmetric case.   

\bigskip

We integrate the Thomas-Fermi nonlinear differential
equations from $ r = 0 $ till the 
boundary $ r = R = R _{200} \sim R_{vir} $ defined as the radius where the 
mass density equals $ 200 $ times the mean DM density [4].

\bigskip

We define the core size $ r_h $ of the halo by analogy with the Burkert density profile as
$   \frac{\rho(r_h)}{\rho_0} = \frac14 \quad , \quad  r_h = l_0 \; \xi_h \; .$
where $ \rho_0 \equiv \rho(0) $ and $ l_0 $ is the characteristic length that emerges from 
the dynamical equations [4]:
\be\label{varsd2}
l_0 \equiv  \frac{\hbar}{\sqrt{8\,G}} \left(\frac{9\pi}{m^8\,\rho_0}\right)^{\! \! \frac16} 
  = R_0 \; \left(\frac{{\rm keV}}{m}\right)^{\! \! \frac43}  \; 
  \left(\rho_0 \; \frac{{\rm pc}^3}{M_\odot}\right)^{\! \! -\frac16} 
  \;,\qquad R_0 = 18.71 \; \rm pc  \; .
\ee
As an example of distribution function $ \Psi(E/E_0) $, we consider the Fermi--Dirac distribution 
$  \Psi_{\rm FD}(E/E_0) = \frac1{e^{E/E_0} + 1} \; $.

\bigskip

We define the dimensionless chemical potential $ \nu(r) $ as  
$ \nu(r) \equiv \mu(r)/E_0 \quad {\rm and} \quad   \nu_0 \equiv \mu(0)/E_0 \quad . $
Large positive values of the chemical potential at the origin $ \nu_0 \gg 1 $ 
correspond to the degenerate 
fermions limit which is the extreme quantum case, and oppositely, $ \nu_0 \ll -1 $ gives 
the diluted regime which is the classical limit. In this classical regime the Thomas-Fermi equations
 become exactly the equations for a self-gravitating Boltzmann gas.

\bigskip

\begin{figure}[h]
\begin{center}
\includegraphics[height=8.cm,width=14.cm]{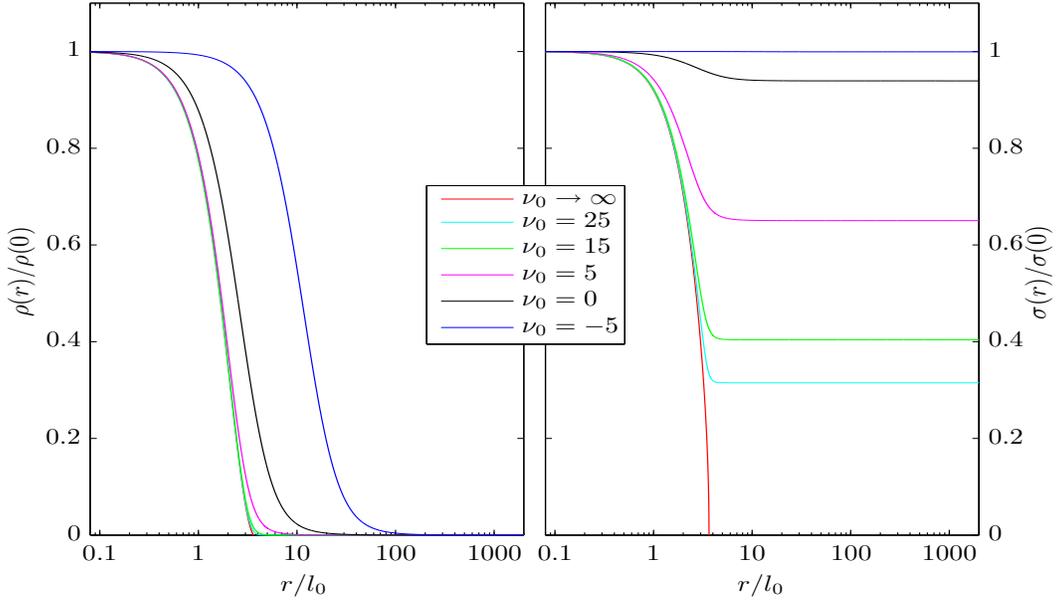}
\caption{Density and velocity profiles, $ \rho(r)/\rho_0 $ and $ \sigma(r)/\sigma(0) $, 
as functions of $ r/l_0 $ for different values of the chemical potential
at the origin $ \nu_0 $ [4]. Large positive values of $ \nu_0 $ correspond
to compact galaxies, negative values of $ \nu_0 $ correspond to the classical regime
describing spiral and elliptical galaxies.
All density profiles are cored. The sizes of the cores $ r_h $ 
are in agreement with the observations, from the compact galaxies where $ r_h \sim 35 $ pc till
the spiral and elliptical galaxies where $ r_h \sim .2 - 60 $ kpc. The larger and positive is 
$ \nu_0 $, the smaller is the core. The minimal one arises in
the degenerate case  $ \nu_0 \to +\infty $ (compact dwarf galaxies).}
\label{deg}
\end{center}
\end{figure}

{\bf Results} We display in fig. \ref{deg} the density and velocity profiles [4]. Namely,
we plot $ \rho(r)/\rho_0 $ and $ \sigma(r)/\sigma(0) $ as functions of 
$ r/R $ for $ \nu_0 \equiv \nu(0) = -5, \; 0, \; 5, \; 15, \;  25 $ and 
for the degenerate fermion limit $ \nu_0 \to +\infty $.
The obtained fermion profiles are always cored. 
The sizes of the cores $ r_h $ 
are in agreement with the observations, from the compact galaxies where $ r_h \sim 35 $ pc till
the spiral and elliptical galaxies where $ r_h \sim 0.2 - 60 $ kpc. The larger and positive is 
$ \nu_0 $, the smaller is the core. The minimal core size arises in
the degenerate case  $ \nu_0 \to +\infty $ (compact dwarf galaxies).

\begin{figure}[h]
\begin{center}
\includegraphics[height=8.cm,width=14.cm]{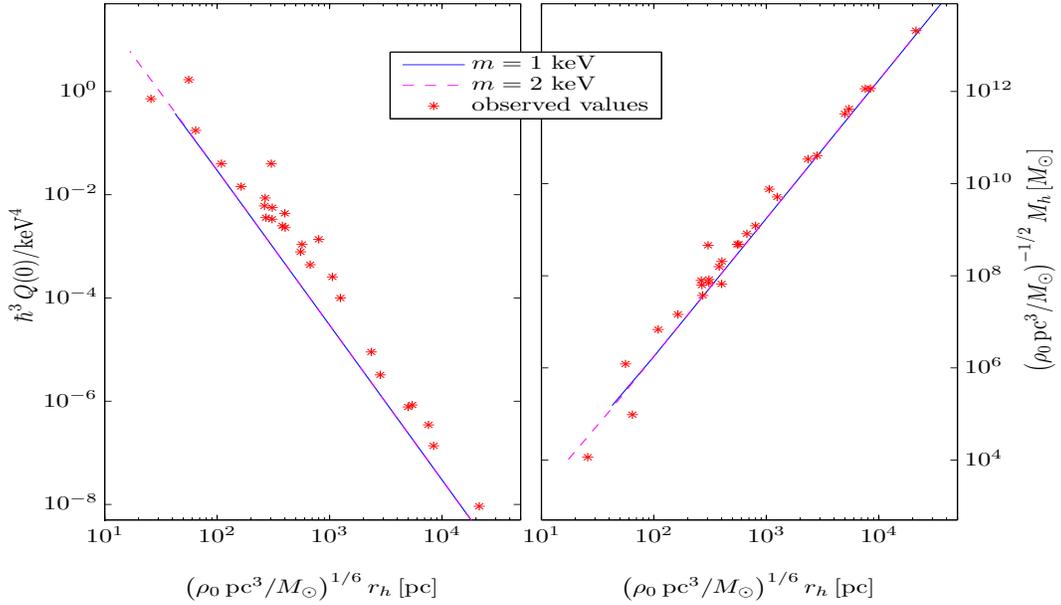}
\caption{In the left panel we display the galaxy phase-space density 
$ \hbar^3 \; Q(0)/({\rm keV})^4 $
obtained from the numerical resolution of the Thomas-Fermi  equations
for WDM fermions of mass $ m = 1 $ and $ 2 $ keV 
versus the ordinary logarithm of the product $ \log_{10}\{r_h \; 
[{\rm pc}^3 \; \rho_0/ M_\odot]^{\frac16} \} $ in parsecs [4].
The red stars $ * $ are the observed values of $ \hbar^3 \; Q(0)/({\rm keV})^4 $ from Table 1.
Notice that the observed values $ Q_h $ from the stars' velocity 
dispersion are in fact upper bounds for the DM $ Q_h $ and therefore the theoretical curve is slightly below them.
In the right panel we display the galaxy mass 
$ (M / M_\odot) \sqrt{M_\odot / [\rho_0 \; {\rm pc}^3]} $ obtained from the numerical resolution of the 
Thomas-Fermi equations for WDM fermions of mass $ m = 1 $ and $ 2 $ keV 
versus the product $ r_h \; [{\rm pc}^3 \; \rho_0/ M_\odot]^{\frac16} $
in parsecs [4].  The red stars $ * $ are the observed values of 
$ (M / M_\odot) \sqrt{M_\odot / [\rho_0 \; {\rm pc}^3]} $ 
from Table  1. Notice that the error bars 
of the observational data are not reported here but they are at least about $ 10-20 \%$.}
\label{halo}
\end{center}
\end{figure}

\bigskip

In the left panel of fig. \ref{halo} we plot the dimensionless phase space density  [4]
$ (\hbar^3/{\rm keV}^4) \; Q(0) \; $. In the right panel of fig. \ref{halo}, 
we plot instead the dimensionless product
$ (M_h/M_\odot) \; \sqrt{M_\odot/(\rho_0 \; {\rm pc}^3)} \; , $
where $ M_h $ is the halo mass, namely the galaxy mass inside the core radius $ r_h$. 

\bigskip

In both cases we consider the two values $ m = 1 $ and $ 2 $ keV and we put in
the abscissa the product $
r_h  \; \left({\rm pc^3}\; \rho_0 /M_\odot \right)^{\! \! \frac16} $
in ~ parsecs, where $ r_h $ is the core radius.

\bigskip

 The phase-space density $ Q(0) $ and the galaxy 
mass $ M_h $ are obtained by solving the Thomas-Fermi eqs.[4].
We have also superimposed the
observed values $ \hbar^3\, Q_h/({\rm keV})^4 $ and $ M_h \sqrt{M_\odot / [\rho_0
  \; {\rm pc}^3]} \; \; \left(m/{\rm keV}\right)^4 $ from Table I.
Notice that the observed values $ Q_h $ from the stars' velocity dispersion are
in fact upper bounds for the DM $ Q_h $. This may explain why the theoretical
Thomas-Fermi curves in the left panel of fig. \ref{halo} appear slightly below
the observational data. Notice also that the error bars of the observational
data are not reported here but they are at least about $ 10-20 \%$.

\bigskip

The phase space density decreases from its maximum value for the
compact dwarf galaxies corresponding to the limit of degenerate fermions till
its smallest value for large galaxies, spirals and ellipticals, corresponding to
the classical dilute regime. On the contrary, the halo radius $ r_h $ and the halo mass $ M_h $
monotonically increase from the quantum (small and compact galaxies) to the classical regime
(large and dilute galaxies). 

\bigskip

{\bf Results on the WDM particle mass}: From figs. \ref{halo} we can extract important 
information on the fermion particle WDM mass.

\bigskip

The small galaxy endpoint of the curves in figs. \ref{halo}  corresponds to
the degenerate fermion limit $ \nu_0 \to +\infty $ and its value depends on the
WDM particle value $ m $. For increasing $ m $, the small galaxy endpoint moves towards
smaller sizes while for decreasing $ m $, it moves towards larger sizes.

We see from figs. \ref{halo} that decreasing the particle mass beyond a given value,
namely for particle masses $ m \lesssim 1 $ 
keV,  the theoretical curves do not reach the more compact galaxy data.
Therefore, $ m \lesssim 1 $ keV is ruled out as WDM particle mass.

\bigskip

For growing $ m \gtrsim $ keV the left part of the theoretical curves corresponding to the lower galaxy 
masses and sizes, will do not have observed galaxy counterpart. Namely, increasing  
$ m \gg $ keV would show an overabundance of small galaxies (small scale structures) 
which do not have counterpart in the data.
This is a further indication that the WDM particle mass is approximately around 2 keV
in agreement with earlier estimations [7].

\bigskip

The overabundance of small scale structures that appears here for $ m \gg $ keV
is a consequence {\bf only} of the DM particle mass value.
This result is independent of the cosmic evolution, namely the primordial 
power spectrum and the structure formation dynamics. 

\medskip

In addition, the galaxy velocity dispersions turn to be fully consistent 
with the galaxy observations in Table 1.

\bigskip

{\bf Conclusion}: To conclude, the galaxy magnitudes: halo radius, galaxy masses and velocity dispersion
obtained from the Thomas-Fermi quantum treatment for WDM fermion masses in the keV scale are
fully consistent with all the observations for all types of galaxies (see Table I). 
Namely, fermionic WDM treated quantum mechanically, as it must be, is able to reproduce
the observed DM cores and their sizes in galaxies [4].

\bigskip

It is highly remarkably that in the context of fermionic WDM, the simple stationary
quantum description provided by the Thomas-Fermi approach is able to reproduce such broad variety of galaxies.

\bigskip

Baryons have not yet included in the present study. This is fully justified for dwarf compact 
galaxies which are composed today by 99.99\% of DM. In large galaxies the baryon fraction can
reach values up to  1 - 5 \%. Therefore, the effect of including 
baryons is expected to be a correction to these pure WDM results which are in 
agreement with observations for all types of galaxies, masses and sizes.

\bigskip

{\bf References}

\begin{description}

\item[1] D. Boyanovsky,  H J de Vega, N. G. Sanchez, 	Phys. Rev.  {\bf D 78}, 063546 (2008). 

H. J. de Vega, N. G. S\'anchez, Phys. Rev. D85, 043516 (2012) and  D85, 043517 (2012).

\item[2] H. J. de Vega, N. G. S\'anchez, 

MNRAS {\bf 404}, 885 (2010) and Int. J. Mod. Phys. {\bf A 26}, 1057 (2011).

\item[3] F. Marinacci, R. Pakmor, V. Springel, arXiv:1305.5360.

\item[4] C. Destri, H. J. de Vega, N. G. Sanchez, arXiv:1204.3090,
New Astronomy {\bf 22}, 39 (2013) and arXiv:1301.1864, Astroparticle Physics, 46, 14 (2013).

\item[5] E. Papastergis et al., Ap J, 739, 38 (2011),
J. Zavala et al., Ap J,	700, 1779 (2009).

\item[6] P. L. Biermann et al. PRL (2009), P. Blasi, P. D. Serpico PRL (2009).

\item[7] H. J. de Vega,  P. Salucci, N. G. Sanchez, 
New Astronomy {\bf 17}, 653 (2012). H. J. de Vega, N. G. S\'anchez, MNRAS {\bf 404}, 885 (2010) and
Int. J. Mod. Phys. {\bf A 26}, 1057 (2011).

\end{description}

\bigskip

\subsection{Jonh KORMENDY and K. C. FREEMAN}

\vskip -0.3cm

\begin{center}

J.K: Department of Astronomy, University of Texas at Austin, 1 University Station C1400, Austin, TX 78712-0259, USA;  kormendy@astro.as.utexas.edu \\        
Max-Planck-Institute for Extraterrestrial Physics, Giessenbachstrasse, D-85748 Garching by Munich, Germany
University Observatory, Ludwig-Maximilians-University, Scheinerstrasse 1, D-81679 Munich, Germany\\   

\medskip

K.C.F: Research School of Astronomy and Astrophysics, Mount Stromlo Observatory, The Australian National University, Cotter Road, Weston Creek, Canberra, ACT 2611, Australia; kcf@mso.anu.edu.au 

\bigskip

 {\bf Scaling Laws for Dark Matter Halos in Late-Type and Dwarf Spheroidal Galaxies}

\end{center}

\medskip

We measure the structural parameter correlations of the dark matter (DM) halos of bulgeless
(pure-disk) galaxies. Figure 1 shows the observed correlations between DM core radius $ r_c $, central
density $ \rho_0 $, particle velocity dispersion $ \sigma $, and total galaxy luminosity 
$ L_B $ as measured by the blue-band absolute magnitude MB. 

\bigskip

For high-luminosity galaxies, DM parameters are from published
mass models fitted to galaxy rotation velocities $ V (r) $ as a function of radius r. In these models, $ V(r) $
is decomposed into the contributions from visible and dark matter under the assumption that the
mass-to-light ratio $ M/L_B $ is constant with radius. ``Maximum disk'' values of $ M/L_B $ are adjusted
to fit as much of the inner rotation curve as possible without making the halo have a hollow core.

\bigskip

Rotation curve decomposition is impossible for dwarf galaxies because $ V \lesssim $
 the velocity dispersion of the gas or stars.

To increase the luminosity range further, we include Jeans equation central densities
of dwarf spheroidal (dSph) and dwarf irregular (dIm) galaxies. Combining these data, we find that
DM halos satisfy well defined scaling laws. Halos in less luminous galaxies have smaller core radii
$ r_c $, higher central densities $ \rho_0 $, and smaller central velocity dispersions 
$ \sigma $. Scaling laws provide new
constraints on the nature of DM and on galaxy formation and evolution. For example:

1. A continuous sequence of DM ($ \rho_0, \; r_c, \; \sigma $) extends at least from dwarf spirals with 
$ M_B \simeq -13 $ to the highest-luminosity disks with $ M_B \simeq -22.4 \; (H_0 = 70 \; km \;s^{-1}
\;  Mpc^{-1}) $.

2. The high DM densities in the dSph companions of our Galaxy imply that they are real galaxies
formed from primordial density fluctuations. They are not tidal fragments. Tidal dwarfs cannot
retain even the low DM densities of their giant-galaxy progenitors. In contrast, dSph galaxies have
higher DM densities than those of typical giant-galaxy progenitors.

3. Dwarfs of lower $ L_B $ have smaller baryon/DM mass ratios (e. g.) because of supernova-driven gas
ejection or because they could not accrete baryons after the Universe was reionized. We estimate
the baryon deficiency as $ \gtrsim  1 $ dex for dIm galaxies and $ \simeq 2 $ 
dex for the dSph galaxies in our sample.

4. Over a large range in rotation velocities $ V_{circ} $  of test particles in DMhalos, we find a linear relation
between $ V_{circ} $ and the maximum rotation velocities  $ V_{circ,disk} $ of baryonic disks. 
Its intercept is robustly at non-zero $ V_{circ} $: baryons become dynamically unimportant 
($ V_{circ,disk} \to 0 $) at $ V_{circ} = 42 \pm 4 $ km/s.
This corresponds to DM  $ \sigma = 30 \pm  3 $ km/s. These measurements are in good agreement with
theoretical estimates of minimum DM halos that can accrete baryons after reionization. Galaxies
with lower-mass halos do exist - they are our dSphs - but they have tiny ratios $ \sim 1/100 $ of central
stellar to DM mass densities limited by our ability to detect low-surface-brightness galaxies.

5. The fact that, as luminosity decreases, dwarf galaxies become much more numerous and much
more nearly dominated by DM suggests that there exists a large population of objects that are
completely dark. Such objects are a canonical prediction of cold DM theory.

6. The slopes of the DM parameter correlations provide a measure on galactic mass scales of the
slope n of the power spectrum $ |\delta_k|^2 \propto  k^n $ of primordial density fluctuations. 
We derive $ n \simeq -2.0 \pm 0.1 $. This is consistent with the theory of cold DM.

\bigskip

We review three of the many problems of cold DM galaxy formation on galaxy scales. The
core-cusp problem and the problem of the predicted but ``missing'' large numbers of dwarf galaxies
may have solutions that involve the messy physics of baryons (in the latter case, point 5, above).
However, we have no candidate solution for the biggest problem faced by our standard theory of
galaxy formation (Kormendy et al. 2010, ApJ, 723, 54): How can hierarchical clustering make so
many giant, pure-disk galaxies with no evidence for merger-built bulges? In a sphere of radius 8 Mpc
centered on our Galaxy, at least 11 of 19 galaxies with $ V_{circ} > 150 $ km/s including our Galaxy
show no evidence for a classical (merger-built) bulge. Four may contain small bulges that contribute
5.12\% of the light of the galaxy. Only 4 of the 19 giant galaxies are ellipticals or have bulges that
contain $ \sim 1/3 $ of the galaxy light. Pure-disk galaxies are so common that thay cannot have formed as
the tail of a distribution of formation properties that involves a few fortuitously mergerless histories.
We emphasize that this problem is a strong function of environment: the Virgo cluster is not a puzzle,
because $ > 2/3 $ of its stellar mass is in merger remnants. Any solution to the problem of forming
pure-disk galaxies must, we suggest, depend critically on galaxy environment.

\begin{figure}[h]
\begin{center}
\begin{turn}{-90}
\includegraphics[height=20.cm,width=18.cm]{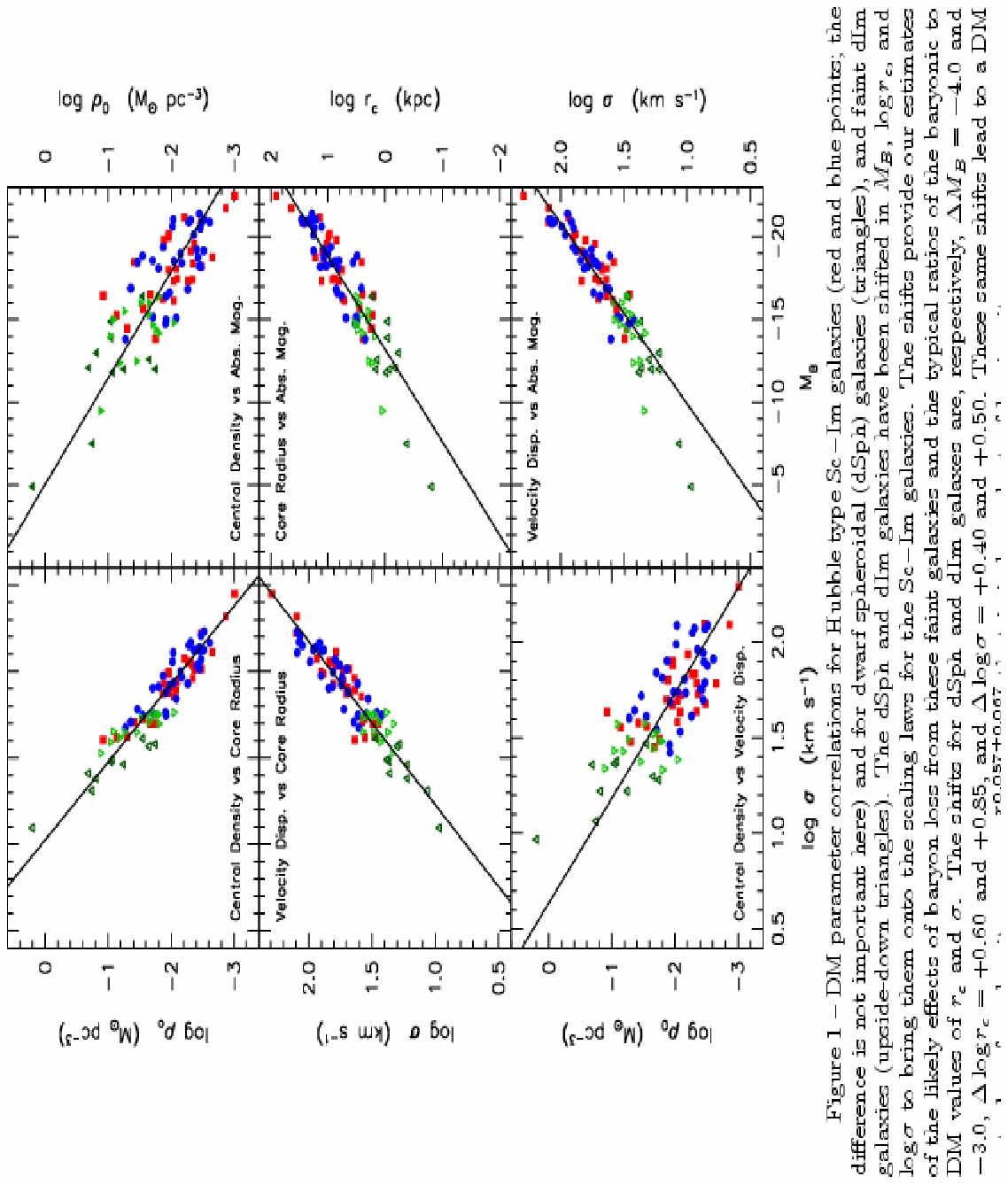}
\end{turn}
\caption{  DM parameter correlations for Hubble type Sc . Im galaxies (red and blue points; the
difference is not important here) and for dwarf spheroidal (dSph) galaxies (triangles), and faint dIm
galaxies (upside-down triangles). The dSph and dIm galaxies have been shifted in $ M_B, \; \log r_c $, and
$ \log \sigma $ to bring them onto the scaling laws for the Sc - Im galaxies. The shifts provide our estimates
of the likely effects of baryon loss from these faint galaxies and the typical ratios of the baryonic to
DM values of $ r_ c $ and $ \sigma $. The shifts for dSph and dIm galaxes are, respectively, 
$ \Delta M_B = -4.0 $ and  $ -3.0, \; \Delta \log r_c = +0.60 $ and $ +0.85 $, 
and $ \Delta \log\sigma  = +0.40 $ and $ +0.50 $. These same shifts lead to a DM
central surface density $ \rho_0 \; r_c \propto L_B^{0.057 \pm 0.067} $, 
 that is independent of luminosity.    }
\end{center}
\end{figure}

\newpage

\subsection{Marco LAVEDER with Carlo GIUNTI}

\vskip -0.3cm

\begin{center}

Department of Physics and Astronomy,
University of Padova
and INFN, Sezione di Padova\\
Via Marzolo 8, I-35131 Padova, Italy
\bigskip

{\bf Sterile Neutrinos: Phenomenology and Fits}

\end{center}

\medskip

Three anomalies affect the short baseline oscillation data
(each of them is at $3\sigma$ level):
the LSND $\bar\nu_{\mu}\to\bar\nu_{e}$ signal [1],
the reactor antineutrino anomaly [2],
the Gallium anomaly [3],
and the MiniBooNE $\nu_{\mu}\to\nu_{e}$ and $\bar\nu_{\mu}\to\bar\nu_{e}$ data [4].

\bigskip

In this lecture I review these indications
in favor of short-baseline neutrino oscillations,
which require the existence of one or more sterile neutrinos.
In the framework of 3+1 neutrino mixing,
which is the simplest extension of the standard three-neutrino mixing 
which can partially explain the data,
there is a strong tension in the interpretation of the data,
mainly due to an incompatibility of the results of
appearance and disappearance experiments [5]-[18].

\bigskip

Testable predictions are given for $m_{\beta}$ and $m_{\beta \beta}$ searches. 

\bigskip

After the MiniBooNE antineutrino results presented this summer at Neutrino 2012,
that show that antineutrino data are similar to neutrino data,
there is no more need of the CP violation allowed
by 3+2 neutrino mixing [18, [19],[20], hep-ph/0705.0107,hep-ph/0906.1997,hep-ph/1007.4171,hep-ph/1103.4570,hep-ph/1107.1452.

\bigskip

Moreover, the tension between the data of appearance and disappearance experiments persists
in the framework of 3+2 neutrino mixing, because the short-baseline disappearance of electron antineutrinos and muon neutrinos
compatible with the LSND and MiniBooNE antineutrino appearance signal has not been observed.

\bigskip

{\bf References}

\begin{description}

\item 1
LSND, A.~Aguilar {\em et~al.},
Phys. Rev. {\bf D64}, 112007 (2001), [hep-ex/0104049].

\medskip

\item 2
G.~Mention {\em et~al.},
Phys. Rev. {\bf D83}, 073006 (2011), [arXiv:1101.2755].

\medskip

\item 3
C.~Giunti and M.~Laveder,
Phys. Rev. {\bf C83}, 065504 (2011), [arXiv:1006.3244].

\medskip

\item 4
MiniBooNE, A.~A. Aguilar-Arevalo,
Phys. Rev. Lett. {\bf 102}, 101802 (2009), [arXiv:0812.2243].

\medskip

\bibitem 5
MiniBooNE, A.~A. Aguilar-Arevalo {\em et~al.},
Phys. Rev. Lett. {\bf 105}, 181801 (2010), [arXiv:1007.1150].

\medskip

\item 6
N.~Okada and O.~Yasuda,
Int. J. Mod. Phys. {\bf A12}, 3669 (1997), [hep-ph/9606411].

\medskip

\item 7
S.~M. Bilenky, C.~Giunti and W.~Grimus,
Eur. Phys. J. {\bf C1}, 247 (1998), [hep-ph/9607372].

\medskip

\item 8
S.~M. Bilenky, C.~Giunti, W.~Grimus and T.~Schwetz,
Phys. Rev. {\bf D60}, 073007 (1999), [hep-ph/9903454].

\medskip

\item 9
M.~Maltoni, T.~Schwetz, M.~A. Tortola and J.~W.~F. Valle,
Nucl. Phys. {\bf B643}, 321 (2002), [hep-ph/0207157].

\medskip

\item 10
M.~Maltoni, T.~Schwetz, M.~Tortola and J.~Valle,
New J. Phys. {\bf 6}, 122 (2004), [hep-ph/0405172].

\medskip

\item 11
M.~Maltoni and T.~Schwetz,
Phys. Rev. {\bf D76}, 093005 (2007), [arXiv:0705.0107].

\medskip

\item 12
G.~Karagiorgi, Z.~Djurcic, J.~Conrad, M.~H. Shaevitz and M.~Sorel,
Phys. Rev. {\bf D80}, 073001 (2009), [arXiv:0906.1997].

\medskip

\item 13
E.~Akhmedov and T.~Schwetz,
JHEP {\bf 10}, 115 (2010), [arXiv:1007.4171].

\medskip

\item 14
C.~Giunti and M.~Laveder,
Phys. Rev. {\bf D83}, 053006 (2011), [arXiv:1012.0267].

\medskip

\item 15
J.~Kopp, M.~Maltoni and T.~Schwetz,
Phys. Rev. Lett. {\bf 107}, 091801 (2011), [arXiv:1103.4570].

\medskip

\item 16
C.~Giunti and M.~Laveder,
Phys.Rev. {\bf D84}, 073008 (2011), [arXiv:1107.1452].

\medskip

\item 17
C.~Giunti and M.~Laveder,
Phys.Rev. {\bf D84}, 093006 (2011), [arXiv:1109.4033].

\medskip

\item 18
C.~Giunti and M.~Laveder,
Phys. Lett. {\bf B706}, 200 (2011), [arXiv:1111.1069].

\medskip

\item 19
M.~Sorel, J.~Conrad and M.~Shaevitz,
Phys. Rev. {\bf D70}, 073004 (2004), [hep-ph/0305255].

\medskip

\item 20
G.~Karagiorgi {\em et~al.},
Phys. Rev. {\bf D75}, 013011 (2007), [hep-ph/0609177].

\end{description}

\bigskip

\subsection{Ernest MA}

\vskip -0.3cm

\begin{center}

Department of Physics and Astronomy, University of California, Riverside, California 92526, USA

\bigskip

{\bf Stable Interacting Majorana Fermion (Scotino) as keV Warm Dark Matter} 

\end{center}

\medskip

Except for the neutrinos, all fundamental fermions (quarks and leptons) and 
bosons ($W^\pm, Z$) obtain their masses through the Higgs boson, which was 
presumably discovered at the Large Hadron Collider (LHC) with a mass of 
about 125 GeV, as announced on July 4, 2012.  The neutrino story is a bit 
more complicated.  Its left-handed component is part of an $SU(2)_L \times 
U(1)_Y$ doublet, i.e. $(\nu,e) \sim (2,-1/2)$, but its right-handed 
component $\nu_R \sim (1,0)$ is a singlet.  As such, $\nu_R$ may have its 
own Majorana mass $m_R$ independent of the Higgs boson.  As $\nu_L$ mixes with 
$\nu_R$ through its Dirac mass $m_D$ resulting from the Higgs boson, the 
neutrino gets a small mass $m_D^2/m_R$ if $m_D << m_R$.  This is the famous 
seesaw mechanism.

\bigskip

Usually, $m_R$ is assumed to be much greater than the electroweak breaking 
scale of about 100 GeV.  However, if it is taken to be of order keV, then 
it is usually called a sterile neutrino $\nu_S$, and has been identified as an 
excellent candidate for warm dark matter.  In that case, it may be 
produced in the early Universe through its mixing with the active neutrinos, 
and even though its lifetime is much greater than the age of the Universe, 
it may still decay ($\nu_S \to \nu_i \gamma$) and be observed.  At present, 
there is an upper bound of 2.2 keV on its mass from galactic X-ray data, 
assuming the canonical (Dodelson-Widrow) production mechanism.  On the 
other hand, Lyman-$\alpha$ forest observations tend to put a lower bound 
of perhaps 5.5 keV on its mass.  This possibly serious tension may be 
relaxed if the dark-matter candidate is stable instead, as in the 
two models I consider in this talk. 

\bigskip

In 2006, I proposed a fundamental connection between neutrino mass and 
dark matter [1], i.e. neutrino masses are quantum effects due to 
the existence of dark matter, i.e. scotogenic from the Greek 'scotos' 
meaning darkness.  The simplest realization of this notion 
is to extend the standard model of particle interactions to include 
3 neutral singlet fermions $N_{1,2,3}$ (analogs of $\nu_R$) and an extra 
scalar doublet $(\eta^+,\eta^0)$ in addition to the Higgs doublet 
$(\phi^+,\phi^0)$, with the important difference that $N_{1,2,3}$ and 
$(\eta^+,\eta^0)$ are odd under an exactly conserved $Z_2$ symmetry, 
whereas all other particles are even.

\medskip

As a result, $N_{1,2,3}$ (scotinos) may have Majorana masses themselves, but 
they do not couple to $\nu$ through $\phi^0$, so there is no $m_D$ and 
neutrinos are massless at the (classical) tree level.  However, at 
the (quantum) one-loop level, the interactions $\nu N \eta^0$ and 
$(\Phi^\dagger \eta)^2$ allow radiative Majorana neutrino masses to 
be generated.  This is possible because 
$\eta^0 = (\eta_R + i \eta_i)/\sqrt{2}$ is split so that 
$m_R \neq m_I$ and 

\bigskip

\begin{equation}
({\cal M}_\nu)_{ij} = \sum_k \frac{h_{ik} h_{jk} M_k}{16 \pi^2} 
\left[ \frac{m_R^2}{m_R^2 - M_k^2} \ln \frac{m_R^2}{M_k^2} - 
\frac{m_I^2}{m_I^2 - M_k^2} \ln \frac{m_I^2}{M_k^2} \right] 
\simeq \frac{\ln (m_R^2/m_I^2)}{16 \pi^2}
\sum_k h_{ik} h_{jk} M_k,
\end{equation}

\bigskip

if $M_k << m_{R,I}$ [2], 
which shows that neutrino masses are linear functions of the scotino masses. 
If the former are of order 0.1 eV, then the latter may be of order 10 keV 
if $h^2_{ik} \sim 10^{-3}$.  The present experimental upper bound of 
$2.4 \times 10^{-12}$ on $B(\mu \to e \gamma)$ implies 
$m_{\eta^+} > 310~{\rm GeV}~(|\sum_k h_{\mu k} h^*_{ek}|/10^{-3})^{1/2}$.

\bigskip

The lightest $N$ (call it $N_1$) is absoultely stable and can be a 
warm dark-matter candidate of 10 keV.  Its effective interactions with 
leptons are of order $h^2/m_\eta^2 \sim 10^{-7}$ GeV$^{-2}$, hence it is 
thermally produced and decouples at a temperature of about 200 MeV. 
Its number density relative to photons is $n_N/n_\gamma = 0.183$, 
resulting in a relic abundance of $\Omega_N h^2 \sim (115/16) 
(3 M_1/{\rm keV})$, where the factor of 3 comes from the assumption that 
$N_{2,3}$ all decay into $N_1$.  For the observed dark-matter abundance 
of $0.1123 \pm 0.0035$, this would mean that $N_1$ is overproduced by a 
factor of about $1.9 \times 10^3$ if $M_1 = 10$ keV.  This is a common 
problem for all warm dark-matter candidates which are thermally produced 
and the solution is to postulate a particle which decouples after $N_1$ 
and decays later as it becomes nonrelativistic, with a large release of 
entropy.  A suitable such particle is a real scalar of mass 20 MeV 
which mixes with the standard-model Higgs boson.

\bigskip

Another model of a scotino is in the context of an $SU(2)_R$ extension [3] of 
the standard model.  Here, instead of $(\nu,e)_R$ transforming as an 
$SU(2)_R$ doublet, $\nu_R$ is a singlet, and $(n,e)_R$ is a doublet, with 
$n_L$ a singlet.  Now $\nu$ gets a canonical seesaw mass, and $n$ does the 
same.  However, they differ by an imposed $Z_2$ symmetry so that 
$\nu$ is even and $n$ is odd.  The model has the symmetry of switching 
$SU(2)_{L,R}$ together with $\nu \leftrightarrow n$.  Assuming that $SU(2)_R$ 
breaks at a scale $10^{2.5}$ that of $SU(2)_L$, $n$ becomes a scotino of mass 
10 keV and $\nu$ becomes a neutrino of mass 0.1 eV.

\bigskip

{\bf References}

\begin{description}

\item[1] E. Ma, Phys. Rev. D 73, 077301 (2006).

\medskip

\item[2] E. Ma, arXiv:1206.1812.

\medskip

\item[3] E. Ma, Phys. Rev. D 85, 091701(R) (2012).

\end{description}

\bigskip

\subsection{John C. MATHER}

\vskip -0.3cm

\begin{center}

NASA Goddard Space Flight Center, Greenbelt, MD 20771 USA

\bigskip

{\bf James Webb Space Telescope and the Origins of Everything - Progress and Promise} 

\end{center}

\medskip

{\bf Abstract.}

James E. Webb built the Apollo program and led NASA to a successful moon landing.  We honor his leadership with the most powerful space telescope ever designed, capable of observing the early universe within a few hundred million years of the Big Bang, revealing the formation of galaxies, stars, and planets, and showing the evolution of solar systems like ours. Under study since 1995, it is a project led by the United States National Aeronautics and Space Administration (NASA), with major contributions from the European and Canadian Space Agencies (ESA and CSA). It will have a 6.6 m diameter aperture (corner to corner), will be passively cooled to below 50 K, and will carry four scientific instruments: a Near-IR Camera (NIRCam), a Near-IR Spectrograph (NIRSpec), a near-IR Imaging Slitless Spectrometer (NIRISS), and a Mid-IR Instrument (MIRI). It is planned for launch in 2018 on an Ariane 5 rocket to a deep space orbit around the Sun - Earth Lagrange point L$_2$, about $1.5 \times 10^6$ km from Earth. The spacecraft will carry enough fuel for a 10 yr mission.

\bigskip

{\bf International Partnership.}

Goddard Space Flight Center leads the NASA team, and is supported by a  prime contract to Northrop Grumman Aerospace Systems and their subcontractors, including ATK, Ball Aerospace, and ITT. The NIRCam comes from the University of Arizona with Lockheed Martin.  The NIRSpec comes from ESA with Astrium, and its microshutter array is provided by Goddard Space Flight Center. The Fine Guidance Sensor and the Near IR Imaging Slitless Spectgrometer come from CSA with Comdev. All of the near IR detectors come from Teledyne. The mid IR instrument is built by a European consortium led by the UK ATC, in partnership with Jet Propulsion Laboratory, and its detectors come from Raytheon. Europe is also providing the Ariane 5 rocket.

\bigskip

{\bf Scientific Objectives.}

A project summary has been published by Gardner et al. [1]. Additional documents about JWST are available here: http://www.jwst.nasa.gov/ and here: http://www.stsci.edu/jwst/doc-archive. A recent meeting ``Frontier Science Opportunities with the James Webb Space Telescope" was held at the Space Telescope Science Institute and the webcast archive is available online: http://www.stsci.edu/institute/conference/jwst2011.
Four key topics were used to guide the design of the observatory:

\medskip

{\bf The end of the dark ages: first light and reionization.} This theme requires the largest feasible infrared telescope, since the first objects of the universe are faint, rare, and highly redshifted. It requires a wide wavelength range, to distinguish  high-redshift objects from cool local objects, and to estimate the ages and photometric redshifts of the stellar contents from colors.  It also requires powerful multi-object infrared spectroscopy, to determine the physical conditions and redshifts of the earliest objects.

\medskip

{\bf The assembly of galaxies.} It is now thought that galaxies form around dark matter concentrations, as products of extensive merger trees, but it is difficult to tell whether simulations match observations.  The interaction between dark matter, ordinary matter, black holes (when and how did they form?), stars, winds, and magnetic fields is extraordinarily complex ``gastrophysics".  It is currently impossible to simulate the full dynamic range from stars and planetary systems to magnetized jets, outflows, dust formation, etc., so the formation and assembly of galaxies is still an observational science. With its infrared imaging and spectroscopy, JWST will reveal details of galactic mergers and show changes of properties with distance (time).

\medskip

The JWST will also address some aspects of dark energy and dark matter. It can extend the Hubble measurements of distant supernovae, achieving greater precision by using rest-frame IR photometry, where the candles are more standard and the dust obscuration is less than at visible wavelengths.  It can also improve the calibration of the Hubble constant, by extending the range of each step of the distance ladder and  eliminating some steps.  It can also extend maps of the dark matter distribution to higher redshift, because it can observe many more faint and higher redshift background galaxies than current observatories.

\medskip

{\bf The birth of stars and protoplanetary systems.}  Local star and planet formation is uniquely observable in the infrared, because of the low temperatures of young objects, and their typical location within obscuring dust clouds. Infrared observations complement radio observations very well.

\medskip

{\bf Planetary systems and the origins of life.} Traces of the formation of the Solar System are found everywhere from the bright planets to the faint comets, asteroids, and dwarf planets of the outer solar system. JWST's image quality, field of view, and ability to track moving targets are essential to detect and analyze many such targets,  In addition, transit spectroscopy and direct imaging of extra-solar planetary systems are feasible, especially now that the Kepler observatory has cataloged 1235 candidate transiting planets.  With the discovery of a favorable nearby target (small star, large Earth), it may be possible to measure the atmospheric composition of an Earth-like planet and to detect the presence of liquid water.

\bigskip

{\bf Observatory Design.} 

The observatory is composed of a deployable telescope, a scientific instrument package, and a spacecraft bus, separated by a deployable sun shield to enable the telescope and instruments to cool to $<$ 50 K.

{\bf Telescope.} 

The telescope uses a three-mirror anastigmat design to provide a wide field of view with diffraction-limited imaging at 2 $\mu$m.  The primary is nearly parabolic with an aperture of 6.6 m, and is composed of 18 hexagonal segments.  These segments are made of beryllium, machined to a few mm thickness with stiffening ribs, and polished at room temperature to a shape that will be nearly ideal (20 nm rms error) at the operating temperature.  All the segments have been polished and coated with gold.  The primary mirror segments are deployed after launch and adjusted to the right position and curvature by actuators having step sizes of a few nm. The secondary mirror is also deployed after launch. The final focus is determined using wavefront sensing algorithms, based on star images taken in and out of focus. These algorithms were derived from those used for the Hubble Space Telescope repair. The predicted image quality (point spread function) is available online at http://www.stsci.edu/jwst/software/webbpsf. A small movable flat mirror located at an image of the primary mirror (a fine steering mirror) is used to null the image motion as sensed by the fine guidance sensor in the instrument package; a signal from its control loop also goes to help maintain the pointing of the whole spacecraft.

\bigskip

{\bf Instrument Package.} 

All of the instruments are mounted to a carbon-fiber structure attached to the back of the telescope. The near IR camera covers 0.6 to 5 $\mu$m in two bands observed simultaneously, with 0.034 and 0.068 arcsec pixels respectively.  The camera  includes the wavefront sensing equipment, and a coronagraph. The near IR spectrometer provides low (R 100) and medium (R 1000 and 3000) spectroscopy.  It can observe 100 simultaneous targets with a microshutter array selector, and also provides fixed slits and an image slicing integral field configuration. The fine guidance sensor is a two-field camera that provides error signals to the pointing control system. The Near IR Imaging Slitless Spectrometer  provides high contrast imaging and spectroscopy in the near IR band.  All of the near IR instruments use HgCdTe detectors. The mid IR instrument provides imaging, coronography, and integral field spectroscopy from 5 to 28 $\mu$m, using Si:As detectors.  The exposure time calculator is available on line at http://jwstetc.stsci.edu/etc. With long exposures, nJy sensitivities are possible (AB magnitude $>$ 31.4).

\bigskip

{\bf Spacecraft.} 

The spacecraft bus includes the command and telemetry system, a deployable telemetry antenna, a pointing control system, a solar power system, and a deployable 5-layer sunshield the size of a singles tennis court. Behind the shield, the instrument package and the telescope radiate heat to the dark sky and reach temperatures of 40-50 K, cold enough for operating all the near IR ($<5$ $ \mu$m) detectors.  A helium compressor provides active cooling for the mid IR instrument to operate below 7 K, without stored cryogens. Pointing control is provided by reaction wheels and thrusters controlled by sun sensors, gyros (no moving parts!) and star trackers.

\bigskip

{\bf Orbit and Orientation.}

The Ariane 5 launch vehicle will ascend from Kourou, French Guiana, to place the observatory in orbit around the Sun-Earth Lagrange point L$_2$, approximately 1.5$\times10^6$ km from Earth. Bi-propellant thrusters will adjust the orbit to stay within several hundred thousand km of the L$_2$ point (which is near the end of the Earth's umbra), and to avoid shadows from the Moon or the Earth.  The orbit is unstable and small adjustments will be required every few weeks.  Fuel is also required to maintain the orientation of the observatory; small torque unbalances from the solar radiation pressure build up and must be compensated with thruster firings.  The fuel tanks are sized to last for 10 years of operation after initial observatory checkout.  

\bigskip

{\bf Operations.}

The scientific operation of the JWST will be similar to that of the Hubble Space Telescope, with observing proposals evaluated by committees.  Approximately 1/2 year of observations are allocated to the instrument teams and the interdisciplinary scientists. The European share of general observation time is to be 15\% and the Canadian share 5\%. It will take about 2 months to reach the L$_2$ orbit and cool down to operating temperature, and 4 more months are allocated for initial setup, focussing, characterization, and optimization of the observatory.

\bigskip

{\bf Project Status.}

One of the flight instruments (MIRI) and was completed and tested in Europe and delivered to NASA, the NIRSpec was tested and repairs are in progress, the FGS and NIRISS are to be delivered to NASA on July 30, 2012 and the NIRCam and NIRSpec are to be delivered to Goddard Space Flight Center within a year, for integration into the Integrated Science Instrument Module, and combined testing with a telescope simulator. Also, all the flight telescope mirrors (18 primary mirror hexagons, and the secondary, tertiary, and fine steering mirror) are completed and awaiting replacement of motor bearings. The near IR detectors (HgCdTe made by Teledyne) were degrading while they were kept warm, the cause has been determined, and new detectors are being produced..

\bigskip

{\bf Replan.}

Following two reports (TAT and ICRP) chaired by John Casani in 2010, the JWST management was replaced at Goddard Space Flight Center and NASA Headquarters, and the project was replanned for a launch date in 2018.  The new plan provides additional funds for risk reduction and testing and raises JWST to one of NASA's top three priorities, along with commercial crew and the space launch system (SLS).  The NASA budget for 2012 was passed by the US Congress in November 2011 and the JWST has kept to its announced schedule ever since the replan.

\bigskip

{\bf Reference}

\begin{description}

\item[1] J. Gardner et al., Space Science Reviews, 123 (\#4), 2006; arXiv:astro-ph/0606175

\end{description}

\bigskip

\subsection{Lyman PAGE}

\vskip -0.3cm

\begin{center}

Department of Physics, Princeton University, Princeton NJ, 08540

\bigskip

{\bf ACT and WMAP} 

\end{center}

\medskip

Measurements of the CMB are remarkable for what they have already been able to tell us about the geometry, age, fundamental processes behind, and contents of the universe. The primary message of this talk is that there is much more to be learned from precise measurements of the CMB. 

\bigskip

We now have a standard model of cosmology that posits that the universe is geometrically flat, homogenous and isotropic on large scales, dominated by a cosmological constant ($\Omega_\Lambda=0.73$), that roughly 80\% of the matter is cold and dark ($\Omega_{cdm}=0.23, \Omega_b=0.046$),  and the universe was reionized around $z=10$. The initial fluctuations were adiabatic and Gaussian-distributed with a small tilt ($n_s$) to the power spectrum such that larger angular scales have a slightly larger amplitude [1].  There are no current cosmological observations in serious disagreement with this model.  This is especially noteworthy because there are so many measurements spanning a wide variety of techniques, wavelengths, and redshifts. 

\bigskip

The model serves a couple of purposes. Perhaps the most important is that it provides a  clear foundation for finding something new. The geometry may not be exactly flat, there may be contributions from warm dark matter (the topic of this meeting), there may be an admixture of isocurvature modes, or perhaps the fluctuations are not Gaussian because of new physics during inflation or what ever process took place.  A persistent hope of many who are making the measurements is that we will find something that does not fit the standard model. 

\bigskip

One forefront is nailing down the scalar spectral index or tilt, another is detecting or limiting large angular scale B-modes. Precisely measuring the tilt is a matter of time. While the departure from unity appears robust, more work is needed on separating all the contributing effects that might bias $n_s$.  Great advances are expected from Planck. It is remarkable that to properly compute $n_s$ to the level achievable by measurement requires modeling hydrogen and helium through the  decoupling process with interacting hundred-level atomic systems. Another under appreciated aspect of measuring $n_s$ is that it cannot be known via the CMB unless one has knowledge of the optical depth to scattering. This means that determinations of $n_s$ prior to WMAP had implicit and untested assumptions about the optical depth. The status of $n_s$ for models in which the B-modes are set to zero is given in Table 1.

\begin{table}[htdp]
\caption{Measurements of the departure of $n_s$ from unity.}
\begin{center}
\begin{tabular}{| l | c | c |}
\hline
Measurements & $1-n_s$ & Reference \\
\hline
WMAP7 & $0.037\pm0.014$  &  [1]\\
WMAP7+BAO+H$_0$ & $0.037\pm0.012$  & [1] \\
WMAP5+ACBAR+QUAD & $0.038\pm0.013$  & [2]\\
WMAP7+ACT &  $0.038\pm0.013$ & [3] \\
WMAP7+SPT &  $0.034\pm0.011$ &[4] \\
\hline
\end{tabular}
\end{center}
\label{tab1}
\end{table}%

The search for primordial B-modes is often motivated as the best test of inflation. While this may be true, a detection of B-modes would have a far greater impact. It would be the first evidence of gravity operating on quantum scales.

\bigskip

The other thing we get from the standard model is  a firm foundation for measuring not-necessarily-cosmological aspects of Nature in new ways. Two such examples are the number of relativistic species and the sum of the neutrino masses, both very relevant for this meeting. The first thing to note about the CMB as a ``neutrino detector" is that it is a blunt instrument.  With the CMB the only handle on neutrinos is how they respond to gravity. We cannot tell that they are fermions, cannot determine flavors, and can't really tell that they are neutrinos.  On the other hand, we know neutrinos exist in large numbers in the universe (113 neutrinos/cc/family) and that they are not completely understood in the context of the standard model of particle physics. Observations of the CMB should be able to add significantly to on-going laboratory based experiments.
In the near term, we will be able to tell the number of relativistic species. The physics is described in [5] and [6] and recent results are summarized in Table 2.
Currently there is reasonable agreement with the expectation from particle physics, ${\rm N}_{eff}=3.04$.
In the longer term, through multiple channels, we should be able to limit the sum of the neutrino masses to about 0.06-0.1 eV, not far from the lower limit set by the atmospheric neutrinos of 0.05 eV.

\begin{table}[htdp]
\caption{Effective number of relativistic species.}
\begin{center}
\begin{tabular}{| l | c | c |}
\hline
Measurements & ${\rm N}_{eff}$ & Reference \\
\hline
WMAP7 & $> 2.7$  &  [1]\\
WMAP7+ACT &  $5.3\pm 1.3$ & [3] \\
WMAP7+ACT+BAO+H$_0$ & $4.56\pm0.75$  & [3] \\
WMAP7+SPT &  $3.85\pm0.62$ &[4] \\
WMAP7+SPT+BAO+H$_0$ & $3.86\pm0.42$  & [4]\\
\hline
\end{tabular}
\end{center}
\label{tab2}
\end{table}%

\bigskip

Both ACT and WMAP are team efforts and the work briefly summarized above is the result of the efforts of many.  The collaboration list for WMAP is at 
http://map.gsfc.nasa.gov/mission/team.html  and the collaboration list for ACT is at  http://www.princeton.edu/act/collaborators/.

\bigskip

{\bf References}

\begin{description}

\item[1] E. Komatsu et al. ApJS, 192:18  (2011).
\medskip

\item[2] M. Brown et al. ApJ, 705:978 (2009).
\medskip

\item[3] J.  Dunkley et al. ApJ, 739:52 (2011).
\medskip

\item[4] R. Keisler et al. ApJ, 743:28 (2011). 
\medskip

\item[5] S. Bashinsky \& U. Seljak,  PRD 69:3002 (2011). 
\medskip

\item[6] Z. Hou et al. arXiv:1104.2333  (2011). 
\medskip

\end{description}

\bigskip

\subsection{Prof. George F. SMOOT}

\vskip -0.3cm

\begin{center}

Lawrence Berkeley National Laboratory \& University of California, Berkeley, USA;  \\
Extreme Universe Laboratory,  Moscow State University, Russia; \\
Institute for the Early Universe, Ewha Womans University, Seoul, Korea; \\ 
University of Paris-Diderot, Paris, France \\

\bigskip
GRB Science, Problems and Prospects: \\
{\bf Early Light from Gamma Ray Bursts} \\
a Probe of the Dawn of the Universe \\
GRBs, their Progenitors \& Science \\ 
\end{center}

\medskip

Gamma ray bursts are the most energetic events in the universe and have been measured to z = 8.2 
(GRB 090423) and perhaps to z = 9.4 (GRB 090429B).
In principle,  GRBs can be seen to $ z \sim 12 $ with large detectors.

\begin{figure}[h]
\begin{center}
 \includegraphics[width=17.5cm]{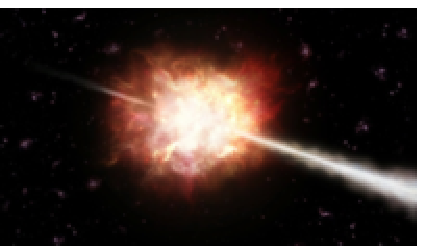}
\caption{Most Distant Object in the Universe to Date: GRB 090423 which has a redshift z = 8.2. 
As soon as Swift detected the gamma-ray burst, other telescopes around the world turned to the same point in the constellation Leo to record the event. 
The telescopes saw a fading afterglow in x-rays but no record of the event in visible light. This figure is an artistic visualization showing the
back-to-back relativisitic jets and the shocked and exploding progenitor. The jet in our direction looks much brighter because of the relativistic boost.}
\end{center}
\end{figure}

Gamma-Ray Bursts (GRB) last between a millisecond  to an hour.
They appear to fall in two categories: (1) the longer duration, softer spectrum GRBs
and (2) shorter duration, harder spectrum GRBs. 
The longer GRBs typically have 90\%\ of their flux in a period longer than 2 seconds, typically 20 seconds,
while the shorter have 90\%\ of their fluence in less than 2 seconds, typically about 0.4 seconds.
Long type GRB are associated with massive star collapse supernovae,
in the sense that a number of them have been observed in the optical and shown to  be at the location of essentially concurrent core collapse supernovae.
For some long-type GRBs,  the afterglow can be detected weeks after burst, with a power law decay light curve in all bands.

\medskip

The short type gamma ray bursts are not well observed as the long type.
Swift, and in smaller numbers HETE-2, have provided the first bona fide short burst X-ray afterglows followed up starting ~ 100 s after the trigger, leading to localizations and redshifts.
However, there is science in being able to see the early light from the GRBs to learn about their mechanisms, including peaking as well as understanding the relativistic jets and internal shocking.
The longer term afterglow for both type GRBs is likely to be similar and due to the external shock mechanism.

\medskip

Most of the detected flux is in the X-rays but some GRBs have gamma-rays measured up to GeV  (rest energy of a proton. They were first discovered in 1973 by their gamma-ray flux and then followed up to find that they provide significantly more photons in the keV range x-rays than in the gamma rays.

\begin{figure}[h]
\begin{center}
\includegraphics[width=17.5cm]{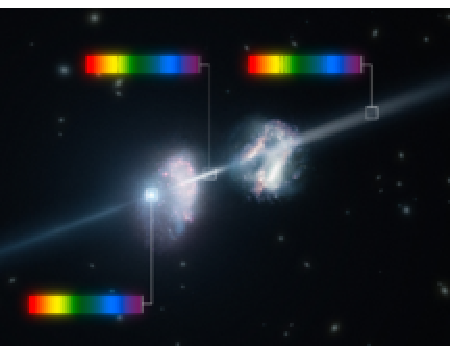}
\caption{An example of how GRB direct and afterglow light can provide information on the early universe. In this artist's illustration, two distant galaxies formed about 2 billion years after the big bang are caught in the afterglow of GRB 090323, a gamma-ray burst seen across the Universe. Shining through its own host galaxy and another nearby galaxy, the alignment of the gamma-ray burst and galaxies was inferred from the afterglow spectrum following the burst's initial detection by the Fermi Gamma Ray Space Telescope in March of 2009. As seen by one of the European Southern Observatory's very large telescope units, the spectrum of the burst's fading afterglow also offered a surprising result - the distant galaxies are richer in heavy elements than the Sun, with the highest abundances ever seen in the early Universe. Heavy elements that enrich mature galaxies in the local Universe were made in past generations of stars. So these young galaxies have experienced a prodigious rate of star formation and chemical evolution compared to our own Milky Way. In the illustration, the light from the burst site at the left passes successively through the galaxies to the right. Spectra illustrating dark absorption lines of the galaxies' elements imprinted on the afterglow light are shown as insets. Of course, astronomers on planet Earth would be about 12 billion light-years off the right edge of the frame. Citation: Savaglio, S., Rau, A., Greiner, J., Krühler, T., McBreen, S., Hartmann, D. H., Updike, A. C., Filgas, R., Klose, S., Afonso, P., Clemens, C., Küpcü Yoldaş, A., Olivares E., F., Sudilovsky, V. and Szokoly, G. (2012), Supersolar metal abundances in two galaxies at z ∼ 3.57 revealed by the GRB 090323 afterglow spectrum. Monthly Notices of the Royal Astronomical Society, 420: 627–636. doi: 10.1111/j.1365-2966.2011.20074.x}
\end{center}
\end{figure}

These bright sources may provide extensive information on the early universe EOR (epoch of re-ionization).  
Therefore, the detection of  significant numbers of GRB in the EOR has an opportunity to make field-changing discoveries about this era - often called the dawn of the universe.
High-z bursts are not extremely faint, i.e. not confined to the faintest 5\%\, 
but they are not so bright that detection would be  automatic.
In order to schedule follow up by large ground-based observatories one has to not only provide accurate time and location (pointing position) but also a strong indication that the GRB is at high redshift.
After four decades of getting priority to interrupt other astronomical observations, traditional astronomy programs, 
are now interested in knowing that there is something important and significant about the GRB and that it is not just another one. 

\bigskip

There are a number of approaches to GRB science in early follow up of the GRB trigger both to see the early UV/optical and IR photons and learn about the GRB mechanism and in detecting the early light from high redshift GRBs which will provide significant understanding about the epoch or reionization.

\bigskip

One approach we have taken is shown by the UFFO-pathfinder (Prof. Il Park PI) [1] to fly on the Lomonosov satellite where a GRB burst detector is combined with a small slewing mirror telescope to allow rapid optical and UV follow up on the order of a second compared to the 100 second scale of SWIFT. There is a certain hit-or-miss to burst detectors in space in coordination with rapid deploying ground-based robotic telescopes. The pathfinder suffers from small size and therefore sensitivity and is intended as a pathfinder and proof of concept for a potentially larger version.  The Lomonosov satellite schedule has been delayed from the original November schedule until at least about June of the following year.

\bigskip

A scaled up set of instruments, named  XIGI, including an infrared camera was proposed to NASA with Dr. Bruce Grossan as PI but no extragalactic astrophysics missions were selected for that AO. 
A new generation concept is being considered. [2]

\bigskip

Another approach is to use a new generation high-resolution GRB burst alert telescopes in conjunction with upgraded robotic telescopes. [3]

\bigskip

A more high-tech and costly approach is to use these high-resolution GRB burst alert telescopes (BAT) in conjunction with large focused X-ray detectors with co-boresighted telescope with UV/optical/IR detectors and get rapid follow up on only a fraction of the GRBs that fortuitously are close by and get to the others more slowly. 
In this manner one gets to deeper reach in red shift and in over all luminosity.
Such an approach obtains as a quality sample.The X-ray and Tracking and Imaging Gamma-Ray Experiment System
(X-TIGRES) [4] is an example.
Such an approach allow for hard Z-ray polarimetry naturally.

\bigskip

There is much science possible in the GRB field. 
Even though we have know of the existence of GRBs for four decades, 
there are still many mysteries about their mechanisms and only now are researchers seriously 
working on using them as probes of our universe.

\bigskip

{\bf References}

\begin{description}

\item[1] P. Chen et al. arXiv:1106.3929,  I. H. Park et al.  arXiv:0912.0773 

\medskip

\item[2] B. Grossan et al. arXiv:1207.5759.

\medskip

\item[3]  Lipunov et al., 2010, Advances in Astronomy, vol. 2010, pp. 1-7, http://observ.pereplet.ru/

\medskip

\item[4]  The X-Tigres Mission Concept  arXiv:1107.5759.

\end{description}

\newpage

\section{Summary and Conclusions of the Colloquium by
H. J. de Vega, M.C. Falvella and N. G. Sanchez}

\subsection{GENERAL VIEW AND CLARIFYING REMARKS}

Participants came from Europe, North, Central and South Americas, Russia, Ukraine, Kazahkstan, Japan, India, Korea. Journalists, science editors and representatives of the directorates of several agencies were present in the Colloquium. Discussions and lectures were outstanding. 

\bigskip

The Standard Model of the Universe was at the center of this Colloquium with the last novelties in the CMB
data, high multipoles, CMB lensing detection, Sunayev-Zeldovich measurements and their
scientific implications, warm dark matter (WDM) advances with both theory and observations, neutrino masses and neutrino oscillations, keV sterile neutrinos as serious WDM candidates, with both theory and experimental search, heliosismology progresses, galaxy observations, clusters  and structure formation, the structure of the interstellar medium and star formation, and the links between these subjects, mainly through gravity and dark matter forming the structures of the cosmic web with the present vaccum energy being the cosmological constant (dark energy). 

\bigskip

Warm dark matter research evolve fastly in both astronomical, numerical, theoretical, particle and experimental research. 
The Colloquium allowed to make visible the work on WDM made by different groups over the world and cristalize 
WDM as the viable component of the standard cosmological model in agreement with CMB + Large Scale
Structure (LSS) + Small Scale Structure (SSS) observations, $\Lambda$WDM, in contrast to 
 $\Lambda$CDM which only agree with CMB+LSS observations and is plagued with SSS problems.  

\bigskip

The participants and the programme represented the different communities doing
research on dark matter:

\begin{itemize}
\item{Observational astronomers}
\item{Computer simulators}
\item{Theoretical astrophysicists not doing simulations}
\item{Physical theorists}
\item{Particle experimentalists}
\end{itemize}

\medskip

WDM refers to keV scale DM particles. This is not Hot DM (HDM). (HDM refers to eV scale DM particles, 
which are already ruled out). CDM refers to heavy DM particles (the so called wimps at the GeV scale or with any 
mass scale larger than the keV). 

\medskip 

It should be recalled that the connection between 
small scale structure features and the mass of the DM particle 
follows mainly from the value of the free-streaming
length $ l_{fs} $. Structures 
smaller than $ l_{fs} $ are erased by free-streaming.
WDM particles with mass in the keV scale 
produce $ l_{fs} \sim 100 $ kpc while 100 GeV CDM particles produce an
extremely small $ l_{fs} \sim  0.1 $ pc. While the keV WDM $ l_{fs} \sim 100 $ kpc
is in nice agreement with the astronomical observations, the GeV CDM $ l_{fs} $ 
is a million times smaller and produces the existence of too many
small scale structures till distances of the size of the Oort's cloud
in the solar system. No structures of such type have ever been observed.

\medskip
 
Also, the name CDM  precisely refers to simulations with heavy DM particles in the GeV scale.
Most of the literature on CDM simulations, do not make explicit
the relevant ingredient which is the mass of the DM particle (GeV scale wimps in the CDM case). 

\medskip

The mass of the DM particle with the free-streaming length naturally enters in the initial power spectrum used in the N-body simulations and in the initial velocity. The power spectrum for large scales 
beyond 100 kpc is identical for WDM and CDM particles,
while the WDM spectrum is naturally cut off at scales below 100 kpc, 
corresponding to the keV particle mass free-streaming length. In contrast, the CDM spectrum 
smoothly continues for smaller and smaller scales till $\sim$ 0.1 pc, which gives rise to
the overabundance of CDM structures at such scales predicted by CDM.

\medskip

CDM particles are always non-relativistic, the initial velocities
are taken zero in CDM simulations, (and phase space density is unrealistically infinity 
in CDM simulations), while all this is not so for WDM.

\medskip

Since keV scale DM particles are non relativistic for $ z < 10^6 $
they could also deserve the name of cold dark matter, although for historical reasons the name WDM
is used. Overall, seen in perspective today, the reasons why CDM does not work are simple: the heavy wimps 
are excessively non-relativistic (too heavy, too cold, too slow), and thus frozen, which preclude them to 
erase the structures below the kpc scale, while 
the eV particles (HDM) are excessively relativistic, too light and fast, (its free streaming length is too large), 
which erase all structures below the Mpc scale; in between, WDM keV particles produce the right answer.

\bigskip

Discussions and lectures were outstanding. Inflection points in several current 
research lines emerged.  
New important issues and conclusions arised and between them, it worths to highlight:

\bigskip

Results and the current state of missions and ongoing projects
were reported by their teams: WMAP, Atacama Cosmology Telescope, QUIET, Planck, Herschel, SPIRE,
ATLAS and HerMES surveys, the James Webb Telescope, UFFO.

\subsection{CONCLUSIONS}

Some conclusions are:

\begin{itemize}

\bigskip

\item{{\it James E. Webb} is honored with the most powerful space telescope ever designed, capable of observing the early universe within a few hundred million years of the Big Bang, revealing the formation of galaxies, stars, and planets, and showing the evolution of solar systems like ours. It is NASA project, with major contributions from the European and Canadian Space Agencies (ESA and CSA). It will have a 6.6 m diameter aperture (corner to corner), will be passively cooled to below 50 K, and will carry four scientific instruments: a Near-IR Camera (NIRCam), a Near-IR Spectrograph (NIRSpec), a near-IR Tunable Filter Imager (TFI), and a Mid-IR Instrument (MIRI). It is planned for launch in 2018 on an Ariane 5 rocket to a deep space orbit around the Sun - Earth Lagrange point L$_2$, about $1.5 \times 10^6$ km from Earth. The spacecraft will carry enough fuel for a 10 yr mission.  Four key topics were used to guide the design of the observatory:  
(i) The end of the dark ages: first light and reionization requires the largest feasible infrared telescope  (first objects of the universe are faint, rare, and highly redshifted), a wide wavelength range and a powerful multi-object infrared spectroscopy, to determine the physical conditions and redshifts of the earliest objects.(ii) The assembly of galaxies around dark matter concentrations. The interaction between dark matter, ordinary matter, black holes (when and how did they form), stars, winds, and magnetic fields is extraordinarily complex ``gastrophysics''. With its infrared imaging and spectroscopy, JWST will reveal details of galaxy formation and evolution. The JWST will also address some aspects of dark energy and dark matter. It can extend the Hubble measurements of distant supernovae; improve the calibration of the Hubble constant;  extend maps of the dark matter distribution to higher redshift.(iii) The birth of stars and protoplanetary systems uniquely observed in the infrared which complement radio observations very well.(iv) Planetary systems and the origins of life from the bright planets to the faint comets, asteroids, and dwarf planets of the outer solar system. JWST's image quality, field of view, and ability to track moving targets are essential to detect and analyze many such targets, among other feasible observations with JWST in this topic.} 

\bigskip

\item{ Is important to recall the results of the $Herschel$ Gould Belt survey  
of the bulk of nearby ($d \sim 500$~pc) molecular clouds 
 in a variety of star-forming environments, allowing 
to clarify the physical mechanisms of prestellar cores's origin out of the diffuse interstellar medium (ISM), and the  stellar initial mass function (IMF) origin. A profusion of parsec-scale filaments in nearby  ISM molecular clouds is found and an intimate connection between the filamentary structure 
 and the dense cloud core formation process. Remarkably, filaments are omnipresent even in unbound, non-star-forming complexes and all appear to share a common width $\sim 0.1$pc:

In active star-forming regions prestellar cores are located within gravitationally unstable filaments which mass per unit length exceeds the critical value $M_{\rm line, crit} = 2\, c_s^2/G \sim 15\, M_\odot$/pc,  
where $c_{\rm s} \sim 0.2$~km/s is the isothermal sound speed for $T \sim 10$~K.  Core formation occurs in two main steps: First, an intricate ISM filaments network is generated by large-scale turbulence ; second, the densest filaments fragment into prestellar cores by gravitational instability. 

This  {\it explains} the star formation threshold at a gas {\it surface density}  $\Sigma_{\rm gas}^{\rm th} \sim $~130~$M_\odot \, {\rm pc}^{-2} $ found recently in both Galactic and extragalactic cloud complexes  : 
Given the typical  $\sim $~0.1~pc width of interstellar filaments, the threshold $\Sigma_{\rm gas}^{\rm th}$ 
corresponds to within a factor of $< 2$ to the critical mass per unit length  $M_{\rm line, crit}$ above which gas filaments at $T \sim 10$~K are gravitationally unstable.}

\bigskip

\item{Impressive CMB observations and their scientific implications have been reported on small angular scales, i.e.\ at the high-$\ell$ part of the CMB power spectrum. Also the Sunyaev-Zeldovich story continues to be interesting. The key modern frontiers being {\em polarization} and {\em the high resolution temperature power spectrum}.
The current direct limit on the primordial tensor to scalar ratio from QUIET is $r<0.9$ (95\%),thus promising well for the future measurements (with more data and a further frequency) from QUIET. Results on $r$ and  $n_s$, are crucial discriminators of the dynamics of inflationary models. Constraints from WMAP7 in the $(r,n_s)$ plane have already effectively ruled out a purely  (monomial) quartic inflation potential, $V(\phi)\propto \phi^4$, and the monomial $V(\phi)\propto \phi^2$ is under pressure. 

\medskip

High-$\ell$ power spectra were released from the Atacama Cosmology Telescope (ACT), and the South Pole Telescope (SPT): The CMB power spectrum impressively displays now 9 peaks clearly discerned. Various degeneracies in cosmic parameters can be now resolved with CMB-alone data. The high-$\ell$ part can also better constraint: (a) the tilt of the primordial spectrum, $n_s$; (b) a possible running  $n_{\rm run}$; (c) the primordial Helium abundance $Y_p$; and (d) the effective number of neutrino species at decoupling, $N_{\rm eff}$. The small-scale CMB can now be used to probe late-time physics  secondary effects.

ACT observed primordial helium at high significance, and detected the CMB lensing (photons gravitationally deflected-a few arcminutes-by large scale structures): lensing directly smooths the acoustic peaks and increases the small-scale power, and couples modes of different scales generating a non-zero lensing 4-point function  whose measure led to a 4$\sigma$ detection of the ACT lensing power: the first direct detection of CMB lensing. The CMB alone can now provide evidence for a dark energy component in the universe.

Amplitudes of the Sunayev-Zeldovich effect in clusters measured by WMAP were below a factor of 0.5-0.7 with respect to the standard X-ray models expectations for temperature and profile. However for the much larger sample of clusters surveyed by Planck (first released results) no such deficit has been found. It is unclear currently how to explain the apparent discrepancy in findings, and suggests that there is still quite a bit to learn in this subject.
ACT cluster number counts and scaling relations between cluster mass and SZ signal: further multi-wavelength observations are expected to better determine cluster masses and constraints. 

\medskip

The maps from the balloon-borne Spider experiment around Antarctica could be of great interest as providing $B$-mode polarisation observations which can be combined with those from the Planck Satellite. As well as results from both the {\em KECK} array, the BICEP2 in the South Pole and ACTPol, this last with improved sensitivity and polarization capabilities to measure the primordial power spectrum in polarization and the improved lensing signal.}

\bigskip

\item{The primordial CMB fluctuations are almost gaussian with negligeable non-gaussianity. The
effective theory of inflation \`a la Ginsburg-Landau, which is a poweful predictive approach allowing
to describe a physically motivated theory-  predicts negligible primordial 
non-gaussianity, negligible running scalar index and the 
tensor to scalar ratio $ r $  in the range $0.021 < r < 0.053$, with the best value $\sim 0.04-0.05$ at reach of the next CMB observations. In this approach all observables turn naturally predicted together in a series expansion in orders of $ 1/N$ , the Number of e-folds of inflation: the running in $k$
of the spectral index $ n_s $ and the non-gaussianity $f_{NL}$ are are both of order
$O(1/N^2)$, ($N ~ 60$)are thus negligeable. The values of $ 1 - n_s $ and $ r $ are both of order $0(1/N)$, Thus, almost non-scale invariance and non-zero amount of gravitons $r$ are directly related among them, and directly and naturally related (and expected to be so) because $N$ is finite and of order $60$ as required by inflation to be non eternal and successfull.
Forecasted  $r$-detection probability for Planck with 4 sky coverages is border line. 
Improved measurements on $n_s$ as well as on TE and EE modes will improve 
these constraints on $r$ even if a detection will be lacking. 
Results from Planck on this primordial polarization are expected.}

\bigskip

\item{Linear polarization of the cosmic microwave background (CMB) provides a direct test of inflation. Gravity waves generated during inflation impart a characteristic curl pattern (B-mode) in the linear CMB polarization.
Satisfying the simultaneous requirements of sensitivity, 
foreground discrimination, and immunity to systematic errors to detect such signal
is a technological challenge. The Primordial Inflation Explorer (PIXIE) is planned to detect and characterize 
such polarization signal. PIXIE will reach the confusion noise 
from the gravitational lensing E-mode signal and has the sensitivity and angular response
to measure even the minimum predicted B-mode power spectrum
at high statistical confidence. PIXIE will measure the CMB linear polarization
to sensitivity of 70 nK per $1\deg \times 1\deg$ pixel,
including the penalty for foreground subtraction. Averaged over the cleanest 75\% of the sky,
PIXIE can detect B-mode polarization to 3 nK sensitivity,
well below the 30 nK predicted from large-field inflation models.
The sensitivity is comparable to the ``noise floor'' of gravitational lensing,
and allows robust detection of primordial gravity waves to limit $r < 10^{-3}$ at more than 5 standard deviations. In addition, PIXIE provides a critical test for keV dark-matter candidates:
 primordial dark matter annihilation distorts the CMB spectrum away from the blackbody shape with 
The resulting chemical potential allows to determine the dark matter mass particle in the keV scale.
PIXIE measurements of the $y$ distortion determine the temperature of the intergalactic medium
at reionization.}

\bigskip

\item{Substantial observation efforts are being made to quantify both the statistical (number counts, correlations, spatial distributions...) and the internal
properties (star formation histories, dark matter density profiles,...) of dSph galaxies. 
Recent discovery surveys have accentuated the {\bf Satellite Problem} of CDM: Although some 25 dSph are now known assciated with each of the Milky way and M31, their numbers remain orders of magnitude too
low compared to simple LCDM predictions. Further, the spatial distributions of these satellites are concentrated in sheets/groups, more so than is predicted. Dominant Baryonic Feedback on galaxy formation is required in LCDM to solve the Satellite Problem, and to reproduce something like the observed
galaxy luminosity function. Feedback is not a free parameter, to be
sub-grid fixed, but it is a consequence of the star formation rate, which
is determinable from the chemical abundance distribution function
of the earliest stars. Substantial progress is being made in quantifying the early histories of the local dSph: In all cases, {\it very low star formation rates are required by observation}. This implies : (i)
{\bf Very low baryonic feedback.} (ii)  The field stars
in the Milky Way  were {\bf not} formed in now tidally destroyed dSph.
Appropriate high-resolution simulation efforts conclude
that {\bf baryon feedback does not affect DM structure} and that CDM
still cannot produce realistic galaxies with realistic feedback and
star formation recipes.

\medskip

Direct kinematic probing of dSph density profiles is making
significant progress: Very precise kinematics for faint
stars in dSph, and interestingly, the total mass enclosed within a
half-light radius is a robust parameter.  The existence of multiple populations inside a single dSph
galaxy is used to determine the mass enclosed with the half-light radius of
each population, thus providing several integrated mass determinations
inside a single profile. This new kinematic and chemical abundance work has considerable potential to
provide direct determinations of primordial DM-dominated density profiles.}

\bigskip

\item{Cosmology and oscillation data show that at least one neutrino mass 
should be in the interval 0.04 - 0.30 eV. The smallness of this neutrino mass,
which is realized in a seesaw mechanism,
can be due to the existence of a large mass scale (as GUT scale). 
Lepton mixing may be explained by the tri-bimaximal mixing scheme (TBM).
The issue of the existence of sterile neutrinos becomes crucial in the theory
of neutrino masses and mixing. Both laboratory experiments and cosmological observations
favour a 3+1 scheme with one sterile neutrino in the $ \sim $eV scale.}

\bigskip

\item{Sterile neutrinos with mass in the keV scale (1 to 4 keV) 
 emerge as leading candidates for the dark matter (DM)
particle from theory combined with astronomical observations.

DM particles in the keV scale (warm dark matter, WDM)
naturally reproduce (i) the observed
galaxy structures at small scales (less than 50 kpc), (ii) the observed
value of the galaxy surface density and phase space density
(iii) the cored profiles of galaxy density profiles seen in
astronomical observations.

Heavier DM particles (as wimps in the GeV mass scale) do not
reproduce the above important galaxy observations and run into
growing and growing serious problems (they produce satellites problem,
voids problem, galaxy size problem, unobserved density cusps and other
problems). 

Minimal extensions of the Standard Model of particle physics 
include keV sterile neutrinos which are very weakly coupled to the standard model particles
and are produced via the oscillation of the light (eV) active neutrinos, with their mixing
angle governing the amount of generated WDM. The mixing angle theta between active and sterile neutrinos
should be in the $ 10^{-4} $ scale to reproduce the average DM density in the
Universe.\\  
Sterile neutrinos are necessarily produced out of thermal equilibrium. 
The production can be non-resonant (in the absence of lepton asymmetries) 
or resonantly ennhanced (if lepton asymmetries are present). The usual X ray bound together
with the Lyman alpha bound forbids the non-resonant mechanism in the $\nu$MSM model.}

\bigskip
 
\item{Warm Dark Matter particles feature a non-vanishing velocity dispersion which leads to a 
cutoff in the matter power spectrum as a consequence of free streaming. 
The essential ingredient is the distribution function of the WDM candidate
solution of the collisionless Boltzmann equation from which the transfer function and power spectrum 
are obtained. 

\medskip

Sterile neutrinos with mass in the $\sim \,\mathrm{keV}$ range are suitable warm dark matter 
candidates. These neutrinos can decay into an active-like neutrino and an X-ray photon. 
Abundance and phase 
space density of dwarf spheroidal galaxies constrain the mass to be in the 
$ \sim $ keV range.  Small scale aspects of sterile neutrinos and different mechanisms of their production 
imprint features in the small scale power spectra as WDM acoustic
oscillations on mass scales $ \sim 10^8-10^{9} \, M_{\odot} $.} 

\bigskip

\item{A right-handed neutrino of a mass of a few keV appears as the most interesting candidate to 
constitute dark matter. A consequence should be Lyman alpha emission and absorption at around a few 
microns; corresponding emission and absorption lines might be visible from molecular Hydrogen H$_2$  
and H$_3$  and their ions, in the far infrared and sub-mm wavelength range.  The detection at very 
high redshift of massive star formation, stellar evolution and the formation 
of the first super-massive black holes would constitute the most striking and testable prediction of 
this dark matter particle. This particle would allow star formation very early, near redshift 80.}

\bigskip

\item{Dark matter may be possibly visible (indirectly) via decay in odd properties of energetic particles
 and photons:  The following discoveries have been reported:  
(i) an upturn in the CR-positron fraction, (ii) 
an upturn in the CR-electron spectrum, (iii) a flat radio emission component near the Galactic Center 
(WMAP haze), (iv) a corresponding IC component in gamma rays (Fermi haze and Fermi bubble), (v) the 
511 keV annihilation line also near the Galactic Center (Integral), and most recently, (vi) an 
upturn in the CR-spectra of all elements from Helium (CREAM), with a hint of an upturn for Hydrogen, 
(vii)  A flat $\gamma$-spectrum at the Galactic Center (Fermi), and (viii) have the complete cosmic ray 
spectrum available through $10^{15}$ to $10^{18}$ eV (KASCADE-Grande). All these features can be 
quantitatively well  explained by the action of cosmic rays accelerated in the magnetic winds of very 
massive stars when they explode. This approach does not require any significant free parameters, it is 
older and simpler than adding Wolf Rayet-star supernova CR-contributions with pulsar wind nebula 
CR-contributions, and implies that Cen A is our highest energy physics laboratory accessible to direct 
observations of charged particles. All this allows with the galaxy data to derive the key properties 
of the dark matter particle: this clearly points to a keV mass particle (keV warm dark matter).}

\bigskip

\item{All the observations of cosmic ray positrons and 
electrons and the like are due to normal astrophysical sources and processes, and do not require 
hypothetical decay or annihilation of heavy DM particles. 
The models of annihilation or decay of cold dark matter (wimps) become increasingly 
tailored and fine tuned to explain these normal astrophysical processes and their ability 
to survive observations is more and more reduced.
Pulsar winds are perfect positron sources to power the positron excess. The spectra inferred 
from observations work fine. The data are improved with AMS-02. As a conclusion on this issue it appeared: 
``Let us be careful to get too excited about spectral features (positrons, nuclei,...): Some of these 
features also appear due to fluctuations in the source activity or locations. There is a lot of work 
to be done before we actually figure out the details of CR propagation and acceleration. Excesses 
should be compared with how well we understand such details. This is especially to be kept in mind 
when invoking unconventional explanations to CR excess, 
such as those based on cold dark matter annihilation.
The CDM explanation to the positron excess was not the most natural: The signal from wimps is naturally 
too small but the theory was contrived (leptophilic DM, boost factors, Sommerfeld enhancement, ...) 
for the sole purpose of fitting one set of data (the positron fraction and the absence of 
antiproton anomalies)''.}

\bigskip

\item{Lyman-$\alpha$ constraints
have been often misinterpreted or superficially invoked in the past
to wrongly suggest on a tension with WDM, but those constraints have been by now clarified
and relaxed, and such a tension does not exist: keV sterile neutrino dark matter (WDM) 
is consistent with Lyman-alpha constraints within a 
{\it wide range} of the sterile neutrino model parameters. {\bf Only} for sterile
neutrinos {\bf assuming} a {\bf non-resonant} (Dodelson-Widrow model) 
production mechanism, Lyman-alpha constraints provide a lower bound for the 
mass of about 4 keV. For thermal WDM relics (WDM particles decoupling at
thermal equilibrium) the Lyman-alpha lower particle mass bounds are 
smaller than for non-thermal WDM relics (WDM particles decoupling out of
thermal equilibrium). The number of Milky-Way satellites indicates lower bounds 
between 1 and 13 keV for different models of sterile neutrinos.

\medskip

WDM keV sterile neutrinos can be copiously produced in the supernovae cores. Supernova stringently 
constraints the neutrino mixing angle squared to be $ \lesssim 10^{-9}$ for sterile neutrino masses
$ m > 100$ keV (in order to avoid excessive energy lost) but for smaller sterile neutrino masses the 
SN bound is not so direct. Within the models 
worked out till now, mixing angles are essentially unconstrained by 
SN in the favoured WDM mass range, namely 
$ 1 < m  < 10 $ keV. Mixing between electron and keV sterile neutrinos could help SN explosions, case 
which deserve investigation.}

\bigskip
        
\item {The possibility of laboratory detection of warm dark matter
is extremely interesting. Only a direct detection of the DM particle can give a clear-cut answer
to the nature of DM and at present. At present,  only the {\bf Katrin and Mare experiments}
have the possibility to do that for sterile neutrinos.
{\bf Mare} bounds on sterile neutrinos are placed from the beta decay of Re187 and EC decay of Ho163,
{\bf Mare} keeps collecting data in both. The possibility that {\bf Katrin} experiment can look to 
sterile neutrinos in the tritium decay was discussed.
Katrin experiment have the potentiality to detect warm dark matter if its set-up 
would be adapted to look to keV scale sterile neutrinos.
KATRIN experiment concentrates its attention right now
on the electron spectrum near its end-point
since its goal is to measure the active neutrino mass.
Sterile neutrinos in the tritium decay will affect the electron
kinematics at an energy about $m$ below the end-point of the
spectrum ($m$ = sterile neutrinos mass). KATRIN in the future
could perhaps adapt its set-up to look to keV scale sterile neutrinos.
It will be a a fantastic discovery to detect dark matter
in a beta decay.}

\bigskip

\item{Astronomical observations strongly indicate that
{\bf dark matter halos are cored till scales below 1 kpc}. 
More precisely, the measured cores {\bf are not} hidden cusps.
CDM Numerical simulations -with wimps (particles heavier than $ 1 $ GeV)-
without {\bf and} with baryons yield cusped dark matter halos.
Adding baryons do not alleviate the problems of wimps (CDM) simulations,
on the contrary adiabatic contraction increases the central density of cups
worsening the discrepancies with astronomical observations. 
In order to transform the CDM cusps into cores, the baryon+CDM simulations
need to introduce strong baryon and supernovae feedback which produces a
large star formation rate contradicting the observations.
None of the predictions of CDM simulations at small scales
(cusps, substructures, ...) have been observed. The discrepancies of CDM 
simulations with the astronomical observations at small scales $ \lesssim 100 $ kpc {\bf is staggering}: 
satellite problem (for example, only 1/3 of satellites predicted by CDM
simulations around our galaxy are observed), the surface density problem (the value 
obtained in CDM simulations is 1000 times larger than the
observed galaxy surface density value),  the voids problem, size problem 
(CDM simulations produce too small galaxies).}

\bigskip

\item{The use of keV scale WDM particles in the simulations instead of the GeV CDM wimps, alleviate all the
above problems. For the core-cusp problem, setting
the velocity dispersion of keV scale DM particles seems beyond
the present resolution of computer simulations. 
However, the velocity dispersion is negligible
for $ z < 20 $ where the non-linear regime and structure formation starts.
Analytic work in the
linear approximation produces cored profiles for keV scale DM particles
and cusped profiles for CDM. Model-independent analysis of DM from phase-space density
and surface density observational data plus theoretical analysis
points to a DM particle mass in the keV scale.
The dark matter particle candidates with high mass (100 GeV, `wimps') 
are strongly disfavored, while cored (non cusped) dark matter halos and warm (keV scale mass) 
dark matter are strongly favoured from theory and astrophysical observations.
As a conclusion, the dark matter particle candidates with large mass 
($ \sim 100$ GeV, the so called `wimps') are strongly disfavored,
while light (keV scale mass) 
dark matter are being increasingly favoured both from theory, numerical 
simulations and a wide set of astrophysical observations.}

\bigskip

\item{Recent $\Lambda$WDM N-body simulations have been performed by different groups. 
High resolution simulations for different types of DM (HDM, WDM or CDM), allow to visualize the effects of 
the mass of the corresponding DM particles: free-streaming lentgh scale, initial velocities and associated 
phase space density properties: for masses in the eV scale (HDM), halo formation occurs top down on all scales 
with the most massive haloes collapsing first; if primordial velocities are large enough, 
free streaming erases 
all perturbations and  haloes  cannot form (HDM). The concentration-mass halo relation for mass of
 hundreds eV  is reversed with respect to that found for
CDM wimps of GeV mass. For realistic keV WDM these simulations deserve investigation: 
it could be expected from 
these HDM and CDM effects that combined free-streaming and velocity effects in keV WDM 
simulations could produce 
a bottom-up hierarchical scenario with the right amount of sub-structures (and some scale at 
which transition 
from top-down to bottom up regime is visualized).

\medskip

Moreover, interestingly enough, recent large high resolution $\Lambda$WDM N-body simulations allow to 
discriminate among thermal and non-thermal WDM (sterile neutrinos): Unlike conventional thermal relics, 
non-thermal WDM has a peculiar velocity distribution (a little skewed to low velocities) which translates 
into a characteristic linear matter power spectrum decreasing slowler across the cut-off free-streaming scale 
than the thermal WDM spectrum. As a consequence, the  radial distribution of the subhalos predicted by WDM 
sterile neutrinos remarkably reproduces the observed distribution of Milky Way satellites in the range above 
$\sim 40$ kpc, while the thermal WDM supresses subgalactic structures perhaps too much, by a factor $2-4$ 
than the observation. Both simulations were performed for a mass equal to 1 keV. Simulations 
for a mass larger than 1 keV  (in the range between 2 and 5 keV, say) should still improve these results.}

\bigskip

\item{The predicted $\Lambda$WDM galaxy distribution in the local universe (as performed by
CLUES simulations with a
a mass of $ m_{\rm WDM}=1 $ keV)  agrees well 
with the observed one in the ALFALFA survey. On the
contrary, $\Lambda$CDM predicts a steep rise in the velocity
function towards low velocities and thus forecasts much  more
sources than the ones observed by the ALFALFA survey (both in Virgo-direction as well as in 
anti-Virgo-direction). These results show again the $\Lambda$CDM problems, also shown in the spectrum of 
mini-voids. $\Lambda$WDM provides a natural solution to these problems.  WDM physics effectively acts as a truncation of the $\Lambda$CDM power spectrum. $\Lambda$WDM CLUES simulations with 
1 keV particles gives much better answer than $\Lambda$CDM when reproducing sizes of local 
minivoids. The velocity function of 1 keV WDM Local Volume-counterpart reproduces the observational 
velocity function remarkably well. 
Overall, keV WDM particles deserve dedicated experimental detection efforts and simulations.}

\bigskip

\item{An `Universal Rotation Curve' (URC) of spiral galaxies emerged from  
3200 individual observed Rotation Curves (RCs) and 
reproduces remarkably well out to the virial radius the Rotation Curve of 
any spiral galaxy. The URC is the observational counterpart of the  circular  
velocity profile from cosmological  simulations.
CDM numerical simulations give the NFW cuspy halo profile. A careful analysis 
from about 100 observed high quality rotation curves has now {\bf ruled out} the disk + NFW halo 
mass model, in favor of {\bf cored profiles}. 
The observed galaxy surface density (surface gravity acceleration) appears to be universal within 
$ \sim 10 \% $ with values around $ 100 \; M_{\odot}/{\rm pc}^2 $
, irrespective of galaxy morphology, luminosity and Hubble types, spanning over 14 magnitudes in 
luminosity and mass profiles determined by several independent methods.}

\bigskip

\item{Interestingly enough, a constant surface density  (in this case column density) with value around 
$ 120 \; M_{\odot}/{\rm pc}^2 $ similar to that found for galaxy systems  is found too for the interstellar 
molecular clouds, irrespective of size and compositions over six order of magnitude; this universal surface 
density in molecular clouds
is a consequence of the Larson scaling laws. This suggests the role of gravity on matter (whatever DM or 
baryonic) as a dominant underlying mechanism to produce such universal surface density 
in galaxies and molecular 
clouds. Recent re-examination of different and independent (mostly millimeter) molecular cloud data sets show 
that interestellar clouds do follow Larson law $Mass \sim (Size)^{2}$ exquisitely well, and therefore very 
similar projected mass densities at each extinction threshold. Such scaling and universality should play a key role in cloud structure formation.}

\bigskip

\item{A key part of any galaxy formation process and evolution involves dark matter. Cold gas accretion and 
mergers became important ingredients of the CDM models but they have little observational evidence. 
DM properties 
and its correlation with stellar masses are measured today up to $z =2$; at $z > 2$ observations are much less certain. Using kinematics and star formation rates, all types of masses -gaseous, stellar and dark- 
are measured 
now up to $z =1.4$. The DM density within galaxies declines at higher redshifts.  Star formation is observed 
to be more common in the past than today. More passive galaxies are in more 
massive DM halos, namely most massive DM halos have lowest fraction of stellar mass. CDM predicts high
overabundance of structure today and under-abundance of structure in the past with respect to observations. 
The size-luminosity scaling relation is the tightest of all purely photometric correlations used to 
characterize galaxies; its environmental dependence have been highly debated but recent findings show that the size-luminosity relation of nearby elliptical galaxies is well defined by a fundamental line and is 
environmental independent. Observed structural properties of elliptical galaxies appear simple and with no environmental dependence, showing that their growth via important mergers -as required by CDM galaxy formation- is not plausible. Moreover, observations in brightest cluster galaxies (BCGs) show little changes in the sizes of most massive galaxies since $z =1$ and this scale-size evolution appears closer to that of radio galaxies over a similar epoch. This lack of size growth evolution, a lack of BCG stellar mass evolution is observed too, demonstrates {\bf that major merging is not an important process}. Again, these observations put in serious trubble CDM `semianalytical models' of BCG evolution which require about $70\%$ of the final BCG stellar mass to be accreted in the evolution and important growth factors in size of massive elliptical massive galaxies.}

\bigskip

\item{The galaxy magnitudes: halo radius, galaxy masses and velocity dispersion
obtained from the Thomas-Fermi quantum treatment for WDM fermion masses in the keV scale are
fully consistent with all the observations for all types of galaxies (see Table I, pag 28). 
Namely, fermionic WDM treated quantum mechanically, as it must be, is able to reproduce
the observed DM cores and their sizes in galaxies.

It is highly remarkably that in the context of fermionic WDM, the simple stationary
quantum description provided by the Thomas-Fermi approach is able to reproduce such broad variety of galaxies.

Baryons have not yet included in the present study. This is fully justified for dwarf compact 
galaxies which are composed today 99.99\% of DM. In large galaxies the baryon fraction can
reach values up to  1 - 3 \%. Fermionic WDM by itself produces galaxies and structures in 
agreement with observations for all types of galaxies, masses and sizes. Therefore, the effect of including 
baryons is expected to be a correction to these pure WDM results, consistent with the fact that dark matter 
is in average six times more abundant than baryons.}

\bigskip

\item {WDM quantum effects play a fundamental role in the inner galaxy regions. WDM quantum pressure -due to the combined effect of the Pauli and the Heinsenberg principle- is crucial in the formation of the galaxy cores. Moreover, dwarf galaxies turn to be quantum macroscopic objects supported against gravity due to the WDM quantum pressure. A measure of the compactness of galaxies and therefore of its quantum character is the phase space density.  The phase space density decreases from its maximum value for the
compact dwarf galaxies corresponding to the limit of degenerate fermions till
its smallest value for large galaxies, spirals and ellipticals, corresponding to
the classical dilute regime. On the contrary, the halo radius $ r_h $ and the halo mass $ M_h $
monotonically increase from the quantum (small and compact galaxies) to the classical regime
(large and dilute galaxies).

Thus, the whole range of values of the chemical potential at the origin $ \nu_0
$ from the extreme quantum (degenerate) limit $ \nu_0 \gg 1 $ to the classical
(Boltzmann) dilute regime $ \nu_0 \ll -1 $ yield all masses, sizes, phase space
densities and velocities of galaxies from the ultra compact dwarfs till the
larger spirals and elliptical in agreement with the observations}.

\bigskip

\item{All evidences point to a dark matter particle mass around 2 keV.
Baryons, which represent 16\% of DM, are expected to give a correction to pure WDM results.
The detection of the DM particle depends upon the particle physics model.
Sterile neutrinos with keV scale mass (the main WDM candidate) can be detected in 
beta decay for Tritium and Renium and in the electron capture in Holmiun.
The sterile neutrino decay into X rays can be detected observing DM
dominated galaxies and through the distortion of the black-body CMB spectrum.
The effective number of neutrinos, N$_{\rm eff}$ measured by WMAP9 and Planck satellites
is compatible with one or two Majorana sterile neutrinos in the eV mass scale.
The WDM contribution to  N$_{\rm eff}$ is of the order $ \sim 0.01 $ and therefore
too small to be measurable by CMB observations.

\bigskip

So far, {\bf not a single valid} objection arose against WDM.}

\end{itemize}

\subsection{OVERALL CONCLUSION}

As an overall conclusion, CDM represents the past and WDM represents the future in the DM research. CDM research is more than 20 years old. CDM simulations and their proposed baryonic solutions, 
and the CDM wimp candidates ($ \sim 100$ GeV)
are strongly pointed out by the galaxy observations as the {\it wrong} solution to DM. 

\medskip

Theoretically, and placed in perspective after more than 20 years, the reason why CDM does not work appears simple and clear to understand and directly linked to the excessively heavy and slow CDM wimp, which determines an excessively small (for astrophysical structures) free streaming length, and unrealistic overabundance of structures at these scales.  
On the contrary, new keV WDM research, keV WDM simulations, and keV scale mass  
WDM particles are strongly favoured by galaxy observations and theoretical analysis, they naturally {\it work} and agree with the astrophysical observations at {\it all} scales, (galactic as well as cosmological scales). Theoretically, the reason why WDM works  so well
is clear and simple, directly linked to the keV scale mass and velocities of the WDM particles,
and free-streaming length.  

\medskip

The quantum nature of the DM fermions must be taken into account for WDM. The quantum effects however are negligeable for CDM: the heavy (GeV) wimps behave classically. The quantum pressure of the WDM fermions solves the core size problem and provides the correct observed galaxy masses and sizes covering from the compact dwarfs to the larger and dilute galaxies, spirals, ellipticals. Dwarf galaxies are natural macroscopic quantum objects supported against gravity by the WDM quantum pressure.

\medskip

The experimental search for serious WDM particle candidates (sterile neutrinos) 
appears urgent and important: it will be a fantastic discovery to detect dark matter in a beta decay. 
There is a formidable WDM work to perform ahead of us, these highlights point some of the directions where 
it is worthwhile to put the effort.

\subsection{THE PRESENT CONTEXT AND FUTURE IN DARK MATTER AND GALAXY FORMATION.}

\begin{itemize}

\item{Facts and status of DM research: Astrophysical observations
point to the existence of DM. Despite of that, proposals to
replace DM  by modifing  the laws of physics did appeared, however
notice that modifying gravity spoils the standard model of cosmology
and particle physics not providing an alternative.
After more than twenty five active years the subject of DM is mature, (many people is involved in this problem, 
different groups perform N-body cosmological simulations and on the other hand direct experimental particle 
searches are performed by different groups, an important
number of conferences on DM and related subjects is held regularly). DM research  
appears mainly in three sets:
(a) Particle physics DM model building beyond the standard
model of particle physics, dedicated laboratory experiments,
annhilating DM, all concentrated on CDM and CDM wimps.
(b) Astrophysical DM: astronomical observations, astrophysical models.
(c) Numerical CDM simulations. The results of (a) and (b)
do not agree and (b) and (c) do not agree neither at small scales.
None of the small scale predictions of CDM simulations 
have been observed: cusps and over abundance of substructures
differ by a huge factor with respect to those observed.
In addition, all direct {\it dedicated} searchs of CDM wimps from more than twenty years 
gave {\bf null results}. {\it Something is going wrong in the CDM research and the right answer is: 
the nature of DM is not cold (GeV scale) but warm (keV scale)}.}

\bigskip

\item {Usually,  (in the litterature, conferences…), CDM is « granted » as 
« the » DM .    And  wimps are granted as  « the » DM particle. 
In most work on CDM galaxies and galaxy formation simulations, 
the problems which face CDM, and CDM + baryons, to agree with observations lead to cyclic CDM crisis,  with more epicyclic type of arguments and recipes.  Each time  CDM is in difficulty, recipes  to make it alive for a while are given and so on. CDM galaxy formation  turns around this situation  from more than 20 years, namely, the subject is turning around itself.
(Moreover, such crisis even led to wrongly replace DM by changing
laws of physics... which spoil the Standard Model of the Universe based mainly on the CMB + LSS + General relativity) . 

\medskip

While over the past 22 years, cosmology,  early and late universe, inflation, CMB , LSS, SSS, made progress and clarifications, Galaxy formation becomes an increasingly confusing and « Ptolomeic » subject , a list of recipes or  ad hoc prescriptions, « termed «astrophysical  solutions » or « baryonic solutions » to CDM, with  which exited the subject itself from a scientific physics framework. The problems faced by CDM confronted to observations are then invoked due to the "baryons complexity", which became each time ``more and more complex'' and so complicated  which no hope for reaching conclusion is seen at each any time, appearing as a `scape to the future'. Namely, in CDM dominated galaxies,  baryons with complexes environments and feedbacks need to make the main work against the wrong features produced by CDM. Basically, baryon effects need to be produced and overwelheimed to `destroy' and elliminate the wrong galactic features already produced by CDM from the beginning. CDM is the wrong solution to Galaxies and its Formation. In WDM, all essential properties of galaxies, as masses, sizes, cored density profiles and small scales turn naturally in agreement with observations and baryons are a correction to WDM, as it should be.}

\item{Many researchers continue to work with heavy CDM candidates
(mass $ \gtrsim 1 $ GeV) despite the {\bf staggering} evidence that these
CDM particles do not reproduce the small scale astronomical observations
($ \lesssim 100 $ kpc). Why? [It is known now that the keV scale DM particles naturally produce the
observed small scale structure]. Such strategic
question is present in many discussions, everyday and off of the record (and on the record) talks in the field. 
The answer deals in large part with the inertia (material, intellectual, social, other, ...) that structured research 
and big-sized science in general do have, which involve huge number of people, huge budgets, longtime planned 
experiments, and the `power' (and the conservation of power) such situation could allow to some of 
the research lines following the trend; as long as budgets will allow to run wimp experimental searches and CDM simulations 
such research lines could not deeply change, although they would progressively decline.

\medskip

Notice that WDM was discussed in the 90's, although without enough broad impact. Is only recently that the differences and clarifications between CDM and WDM are being clearly recognized 
and acknowledged.  

\medskip

While wimps were a testable hypothesis at the beginning of the CDM research, 
today the question emerges on the reason(s) why wimps continue to be worked out and `searched' experimentally in spite of the strong astronomical and astrophysical  evidence against them. 

\medskip

Similar situations (although not as such extremal as the today CDM and wimp situation) happened in other branches of physics and cosmology:
Before the CMB anisotropy observations, the issue of structure formation was plugged with several alternative 
proposals which were afterwards ruled out. Also, string theory passed from being considered "the theory of 
everything" to "the theory of nothing" (as a physical theory), as no physical experimental evidence have been obtained and its cosmological implementation and predictions desagree with observations. (In despite of all that, papers on such proposals continue -and probably will continue- to appear. 
But is clear that  big dedicated 
experiments are not planned or built to test such papers).  In science, what is today `popular' 
can be discarded 
afterwards; what is today `new' and minoritary can becomes `standard' and majoritarily accepted if verified 
experimentally. }

\end{itemize}

\bigskip

\begin{center}

 {\bf \em ``Things are beginning to hang together, and we can now make quite specific 
predictions as a consequence of the keV DM model.  
If the right-handed neutrino were this particle, star formation and the first super-massive 
black holes could be formed quite early, possibly earlier than redshift 50.  
A confirmation would be spectacular.''}

\medskip

[Peter Biermann, his conclusion of the 14th Paris Cosmology Colloquium Chalonge 2010 in
 `Live minutes of the Colloquium',  arXiv:1009.3494].  

\bigskip

\bigskip

 {\bf \em  ``Examine the objects as they are and you will see their true nature;
look at them from your own ego and you will see only your feelings;
because nature is neutral, while your feelings are only prejudice and obscurity.''}

\medskip

[Gerry Gilmore quoting Shao Yong, 1011-1077 in the 14th Paris Cosmology Colloquium Chalonge 2010
http://chalonge.obspm.fr/Programme\_Paris2010.html, arXiv:1009.3494].

\bigskip

\bigskip

 {\bf \em ``Let us be careful to get too excited about spectral features (positrons, nuclei,...): 
This is especially to be kept in mind when invoking unconventional explanations to CR excess, 
such as those based on cold dark matter annihilation.
The CDM explanation to the positron excess was not the most natural: The signal from wimps is naturally 
too small but the theory was contrived (leptophilic DM, boost factors, Sommerfeld enhancement, ...) 
for the sole purpose of fitting one set of data (the positron fraction and the absence of 
antiproton anomalies).''}

\medskip

[Pasquale Blasi, his conclusion in his Lecture at the  15th Paris Cosmology Colloquium Chalonge 2011.]

\end{center}

\bigskip

\bigskip

\begin{center}

The Lectures of the Colloquium can be found at:

\bigskip

{\bf http://chalonge.obspm.fr/Programme$_{}$Paris2012.html}

\bigskip

The photos of the Colloquium can be found at:

\bigskip

{\bf http://chalonge.obspm.fr/album2012/album/index.html}

\bigskip

The photos of the Open Session of the Colloquium can be found at:

\bigskip

{\bf http://chalonge.obspm.fr/albumopensession2012/index.html}

\end{center}

\bigskip

Best congratulations and acknowledgements to all lectures and participants which 
made the 16th Paris Cosmology Colloquium 2012 so fruitful and interesting, the 
Ecole d'Astrophysique Daniel Chalonge looks forward to you for the next Colloquium of this series:

\begin{center}

{\bf The 17th Paris Cosmology Colloquium 2013 devoted to 

\bigskip

THE NEW STANDARD MODEL OF THE UNIVERSE: LAMBDA WARM DARK MATTER ($\Lambda$WDM) THEORY versus OBSERVATIONS

\bigskip

Observatoire de Paris, historic Perrault building,  24, 25, 26 JULY 2013.

\bigskip

http://www.chalonge.obspm.fr/colloque2013.html}

\end{center}

\section{Photos of the Colloquium}

\bigskip

Photos of the Colloquium are available at:

\bigskip

http://www.chalonge.obspm.fr/colloque2012.html

\bigskip

\begin{figure}[ht]
\includegraphics[height=14cm,width=18cm]{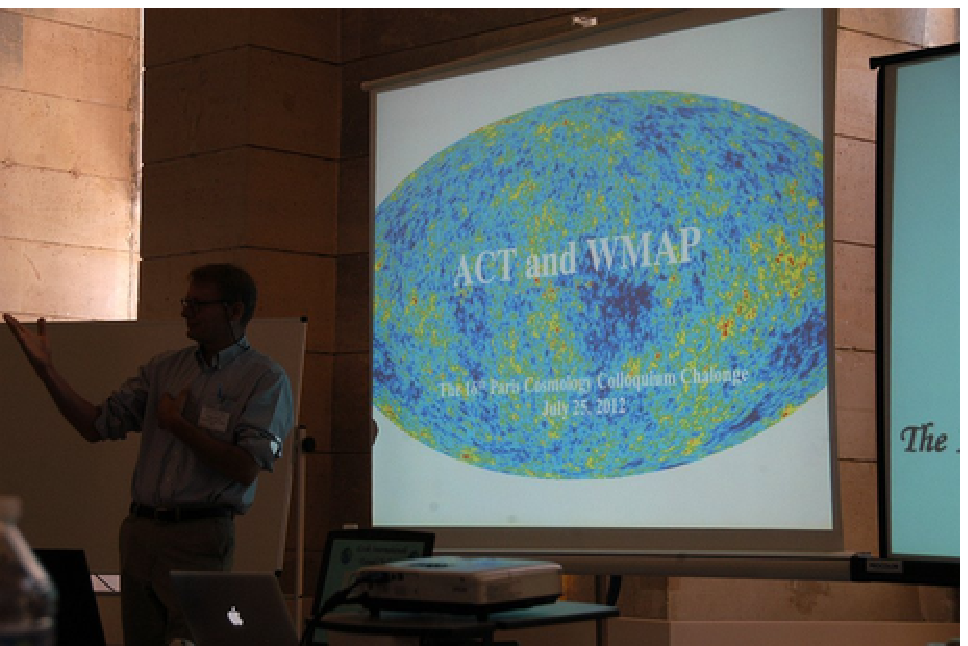}
\caption{Session of the  Chalonge School at the  Salle Cassini}
\end{figure}

\begin{figure}[ht]
\includegraphics[height=14cm,width=18cm]{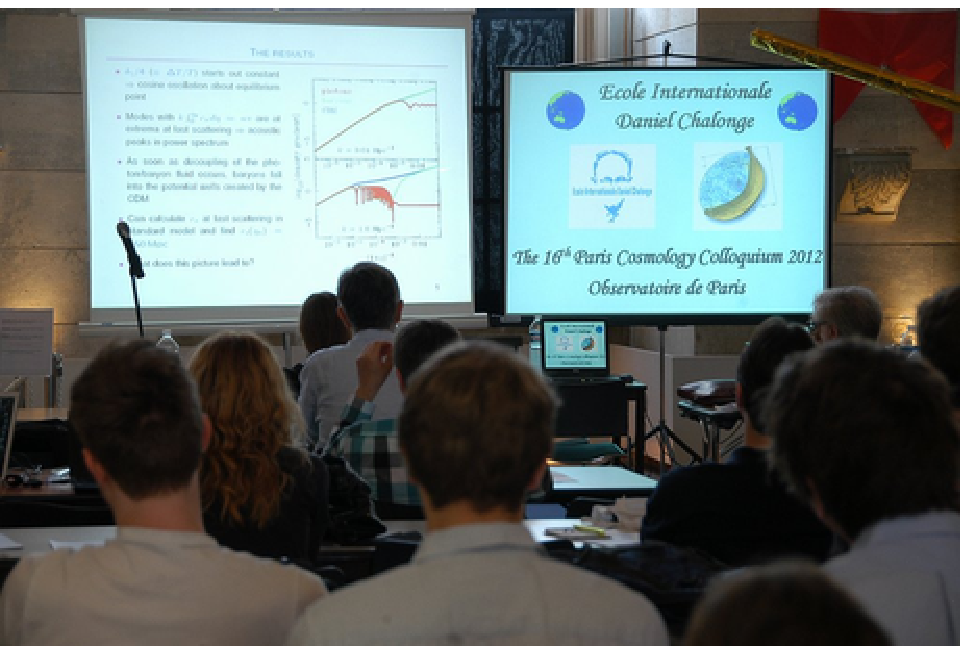}
\caption{Lectures at the Salle Cassini}
\end{figure}

\begin{figure}[ht]
\includegraphics[height=14cm,width=18cm]{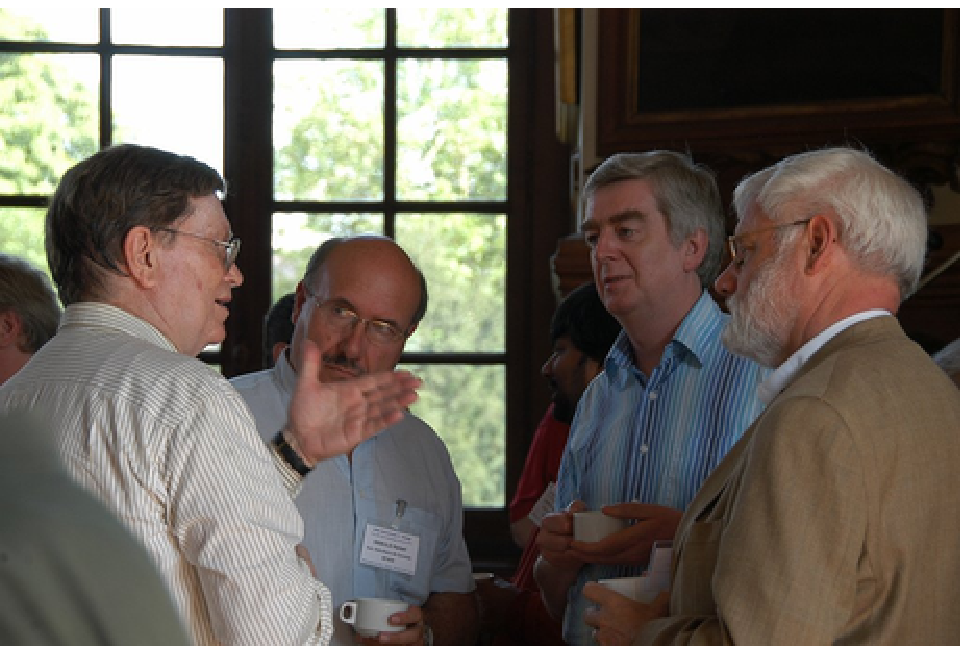}
\caption{Discussion over the Coffee break at the Salle du Conseil}
\end{figure}

\begin{figure}[ht]
\includegraphics[height=14cm,width=18cm]{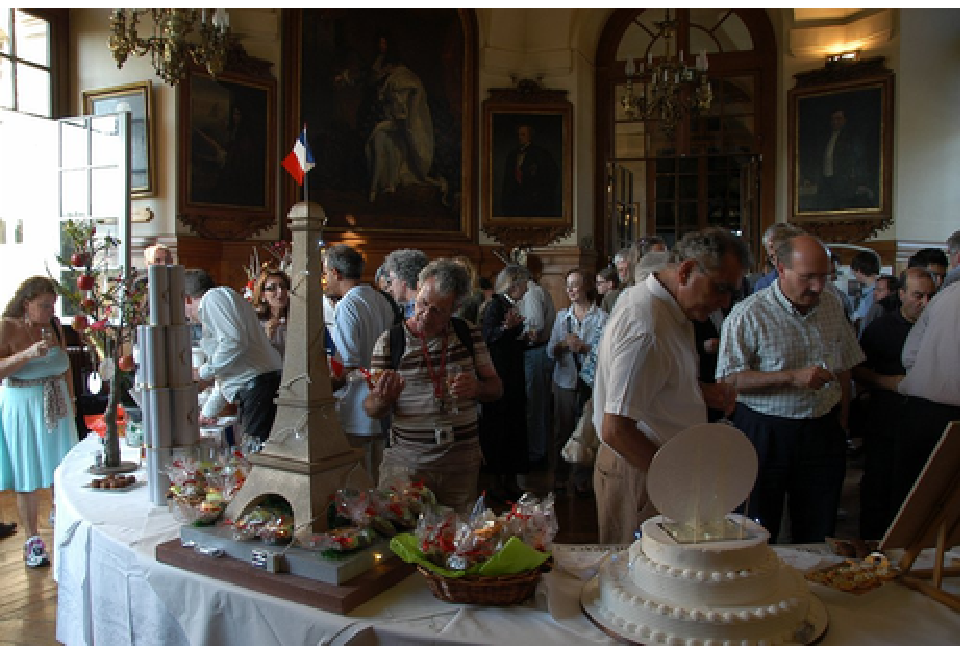}
\caption{Coffee break at the Salle du Conseil}
\end{figure}

\begin{figure}[ht]
\includegraphics[height=14cm,width=18cm]{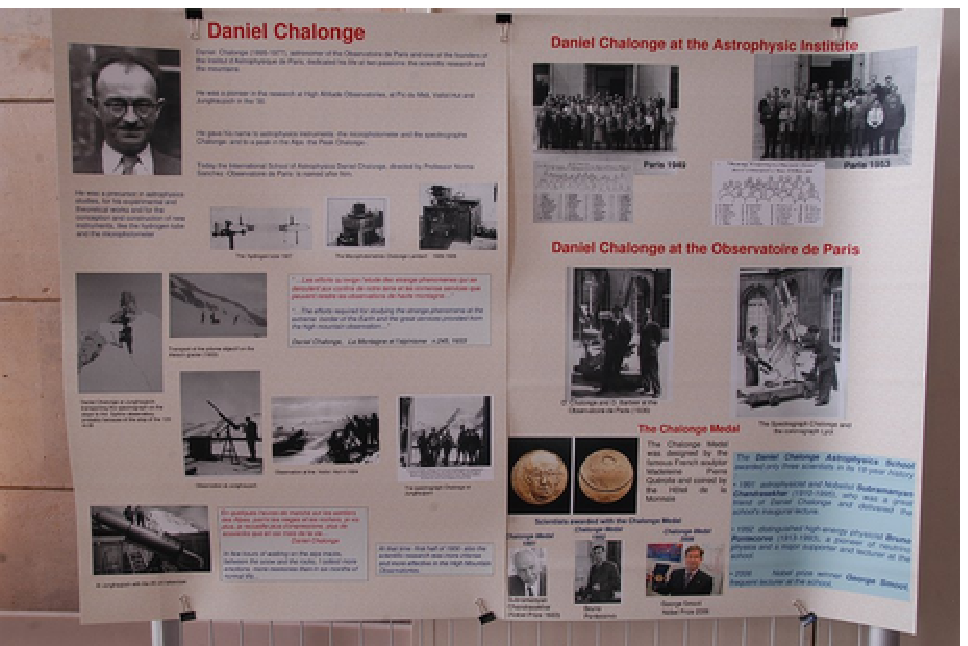}
\caption{A pannel of Chalonge Exhibition}
\end{figure}

\begin{figure}[ht]
\includegraphics[height=14cm,width=18cm]{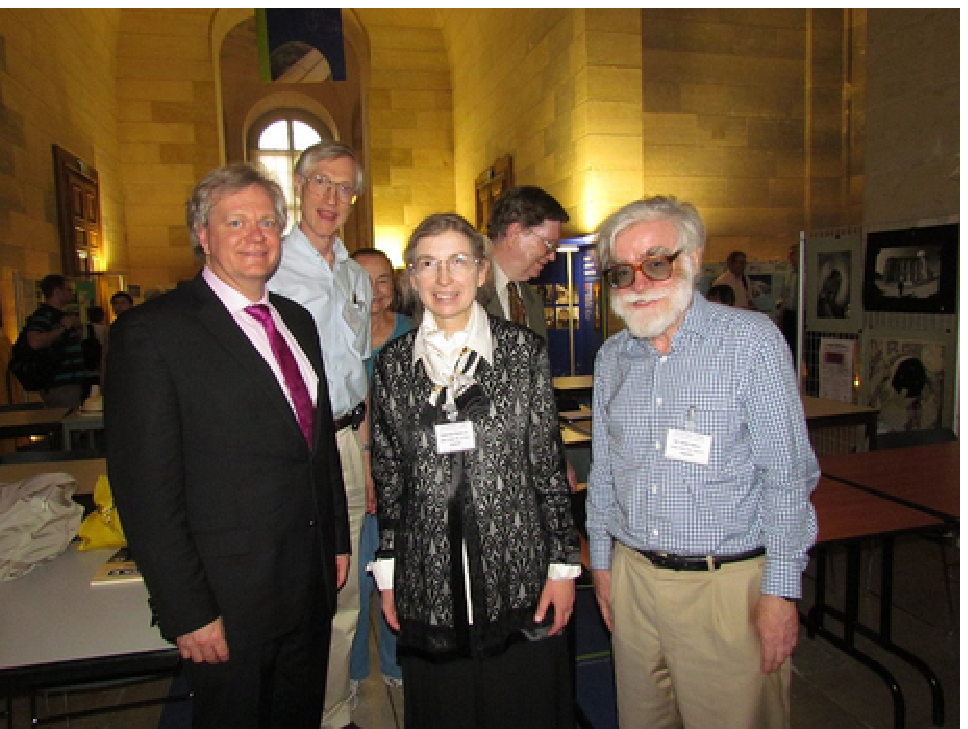}
\caption{Open Session of the Chalonge School at the Salle Cassini}
\end{figure}

\section{List of Participants}

\bigskip

\noindent
\medskip
ALEXANDROVA  Alexandra,	Karl-Francens University Graz,	Graz,	Austria \\
\medskip
BAIDOLDA 	Farida,	IMCCE	Paris	, FRANCE\\
\medskip
BAUYRZHAN 	Mukhametkali,	Universite Nationale Al-Farabi, Kazakh	Almaty,	KAZAHKSTAN \\
\medskip
BIERMANN 	Peter,	MPIfR Bonn, Germany and Univ. Alabama, Tuscaloosa,	Bonn/Tuscaloosa, GERMANY \\
\medskip
BIERMANN 	Joanna,	Accompanying	Bonn/Tuscaloosa, GERMANY \\
\medskip
BOUCHARD	Philippe,	Self	Gatineau, CANADA \\
\medskip
BOULKROUNE	Nassima	Laboratoire de Physique Théorique de l'Université 	Bejaia	ALGERIA \\
\medskip
BOWMAN	Lorraine,	NRAO	Socorro	,USA \\
\medskip
BOZEK	Noemi	AGH University of Science and Technology	Cracow	POLAND \\
\medskip
BURIGANA	Carlo, 	INAF-IASF-Bologna,	Bologna,	ITALY \\
\medskip
CHAKRABARTI	 Sukanya,	Florida Atlantic University,	Boca Raton,	USA \\
\medskip
CHINAGLIA	Mariana,	Universidade Federal de Sao Carlos,	Sao Carlos, BRAZIL \\
\medskip
CNUDDE	Sylvain,	LESIA Observatoire de Paris	Meudon,	FRANCE \\
\medskip
CONSELICE	Christopher, 	Univ. of Nottingham, School of Physics \& Astronomy, Nottingham, UK \\
\medskip
COORAY	Asantha,	University of California, Dept of Physics \& Astronomy	Irvine,	USA \\
\medskip
COURRIER	Christophe, 	IPGP,	Paris,	FRANCE \\
\medskip
DAS	Moumita,	Physical Research Laboratory, Ahmedabad, 	INDIA \\
\medskip
DAS	Saurabh, 	University of Calcutta,	Kolkata,	INDIA \\
\medskip
DAS	Subinoy,	RWTH, Univ of Aachen,	Aachen,	GERMANY \\
\medskip
DAS	Sudeep, 	Berkeley Center Cosmological Physics, BCCP-LNBL	Berkeley, USA \\	
\medskip
DE ROSSI	Maria Emilia,	IAFE (CONICET-UBA), FCEyN 	Buenos Aires,	 ARGENTINA \\
\medskip
DE VEGA	Hector,	CNRS LPTHE Universite Pierre \& Marie Curie Paris VI,	Paris,	FRANCE \\
\medskip
DESTRI	Claudio,	Universita di Milano-Bicocca,	Milano,	ITALY \\
\medskip
DIXIT	Satish Kumat, Raman Institute Raipur,	Raipur CG,	INDIA \\
\medskip
DJOUHRI	Nawel, CRAAG-Centre de Recherches en Astronomie Astrophysics, Bouzareah,	ALGERIA \\
\medskip
DVOEGLAZOV	Valeriy,	Universidad de Zacatecas,	Zacatecas,	MEXICO \\
\medskip
ESCUDERO	Carlos,	Universidad Autónoma de Madrid,	Madrid,	SPAIN \\
\medskip
EVOLI	Carmelo, 	Universität Hamburg,	Hamburg,	GERMANY \\
\medskip
FALVELLA	Maria Cristina,	Italian Space Agency,	Rome,	ITALY \\
\medskip
FAZLOLLAHPOUR	  Sarah,	IPM - Institute for Research in Fundamental Sciences,	Tehran,	IRAN \\
\medskip
FERRARA	Andrea,	Scuola Normale Superiore,	Pisa,	ITALY \\
\medskip
FLIN	Piotr,	Jan Kochanowski Uniwersity, Institute of Physics,	KIELCE,	POLAND \\
\medskip
GIEBELS	Berrie,	LLR Ecole Polytechnique,	Palaiseau,	 FRANCE \\
\medskip
GOMEZ	Liliana,	UIS	Bucaramanga, Bucaramanga,	COLOMBIA \\
\medskip
HAGSTOTZ	Steffen,	Heidelberg University,	Heidelberg,	GERMANY \\
\medskip				
JIMENEZ	Noelia,	UNLP- CONICET,	La Plata,	ARGENTINA \\
\medskip
JIMENEZ-PEREZ	Hugo,	IAP,	Paris,	 FRANCE \\
\medskip
KAO	W.F.,	Chiao Tung University,	Hsin Chu,	TAIWAN \\
\medskip
KIM	Jaisam,	Pohang University of Science and Technology,	Pohang,	South KOREA \\
\medskip
KOEGLMEIER	Julia,	Infineon,	Regensburg,	  GERMANY \\
\medskip
KOOPMANS	Leon,	Kapteyn Astronomical Institute,	Groningen,	NETHERLANDS\\
\medskip
KARIBAYEV	Beibit,	Universite Nationale Al-Farabi, Kazakh	Almaty,	KAZAHKSTAN \\
\medskip
KORMENDY	John, 	University of Texas, Dept of Astronomy, 	Austin,Texas, USA \\
\medskip
KORMENDY 	Mary,	 Accompanying,	Austin, Texas, 	USA \\
\medskip				
LACY-MORA	Gerardo,	Tuorla Observatory, University of Turku,	Turku,	FINLAND \\
\medskip
LALOUM	Maurice,	LPNHE-IN2P3-Paris,	Paris,	FRANCE \\
\medskip
LASENBY	Anthony,	Cavendish Laboratory, University of Cambridge,	Cambridge,	UNITED KINGDOM \\
\medskip
LASENBY	Joan,	Accompanying,	Cambridge,	UNITED KINGDOM \\
\medskip
LATTANZI	Massimiliano,	INFN and Universita' Milano-Bicocca,	Milan,	ITALY \\
\medskip
LAVEDER	Marco,	University of Padua and INFN	Padua,	Padua, ITALY \\
\medskip
LAVEDER	Emilia,	Accompanying,	Padua,	ITALY \\
\medskip
LEMETS	Oleg,	National  Science  Center Kharkov,  Institute of  Physics,	Kharkov,	UKRAINE \\
\medskip
LETOURNEUR 	 Nicole,	CIAS-LESIA Observatoire de Paris, 	 Meudon,	FRANCE \\
\medskip
LIAO	Wei	East China University of Science and Technology	Shanghai	CHINA \\
\medskip
MA	Ernest,	University of California Riverside,	Riverside	USA \\
\medskip
MASTACHE	Jorge,	Instituto de Fisica - UNAM,	Mexico City,	MEXICO \\
\medskip
MATHER	John C.,	NASA Goddard Space Flight Center, 	Greenbelt, MD 	USA \\
\medskip
MATHER, Jane, Accompanying, Greenbelt, MD, USA \\
\medskip
MERCIER	Elsa,	Univ Charles de Gaulle,	Arras,	FRANCE \\
\medskip
MIRABEL 	Félix, 	CEA-Saclay \& IAFE-Buenos Aires 	Gif-sur-Yvette	FRANCE \\ 
\medskip
MONTES	Virginie,		Paris,	FRANCE \\
\medskip
MORFINO	Paolo,	FIDIA,	Turin,	ITALY \\
\medskip
NAURZBAYEVA	Aisha,	Universite Nationale Al-Farabi Kazakh,	Almaty,	KAZAHKSTAN \\
\medskip
NICOLAIDIS	Argyris,	Aristotle University of Thessaloniki,	Thessaloniki,	GREECE \\
\medskip
NURZAT	Shamgun,	Universite Nationale Al-Farabi Kazakh,	Almaty,	KAZAHKSTAN \\
\medskip
OLSON	Elizabeth S., 	Columbia University, 	New York, NY,	USA \\
\medskip
PADOLFI	Stefania,	Dark Cosmology Centre,	Copenhagen,	DENMARK \\
\medskip				
PADUROIU 	Sinziana,	Geneva Observatory,	Geneva,	SWITZERLAND \\
\medskip
PAGE	Lyman, 	Princeton University, Dept. of Physics, 	Princeton,  NJ,	USA \\
\medskip
OLSON 	Elizabeth S.,	Accompanying,	New York,  NY	USA \\
\medskip
PAYEN	Marion,	Villeneuve,	FRANCE \\
\medskip
PULIDO	Victor Alfonso,	Universidad Industrial de Santander,	Bucaramanga,	COLOMBIA \\
\medskip
RAMON MEDRANO Marina,	Univ. Complutense	Madrid, Madrid	SPAIN \\
\medskip
RAUCH	Ludwig,	University of Heidelberg,	Heidelberg,	GERMANY \\
\medskip
REBOLO	Rafael,	Instituto Astrofisico de Canarias, IAC	Tenerife,	SPAIN \\
\medskip
ROSEN	Amnon,	Hemda Centre for Science Education,	Tel Aviv,	ISRAEL \\
\medskip
SARIDAKIS	Emmanuel,	Baylor University, USA	Waco, TX,	USA \\
\medskip
SANCHEZ	Norma G., CNRS LERMA Observatoire de Paris, Paris,	FRANCE \\
\medskip
SCHMIDT	Brian P., Research School Astron.\& Astrophys., Australian Nat.Univ. Weston Creek, AUSTRALIA \\ 
\medskip
SERENELLI Aldo, 	Inst. de Ciències de l'Espai, ICE-CSIC	Barcelone, Belaterra,	SPAIN \\
\medskip				
SINGH	Gajendra Pal,	  M.L.S.University Udaipur,	Udaipur,	INDIA \\
\medskip
SMOOT	George F.,	BCCP LBL Berkeley, IEU Seoul, Univ Paris Diderot, 	Berkeley,	USA \\
\medskip
STEINBRINK	Nicholas	Institut für Kernphysik, WWU Münster,	Münster,	GERMANY \\
\medskip
SUDEEP	DAS,	Berkeley Center Cosmological Physics BCCP-LNBL,	Berkeley, USA	USA \\
\medskip
TEDESCO	Luigi,	Dipartimento di Fisica di Bari  \& INFN di Bari,	 Bari,	ITALY \\
\medskip
TIWARI	Rishi Kumar,	  Dept. of  Mathematics, Govt.P.G.College, Shahdol,	INDIA \\
\medskip
TOGNINI	Michel Ange,	 Général de Brigade de l'Armée de l'Air,	Paris,	FRANCE \\ 
\medskip
TREGUER	Julien,	Observatoire de Paris,	Paris,	FRANCE \\
\medskip
TROMBETTI	Tiziana,	INAF-IASF-Bologna,	Bologna,	ITALY \\
\medskip
VANDECASTEELE	   Lucile,	Université Charles de Gaulle, English  Department, Arras	FRANCE \\
\medskip
VANDECASTEELE	 Mathilde,	Université d'Artois Art,   Department	Arras,  FRANCE \\
\medskip
VELTEN	Hermano,	Bielefeld  Universität,	Bielefeld,	GERMANY \\
\medskip
VENNIK	Jaan,	Tartu Observatory,	Tartumaa,	ESTONIA \\
\medskip
VERMA	Murli,	Lucknow University,	Lucknow,	INDIA \\
\medskip
VILAVINAL 	John Moncy,	St. Thomas College, Kozhencherry, Kerala   	Kozhencherry,	INDIA \\
\medskip
VAANANEN	David,	IPN ORSAY,	Paris,	FRANCE \\
\medskip
WALKER	Matthew,	Harvard University,	Cambridge,	UNITED STATES\\
\medskip
WEHUS	Ingunn Kathrine,	University of Oxford,	Oxford,	UK\\
\medskip
WEINHEIMER	Christian, 	Institut für Kernphysik Universität Münster,	Münster,	GERMANY \\
\medskip
ZANINI	Alba,	INFN-Sezione di Torino,	Turin,	ITALY\\
\medskip
ZIAEEPOUR	Houri,	Max-Planck Institute  E	Garching,	GERMANY\\
\medskip
ZIDANI	Djilali,	Observatoire de Paris - CNRS,	Paris	FRANCE\\

\end{document}